\def\MyApJ#1{}
\def\MyMNRAS#1{#1}

\RequirePackage{rotating}
\documentclass[a4paper,fleqn,usenatbib,amsmath,amssymb]{mnras}

\usepackage{graphicx}
\usepackage{dcolumn}
\usepackage{bm}
\usepackage{epsf}
\usepackage{verbatim}
\usepackage{amsmath}
\usepackage{amssymb}
\usepackage{amsfonts}
\usepackage{ifthen}
\usepackage{epsfig}
\usepackage{verbatim}
\usepackage{dcolumn}
\usepackage{epstopdf}
\usepackage{subfigure}
\usepackage{threeparttable}
\usepackage{booktabs}
\usepackage{calc}
\usepackage{anyfontsize}
\usepackage{relsize}
\usepackage{multirow}
\usepackage{rotating}
\usepackage{comment}

\MyMNRAS{\usepackage{pdflscape}}

\def\myfig#1{./Figures/#1}
\def\mybib#1{./#1}

\def\DrawFig#1{#1}

\newcommand{\njump}{q_{\text{\fontsize{5}{6}\selectfont{$\mbox{n}$}}}}

\newcommand{\Tjump}{q_{\text{\fontsize{5}{6}\selectfont{$\mbox{T}$}}}}

\newcommand{\TjumpP}{q_{\text{\fontsize{5}{6}\selectfont{$\mbox{T,proj}$}}}}
\newcommand{\Zjump}{q_{\text{\fontsize{5}{6}\selectfont{\mbox{$Z$}}}}}
\newcommand{\Pjump}{q_{\text{\fontsize{5}{6}\selectfont{\mbox{$P$}}}}}

\newcommand{\barnjump}{\bar{q}_{\text{\fontsize{5}{6}\selectfont{$\mbox{n}$}}}}
\newcommand{\barTjump}{\bar{q}_{\text{\fontsize{5}{6}\selectfont{$\mbox{T}$}}}}

\newcommand{\barZjump}{\bar{q}_{\text{\fontsize{5}{6}\selectfont{\mbox{$Z$}}}}}
\newcommand{\barPjump}{\bar{q}_{\text{\fontsize{5}{6}\selectfont{\mbox{$P$}}}}}

\newcommand{\myPsi}{\psi}

\newcommand{\RCF}{R_{\mbox{\tiny cf}}}
\newcommand{\RCFbar}{\bar{R}_{\mbox{\tiny cf}}}

\newcommand{\Sx}{S_x}
\newcommand{\MHE}{M_{h}}
\newcommand{\MHEi}{M_{h,i}}
\newcommand{\MHEo}{M_{h,o}}

\newcommand{\PjumpP}{q_{\text{\fontsize{5}{6}\selectfont{\mbox{${P,proj}$}}}}}
\newcommand{\PjumpPbar}{\bar{q}_{\text{\fontsize{5}{6}\selectfont{\mbox{${P,proj}$}}}}}

\newcommand{\rcorei}{r_{\tiny \text{c,i}}}
\newcommand{\rcoreo}{r_{\tiny \text{c,o}}}
\newcommand{\rcf}{r_{\tiny \text{cf}}}

\newcommand{\TtwoDi}{T_{\text{\fontsize{5}{6}\selectfont proj,i}}}
\newcommand{\TtwoDo}{T_{\text{\fontsize{5}{6}\selectfont proj,o}}}

\newcommand{\TtwoD}{T_{\text{\fontsize{5}{6}\selectfont proj}}}
\newcommand{\ZtwoD}{Z_{\text{\fontsize{5}{6}\selectfont proj}}}

\newcommand{\Mydelta}{\Upsilon}
\newcommand{\barMydelta}{\bar{\Upsilon}}
\newcommand{\PTh}{P_t}
\newcommand{\PThi}{P_{t,i}}
\newcommand{\PTho}{P_{t,o}}
\newcommand{\Pnt}{P_{nt}}

\newcommand{\myu}{\mbox{u}}
\newcommand{\alphani}{\alpha_{\tiny \text{n,i}}}
\newcommand{\alphano}{\alpha_{\tiny \text{n,o}}}
\newcommand{\alphaTi}{\alpha_{\tiny \text{T,i}}}
\newcommand{\alphaTo}{\alpha_{\tiny \text{T,o}}}
\newcommand{\alphaZi}{\alpha_{\tiny \text{Z,i}}}
\newcommand{\alphaZo}{\alpha_{\tiny \text{Z,o}}}
\newcommand{\corl}{\mathcal{R}}

\newcommand{\incAng}{\mathsf{i}}
\newcommand{\myz}{\mathsf{z}}

\newcommand{\MyMach}{\mathcal{M}}

\MyApJ{\bibliographystyle{apj}}
\MyMNRAS{\bibliographystyle{mnras}}

\defcitealias{ReissKeshet2014}{RK14}
\newcommand{\RK}{{\citetalias{ReissKeshet2014}}}
\defcitealias{KeshetEtAl2010}{K10}
\newcommand{\UK}{{\citetalias{KeshetEtAl2010}}}
\defcitealias{Keshet2012}{K12}
\newcommand{\UKspirals}{{\citetalias{Keshet2012}}}

\newcommand{\ie}{\emph{i.e.} }
\newcommand{\eg}{\emph{e.g.,} }

\newcommand{\be}{\begin{equation}}
\newcommand{\ee}{\end{equation}}
\newcommand{\bea}{\begin{equation*}}
\newcommand{\eea}{\end{equation*}}
\newcommand{\beqr}{\begin{eqnarray} \nonumber}
\newcommand{\eeqr}{\end{eqnarray}}
\newcommand{\beqrb}{\begin{eqnarray}}
\newcommand{\eeqrb}{\nonumber \end{eqnarray}}
\newcommand{\fin}{\mbox{ .}}
\newcommand{\coma}{\mbox{ ,}}

\newcommand{\cm}{\mbox{ cm}}

\newcommand{\se}{\mbox{ s}}

\newcommand{\Myr}{\mbox{ Myr}}
\newcommand{\Gyr}{\mbox{ Gyr}}
\newcommand{\erg}{\mbox{ erg}}

\newcommand{\kpc}{\mbox{ kpc}}

\newcommand{\keV}{\mbox{ keV}}

\newcommand{\const}{\mbox{const.}}

\newcommand{\grad}{\bm{\nabla}}
\newcommand{\vect}[1]{\mathbf{#1}}
\newcommand{\unit}[1]{\bm{\hat{#1}}}

\usepackage{xcolor}
\definecolor{darkgreen}{rgb}{0.0,0.5,0.0}
\MyApJ{\usepackage[colorlinks,citecolor=darkgreen]{hyperref} 
}

\MyApJ{}
\MyMNRAS{}

\MyApJ{
\begin{document}}

\MyApJ{
\title{
    Deprojecting galaxy-cluster cold fronts: evidence for bulk, magnetised spiral flows
	}
\shorttitle{Magnetized GC spiral flows}}

	\MyApJ{
	\shortauthors{Naor et al.}
	\author{Yossi Naor$^{a,\dagger}$}
	\author{Uri Keshet$^{a,\ddagger}$}
	\author{Qian H. S. Wang$^b$}
	\author{Ido Reiss$^{a,c}$}
	\affil{$^a$Physics Department, Ben-Gurion University of the Negev, POB 653, Be'er-Sheva 84105, Israel}
	\affil{$^b$Department of Physics and Astronomy, University of Utah, 201 James Fletcher Bldg., 115 S 1400 E, Salt Lake City, UT 84112, USA}
	\affil{$^c$Physics Department, Nuclear Research \MyMNRAS{Centre}\MyApJ{Center} Negev, POB 9001, Be'er-Sheva 84190, Israel}
	\thanks{$^{\dagger}$Electronic address: naoryos@post.bgu.ac.il}
	\thanks{$^{\ddagger}$Electronic address: ukeshet@bgu.ac.il}
	}
	
	\MyMNRAS{
    \title[Magnetized GC spiral flows]{Deprojecting galaxy-cluster cold fronts: evidence for bulk, magnetised spiral flows}
	\author[Naor et al.]{
	Yossi Naor$^{a}$\thanks{E-mail: naoryos@post.bgu.ac.il},
	Uri Keshet$^{a}$\thanks{E-mail: ukeshet@bgu.ac.il},
	Qian H. S. Wang$^b$,
    \& Ido Reiss$^{a,c}$
	\\
	$^a$Physics Department, Ben-Gurion University of the Negev, POB 653, Be'er-Sheva 84105, Israel\\
	$^b$Department of Physics and Astronomy, University of Utah, 201 James Fletcher Bldg., 115 S 1400 E, Salt Lake City, UT 84112, USA\\
    $^c$Physics Department, Nuclear Research \MyMNRAS{Centre}\MyApJ{Center} Negev, POB 9001, Be'er-Sheva 84190, Israel
	}
	}

\MyApJ{\date{\today}}

\MyMNRAS{
\pubyear{2019}
\begin{document}
\label{firstpage}
\pagerange{\pageref{firstpage}--\pageref{lastpage}}
\maketitle
}	
	
\begin{abstract}
Tangential discontinuities known as cold fronts (CFs) are abundant in groups and clusters of galaxies (GCs).
The relaxed, spiral-type CFs were initially thought to be isobaric, but a significant, $10\%$--$20\%$ jump in the thermal pressure $P_t$ was reported when deprojected CFs were stacked, interpreted as missing $P_t$ below the CFs (\ie at smaller radii $r$) due to a locally-enhanced nonthermal pressure $P_{nt}$.
We report a significant ($\sim4.3\sigma$) deprojected jump in $P_t$ across a single sharp CF in the Centaurus cluster.
Additional seven CFs are deprojected in the GCs A2029, A2142, A2204, and Centaurus, all found to be consistent (stacked: $\sim1.9\sigma$) with similar pressure jumps.
Combining our sample with high quality deprojected CFs from the literature indicates pressure jumps at significance levels ranging between $2.7\sigma$ and $5.0\sigma$, depending on assumptions.
Our nominal results are consistent with $P_{nt}\simeq (0.1\mbox{--}0.3)P_t$ just below the CF.
We test different deprojection and analysis methods to confirm that our results are robust, and show that without careful deprojection, an opposite pressure trend may incorrectly be inferred.
Analysing all available deprojected data, we also find:
(i) a nearly constant CF contrast $q$ of density and temperature within each GC, monotonically increasing with the GC mass $M_{200}$ as $q\propto M_{200}^{0.23\pm0.04}$;
(ii) hydrostatic mass discontinuities indicating fast bulk tangential flows below all deprojected CFs, with a mean Mach number $\sim0.76$;
and (iii) the newly deprojected CFs are consistent (stacked: $\sim2.9\sigma$) with a $1.25^{+0.09}_{-0.08}$ metallicity drop across the CF.
These findings suggest that GCs quite generally harbor extended spiral flows.
\end{abstract}

\date{Accepted ---. Received ---; in original ---}
	
\MyApJ{\label{firstpage}}

\MyMNRAS{
}
	
\MyApJ{	
}

\MyApJ{	
\maketitle
}
	
\section{Introduction}
\label{sec:Introduction}

In the past two decades, high resolution X-ray imaging of the intracluster medium (ICM) of galaxy clusters (GCs) revealed an abundance of thermal discontinuities known as cold fronts \citep[CFs; for a review, see][]{MarkevitchVikhlinin2007}.
CFs are broadly classified \citep{MarkevitchVikhlinin2007} into merger CFs, such as the contact discontinuity trailing a shock in the Bullet cluster \citep{MarkevitchEtAl02_Bullet}, and core CFs (also known as spiral CFs), identified as tangential discontinuities of substantial shear \citep[][henceforth \UK]{KeshetEtAl2010} found inside or near the core, probably as part of an extended spiral flow \citep[][henceforth \UKspirals]{AscasibarMarkevitch2006, Keshet2012}.
Other, putative types of CFs \citep[\eg][]{BirnboimEtAl10} were not yet detected.
We henceforth focus on core CFs.

As one crosses such a CF outward (\ie with increasing distance from the \MyMNRAS{centre}\MyApJ{center} of the GC), the plasma density sharply drops, and its temperature sharply increases (jumps), over a length scale shorter than the Coulomb mean free path \citep[\eg][]{EttoriFabian2000,VikhlininEtAl2001,WernerEtAl2016B}.
This \MyMNRAS{behaviour}\MyApJ{behavior}, combined with evidence for shear (\UK), is the reason for classifying these CFs as tangential discontinuities.
Note that an opposite discontinuity, in which the density jump with increasing radius, would be Rayleigh-Taylor unstable.
Assuming hydrostatic equilibrium and no additional pressure components, the thermal pressure, $\PTh$, should then be equal on both sides of the CF.
Indeed, such core CFs were initially thought to be isobaric \citep[\eg][]{MarkevitchEtAl2001,TanakaEtAl2006,SandersEtAl2009,NulsenEtAl2013}.

However, some nonthermal pressure $\Pnt$ is nevertheless expected, due to magnetisation and turbulence.
Magnetic fields, in particular, can exert different levels of pressure on each side of the discontinuity.
Such an effect is expected due to shear magnetisation in the fast flows below CFs,
inferred from discontinuities in the hydrostatic mass $M_h$ (\UK).
Indeed, $\gtrsim 10\%$ magnetisation is needed to explain why these CFs are thinner than the Coulomb mean free path and to \MyMNRAS{stabilise}\MyApJ{stabilize} them against Kelvin-Helmholtz instabilities (\UK).
Therefore, as the total pressure $P_{tot}=\PTh+\Pnt$ should be continuous across a CF, some discontinuity in $\PTh$ is expected.
Measuring such a $\PTh$ discontinuity would thus imply an opposite discontinuity in $\Pnt$, gauging in particular the magnetic field \citep[][henceforth \RK]{ReissKeshet2014}.

An accurate determination of the thermal pressure near the CF requires a careful deprojection of the X-ray data, from photon counts in two-dimensional sky coordinates to three-dimensional ($3$D) thermal distributions.
A survey of previously deprojected profiles (\RK), interpolated to the CF position, indicated small discontinuities in $\PTh$.
Namely, the thermal pressure $\PTho$ just outside (\ie above) the CF was found to somewhat exceed its value $\PThi$ just inside (\ie below) the CF.
Here, subscript $i$ ($o$) denotes regions inside (outside) the CF, namely below (above) it.
Let $\Pjump\equiv\PTho/\PThi$ denote the ratio between the thermal pressure just outside and just inside the CF.
(Notice that {\RK} used the notation  $\xi\equiv\Pjump^{-1}$, instead.)
We henceforth refer to a sudden rise in the thermal pressure with increasing distance from the \MyMNRAS{centre}\MyApJ{center} (\ie $\Pjump>1$) as a pressure jump.
{\RK} found pressure jumps with a mean value in the range $1.1\lesssim \barPjump\lesssim 1.3$, depending on the sample selection.
Stacking a sample of deprojected CFs indicated a significant jump, $\barPjump=1.23\pm0.04$.

However, the {\RK} sample of deprojected profiles was insufficient for further studying this effect.
First, the pressure jumps across individual CFs were typically identified at low confidence levels, so even establishing the presence of the effect necessitated a stacking analysis.
It should be pointed out that a CF in the Virgo cluster did show a high significance jump, but this was based on a narrow sector with substructure, and so deemed unreliable by {\RK}. The CF closest to the \MyMNRAS{centre}\MyApJ{center} of the GC A133 also appeared to show a significant ($>4\sigma$) jump; however, the deprojected temperature profile \citep{RandallEtAl2010} was misaligned with the CF, introducing a systematic error in the inferred $\PTh$ jump and its significance.
Second, measuring a $\PTh$ jump at the $\sim10\%$ level may be sensitive to the simplifying assumptions often made when deprojecting CFs.
Such assumptions include adopting a homogeneous or effective \citep[\eg][]{MazzottaEtAl2004} cooling function, and neglecting metallicity gradients, in particular discontinuities in the deprojected metallicity that were demonstrated \citep{SandersEtAl2005} across two CFs in Abell 2204 \citep[and recently, in one CF in the Centaurus cluster;][]{SandersEtAl2016}.
And third, the previous deprojected CF sample was too noisy to test subtle properties, such as the anticipated correlation between the magnetic pressure (inferred from the $\PTh$ jump) and the shear velocity (inferred from the $M_h$ jump; \UK).
In conclusion, a dedicated deprojection of the thermal profiles across CFs is needed in order to critically test the pressure jumps and study their related properties.

We carry out a careful deprojection analysis of CFs from raw X-ray observations, targeting both the local thermal discontinuities and the global thermal structure.
We focus on the GCs Abell 2029, 2142, 2204 (hereafter A2029, A2142, and A2204, respectively), and Centaurus, all of which  present CFs in multiple sectors.
To our knowledge, the spiral CF in A2029 was not previously deprojected.
The analysis of these CFs is supplemented by a study of the available deprojected discontinuities across all CFs in the literature.
We examine the contrast values of the deprojected properties --- density, temperature, pressure, metallicity, hydrostatic mass, and tangential velocity --- across all the analysed CFs, and the correlations between them.

The paper is \MyMNRAS{organised}\MyApJ{organized} as follows.
In \S\ref{sec:Data Reduction}, we describe the observational targets and the data reduction.
In \S\ref{sec:Spectral Analysis}, we outline the spectral analysis methods used to extract the deprojected thermodynamic profiles.
The resultant deprojected thermodynamic properties are discussed in \S\ref{sec:DeproProp}.
The newly deprojected thermal pressure profiles are presented and \MyMNRAS{analysed}\MyApJ{analyzed} in \S\ref{sec:PjumpsSec}.
In addition, in this section we \MyMNRAS{analyse}\MyApJ{analyze} the deprojected thermal pressure discontinuities, presently available in the literature.
In \S\ref{sec:PjumpsProj}, we test how projection effects the inferred pressure discontinuities.
In \S\ref{sec:delta} we discuss the shear flows that are found along the analysed CFs.
Our results are \MyMNRAS{summarised}\MyApJ{summarized} and discussed in \S\ref{sec:Discussion}.
In Appendix~\ref{app:deproj_models}, we examine different deprojection models.
We discuss the correlations between the different CF diagnostics in Appendix~\ref{app:Corl}.

We adopt a concordance $\Lambda$CDM model with a Hubble parameter $H_0=70\,\mbox{km}\,\mbox{s}^{-1}\,\mbox{Mpc}^{-1}$
and a matter fraction $\Omega_m=0.3$.
An angular separation of $1''$ is thus equivalent to a proper distance separation of $1.45$, $1.68$, $2.65$, and $0.22\kpc$, respectively, in A2029, A2142, A2204, and Centaurus.
A $76\%$ hydrogen mass fraction is assumed. Errors are quoted at the single-parameter $1\sigma$ confidence level, unless otherwise stated.

\section{Sample and data reduction}\label{sec:Data Reduction}

The GCs A2029, A2142, A2204, and Centaurus were observed with the \emph{Chandra} Advanced CCD Imaging Spectrometer (ACIS).
Table~\ref{tab:Obsids} presents the observations parameters: the ID (ObsIDs) of each observation, the observation starting date, the total exposure time, and the cleaned total exposure time, as described below.

The GC A2029 (at a redshift $z \simeq 0.0765$) is regarded as one of the most relaxed known clusters \citep[\eg][]{AscasibarMarkevitch2006}. The mass of the cluster, parameterised as the mass within a radius enclosing a mean density $200$ times the critical density of the Universe, is $M_{200}=(9.1\pm5.6)\times10^{14}M_{\sun}$ \citep{GonzalezEtAl2018} for a Navarro-Frenk-White model \citep[NFW; henceforth;][]{NavarroEtAl1997}. It was the first GC to present a clear, continuous spiral pattern, extending far from the \MyMNRAS{centre}\MyApJ{center} \citep{ClarkeEtAl2004,PaternoMahlerEtAl2013}, when subtracting a model from the surface brightness ($\Sx$) image. Across this spiral, the thermal pressure was reported as continuous, motivating its classification as a core CF \citep{PaternoMahlerEtAl2013}.
To our knowledge, this GC was not previously deprojected.
As explained in \S\ref{subsubsec:Sx}, we are able to analyse only the south-western sector, A2029SW, where a CF is observed at a radius $\sim 30\kpc$ from the X-ray peak.
(We use such sector notations --- the cluster identifier with letters designating the cardinal or intercardinal sector direction, henceforth.)

Abell 2142 ($z \simeq 0.0898$) is a relaxed cluster, with no evidence for a recent major merger.
In the outskirts, however, multiple minor mergers are suggested from the complex dynamics, as inferred from a combination of X-ray and the optical band observations \citep[][]{LiuEtAl2018}.
Such minor mergers were previously suggested as responsible for  sloshing in the core of the GC \citep[\eg][]{AscasibarMarkevitch2006,OwersEtAl2011}.
A2142 is massive, with $M_{200}=(1.2\pm0.2)\times 10^{15}M_{\sun}$  \citep{MunariEtAl2014}.
A spiral pattern is visible in this GC, as seen in the projected temperature map \citep{RossettiEtAl2013}.
The GC presents a narrow CF that stretches from west to north (A2142NW) at an angular separation $\myPsi\sim3'$ \citep[\eg][]{MarkevitchEtAl2000}, a southeast (A2142SE) CF at $\myPsi\sim0'.7$  \citep[\eg][]{MarkevitchEtAl2000}, and another CF in the A2142SE sector at $\myPsi\sim10'$, outside the field of view of present \emph{Chandra} observations \citep[see][]{RossettiEtAl2013}.
As explained below, in this GC we can only deproject the A2142NW sector.
This sector was recently deprojected \citep{WangEtAl2018}, however, the metallicity profile was not analysed, and the method utilised (in particular, its density and temperature weights; see \S\ref{subsubsec:Weights}) may be inappropriate for the accurate determination of a CF pressure jump; see discussion in Appendix \ref{app:deproj_models}.

Abell 2204 ($z\simeq0.1524$) is regarded as a very relaxed cluster \citep{ReiprichEtAl2009}, with mass $M_{200}=(7.1^{+3.8}_{-2.6})\times 10^{14}M_{\sun}$  \citep{CorlessEtAl2009}. It presents a spiral pattern seen directly in the $\Sx$ and projected temperature images.
The cluster shows two distinct CFs \citep{SandersEtAl2005}, one to the west (A2204W) and one to the northeast (A2204NE), at angular separations $\myPsi\sim0'.4$ and $\sim0'.2$ from the \MyMNRAS{centre}\MyApJ{center}, respectively.
We analyse all four sectors, A2204W, A2204E, A2204S, and A2204N, each showing a different part of the same spiral CF.
The south and northwest sectors are deprojected here, to our knowledge, for the first time.

The Centaurus cluster (A3526; hereafter Centaurus; $z \simeq 0.0109$) is one of the most well-observed GCs, at various wavelengths, with a mass $M_{200}=(1.6^{+0.3}_{-0.2})\times 10^{14}M_{\sun}$  \citep{WalkerEtAl2013}.
It is a nearby relaxed GC which presents two CFs to the west (A3526W), at $\myPsi\sim1'$ and $\sim3'.3$, and two CFs to the east (A3526E), at $\myPsi\sim1'.5$ \citep[\eg][]{SandersFabian2002,SandersEtAl2016} and one CF that we identify (see \S\ref{subsubsec:Sx}) in at $\psi\sim6'.6$.
Below the inner CF, a complex X-ray structure is visible, strongly correlated with the radio source found in the \MyMNRAS{centre}\MyApJ{center} \citep[][]{TaylorETAl2007}.
A model-subtracted $\Sx$ image and the projected temperature map each indicates the presence of a spiral pattern \citep{SandersEtAl2016}.
We are able to deproject the outer CFs in the A3526E and A3526SW sectors, at $\psi\sim3'.3$ and $\psi\sim6'.6$.
The A3526SW CF was previously deprojected \citep{SandersEtAl2016}, but using data only at small distances above the CF;
to our knowledge, the A3526E sector was not previously deprojected.

In all these GCs, the spiral pattern can be directly inferred from the morphology and position of the CFs, which themselves
are apparent in the $\Sx$ images.

\MyApJ{\begin{table}[h]}
\MyMNRAS{\begin{table}}
\caption{Observations used in this work} 
\centering
\setlength{\tabcolsep}{0.5em} 
{\renewcommand{\arraystretch}{1.5}
\begin{tabular}{| c | c |c| c |c |} 
\hline 
\multirow{3}{*}{GC}& \multirow{3}{*}{ObsID}& Observation&  Total& Cleaned  \\
& & start& exposure  & exposure \\
\multirow{1}{*}{(redshift)}& & date&  (ks)  &  (ks) \\
\hline 
\multirow{1}{*}{A2029} & $891$& $2000$-$04$-$12$& $20.1$& $19.7$\\ (0.0765)
& $4977$& $2004$-$01$-$08$& $78.9$& $78.8$\\
\hline
\multirow{ 3}{*}{A2142} & $5005$& $2005$-$04$-$13$& $45.2$& $44.6$\\\multirow{3}{*}{(0.0898)}
& $15186$& $2014$-$01$-$19$& $91.1$& $90.2$\\
& $16564$& $2014$-$01$-$22$& $45.1$& $44.6$\\
& $16565$& $2014$-$01$-$24$& $21.1$& $20.7$\\
\hline
\multirow{2}{*}{A2204} & $499$& $2000$-$07$-$29$& $10.4$& $9.3$\\\multirow{2}{*}{(0.1524)}
& $6104$& $2004$-$09$-$20$& $9.7$& $9.3$\\
& $7940$& $2007$-$06$-$06$& $78.5$& $77.7$\\
\hline
\multirow{ 12}{*}{Centaurus} & $504$& $2000$-$05$-$22$& $32.5$& $31.1$\\\multirow{12}{*}{(0.0109)}
& $505$& $2000$-$05$-$22$& $10.9$& $9.3$\\
& $4954$& $2004$-$04$-$01$& $90.2$& $89.2$\\
& $4955$& $2004$-$04$-$02$& $45.3$& $44.6$\\
& $5310$& $2004$-$04$-$04$& $50.0$& $49.8$\\
& $16223$& $2014$-$05$-$26$& $181.3$& $180.5$\\
& $16224$& $2014$-$04$-$09$& $42.9$& $42.5$\\
& $16225$& $2014$-$04$-$26$& $30.5$& $30.1$\\
& $16534$& $2014$-$06$-$05$& $56.2$& $56.0$\\
& $16607$& $2014$-$04$-$12$& $46.3$& $45.6$\\
& $16608$& $2014$-$04$-$07$& $34.6$& $34.2$\\
& $16609$& $2014$-$05$-$04$& $83.4$& $83.0$\\
& $16610$& $2014$-$04$-$27$& $17.6$& $17.6$\\
\hline \end{tabular}}\label{tab:Obsids}\vspace{0.2cm}
\begin{tablenotes}
\item	Columns: (1) The GC; (2) The observation ID number; (3) The observation start date; (4) Total exposure; (5) Total exposure after the cleaning process, described in the text.
\end{tablenotes}
\end{table}

The raw data of these GCs is processed in the same way as described in \citet{WangEtAl2016}. For each of the ObsIDs, we reprocess Level = 1 event files, by using the tool
\texttt{acis\_process\_events} of the \emph{Chandra} X-ray \MyMNRAS{Centre}\MyApJ{Center} (CXC) software, CIAO ($4.8$)\footnote{http://cxc.harvard.edu/ciao}. We apply the standard event filtering procedure of masking bad pixels, grade filtering, removal of cosmic-ray afterglow and streak events, and the detector background events identified using the VFAINT mode data.
The ObsIDS are then aligned in the sky coordinates by running the CIAO wavelet source detection routine \texttt{wavedetect} (see CIAO science threads).
Periods of elevated background are identified by using the $2.5 –- 7$ keV light curve in a background region free of cluster emission on the ACIS chip. This is done by excluding circles of radii $6'.6$, $8'.7$,$4'$, and $6'.5$, respectively, for A2029, A2142, A2204, and Centaurus, around the \MyMNRAS{centre}\MyApJ{center} of each GC. We use time bins of $1$ ks and discard bins that have count rates that are different by more than $20\%$ from the mean value. The resulting clean exposures of the above ObsIDs are provided in Table~\ref{tab:Obsids}.

To model the detector and sky background, we use the ACIS blank-sky background data set of the corresponding
epoch \citep[see][]{MarkevitchEtAl2003,HickoxMarkevitch2006}. The VFAINT mode filter and bad pixel masking are applied. The events are then projected onto the sky for each observation, using the task \texttt{make\_acisbg}\footnote{http://cxc.harvard.edu/contrib/maxim/acisbg/}.
The $41$ ms readout time of the CCD introduces \MyMNRAS{artefacts}\MyApJ{artifacts} in the CHIPY direction. These \MyMNRAS{artefacts}\MyApJ{artifacts} are \MyMNRAS{modelled}\MyApJ{modeled} by the tool \texttt{make\_readout\_bg}\footnote{http://cxc.harvard.edu/contrib/maxim/make\_readout\_bg}.
The count rate derived from the background data is then scaled so that it has the same $9.5 -– 12$ keV counts as the observed data. This value is further reduced by $1.32\%$ or $1.28\%$, depending if the observation included five or six chips, respectively, to accommodate the amount of background contained in the readout \MyMNRAS{artefact}\MyApJ{artifact}.
Exposure maps are then created using
Alexey Vikhlinin's tools\footnote{http://hea-www.harvard.edu/~alexey/CHAV/}.
These exposure maps account for the position- and energy-dependent variation in effective area and
detector efficiency \citep{WeisskopfEtAl2002} using a MEKAL model.

The exposure maps and the images of the cleaned event, background, and readout files of the different ObsIDs are co-added in sky coordinates.
After cleaning and co-addition, $\Sx$ images in the $0.8-–7.0$ keV energy band are created by subtracting both background and readout files from the processed event files, and dividing the outcome by the co-added exposure map. The surface brightness in the energy band $[\epsilon_a,\epsilon_b]$ can be expressed as
\begin{equation}
\Sx(\epsilon_a,\epsilon_b)=\frac{N_{\mbox{\scriptsize event}}-N_{\mbox{\scriptsize readout}}-N_{\mbox{\scriptsize background}}}{t_{\mbox{\scriptsize exp}}}\coma
\end{equation}
where $N_f$ is the number of counts in some pixel in file $f$, and $t_{\mbox{\scriptsize exp}}$ is the total exposure time.
In order to account for the energy-dependence of the chip effective area, $\Sx$ images in narrower energy bands are created (as described above) and then summed to give an $\Sx$ image in the $0.8-–7.0$ keV band,
\begin{equation}
    \Sx\left(0.8\, \mbox{keV},7.0\, \mbox{keV}\right)=\sum_k \Sx\left(\epsilon_k,\epsilon_{k+1}\right)\fin
\end{equation}
\MyApJ{Figures}\MyMNRAS{Figs.}~\ref{fig:A2029Figs}, \ref{fig:A2142Figs}, \ref{fig:A2204Figs}, and \ref{fig:A3526Figs} present the cleaned $\Sx$ images in the $0.8 -– 7.0$ keV band of the GCs A2029, A2142, A2204, and Centaurus, respectively.

In each GC, we exclude point sources from our analysis by visually inspecting the $0.8 -– 7.0$ keV $\Sx$ images.

\MyMNRAS{\begin{figure*}
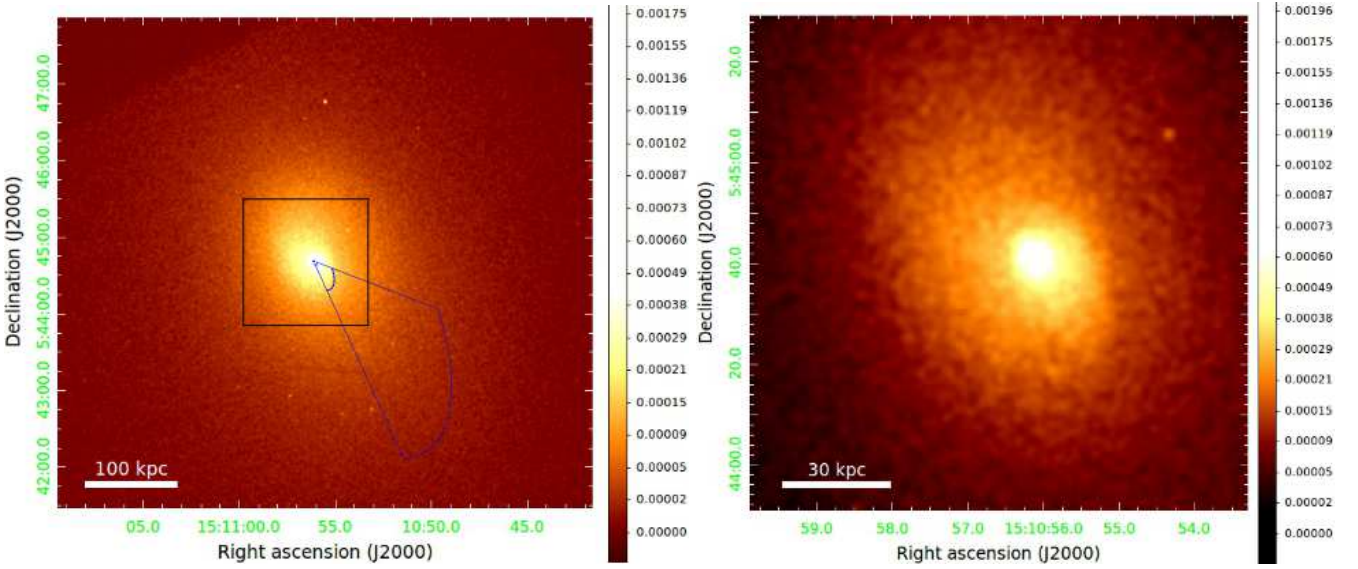

\centering
\DrawFig{
\includegraphics[width=9.0cm]{\myfig{A2029B.eps}}
\includegraphics[width=8.4cm]{\myfig{A2029_ZoomIn.eps}}
}
\caption{\emph{Chandra} surface brightness image of A2029 in the $0.8-7.0$ keV band. The image was exposure corrected, background subtracted, and Gaussian smoothed with a $3''$ radius. A sharp CF is present to the southwest of the GC \MyMNRAS{centre}\MyApJ{center}.  \emph{Left panel}: elliptical fit to the CF morphology (blue dashed line), with its \MyMNRAS{centre}\MyApJ{center} at the X-ray \MyMNRAS{centre}\MyApJ{center}.
The projected temperature and metallicity profiles ($\TtwoD$ and $\ZtwoD$, respectively) are derived within the radial bounds of the sector.
\emph{Right panel}: zoom in to the CF region (marked by black frame on the left panel).}
\label{fig:A2029Figs}
\end{figure*}
}

\MyMNRAS{\begin{figure*}
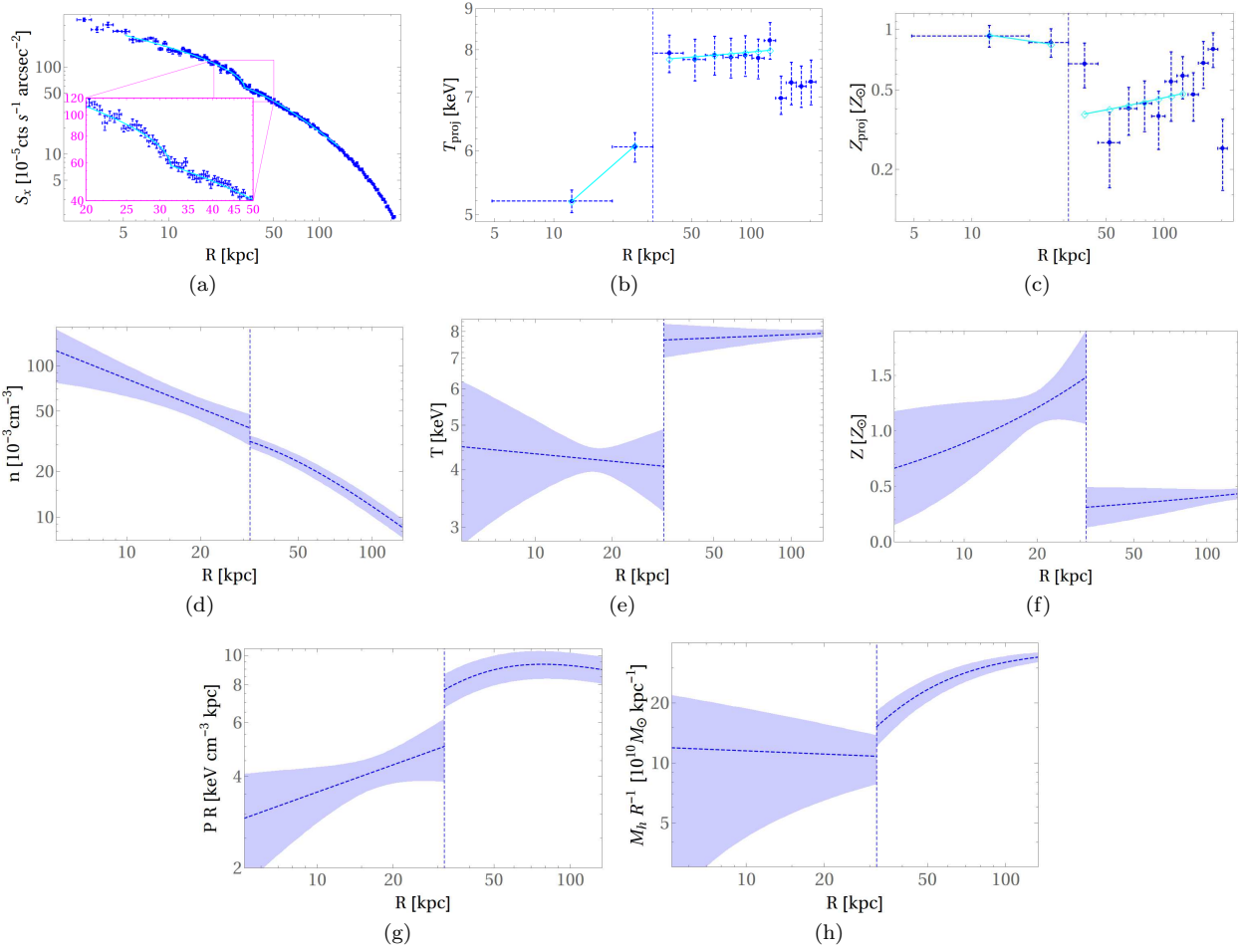

\centering
\subfigure[]
{
\DrawFig{
\includegraphics[width=2.035in]{\myfig{A2029Sx.eps}}
}
\label{fig:A2029Sx}
}
\subfigure[]
{
\DrawFig{
\includegraphics[width=1.995in]{\myfig{A2029T2d.eps}}}
\label{fig:A2029ProjT}
}
\subfigure[]
{
\DrawFig{\includegraphics[width=2.0235in]{\myfig{A2029Z2d.eps}}}
\label{fig:A2029ProjZ}
}\\
\subfigure[]
{
\DrawFig{\includegraphics[width=2.095in]{\myfig{A2029n1z_newncf.eps}}}
\label{fig:A2029n}
}
\subfigure[]
{
\DrawFig{\includegraphics[width=2.065in]{\myfig{A2029T.eps}}}
\label{fig:A2029T}
}
\subfigure[]
{
\DrawFig{\includegraphics[width=2.025in]{\myfig{A2029Z.eps}}}
\label{fig:A2029Z}
}\\
\subfigure[]
{
\DrawFig{\includegraphics[width=2.1in]{\myfig{A2029rP_newncf.eps}}}
\label{fig:A2029rP}
}
\subfigure[]
{
\DrawFig{\includegraphics[width=2.15in]{\myfig{A2029rM_CFside1.eps}}}
\label{fig:A2029rM}
}
\caption{A2029 - radial thermodynamic properties on the plane of the sky, in the SW (dashed blue) sector (see \MyApJ{Figure}\MyMNRAS{Fig.}~\ref{fig:A2029Figs}): (a) the surface brightness; (b) the projected temperature; (c) the projected metallicity; (d) the electron deprojected number density; (e) the deprojected temperature; (f) the deprojected metallicity; (g) the deprojected electron thermal pressure times the radial distance from the \MyMNRAS{centre}\MyApJ{center}; and (h) the deprojected hydrostatic mass divided by the radial distance from the \MyMNRAS{centre}\MyApJ{center}.
Shaded regions represent the confidence intervals of the corresponding curves. Vertical, dot-dashed blue line represents the position of the CF. The contrast values of the deprojected properties at the CF are given in Table~\ref{tab:jumps}. The best-fit models of $\TtwoD$ and $\ZtwoD$ at each bin are presented by a cyan rhombus. The best-fit model of $\Sx$ is presented by a cyan line. }\label{fig:A2029Thermal}
\end{figure*}}

\MyMNRAS{\begin{figure*}
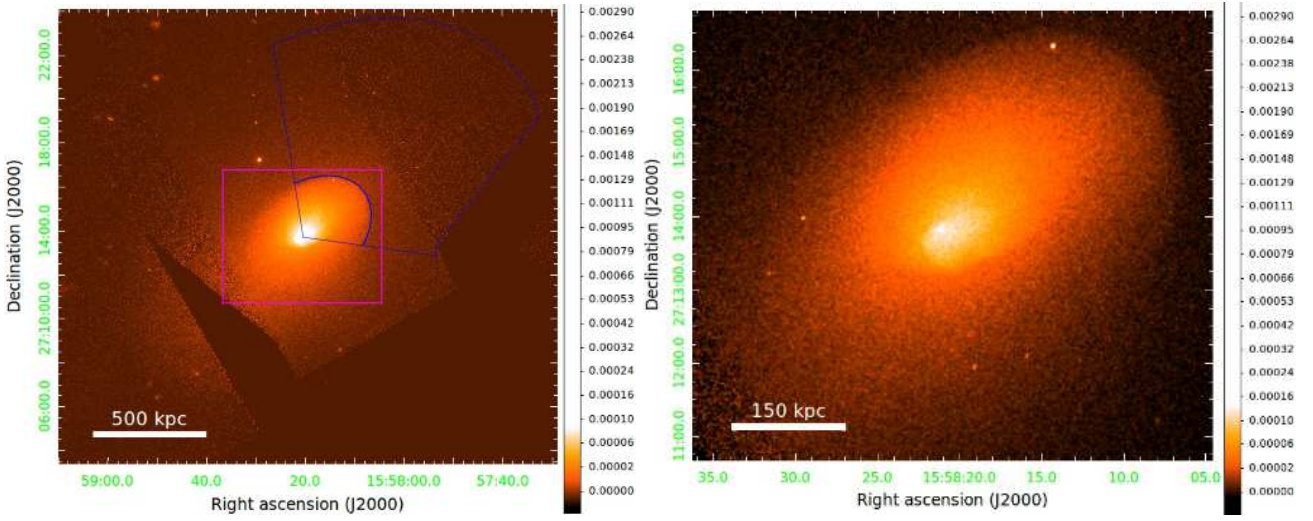

\centering
\DrawFig{
\includegraphics[width=8.3cm]{\myfig{A2142B.eps}}
\includegraphics[width=8.6cm]{\myfig{A2142_ZoomIn.eps}}
}
\caption{\emph{Chandra} surface brightness image of A2142 in the $0.8-7.0$ keV band. The image was exposure corrected, background subtracted, and Gaussian smoothed with a $3''$ radius. A wide and sharp CF is present to the north west of the GC
\MyMNRAS{centre}\MyApJ{center}.
Notations are as in Figure \ref{fig:A2029Figs}.}
\label{fig:A2142Figs}
\end{figure*}}

\MyMNRAS{\begin{figure*}
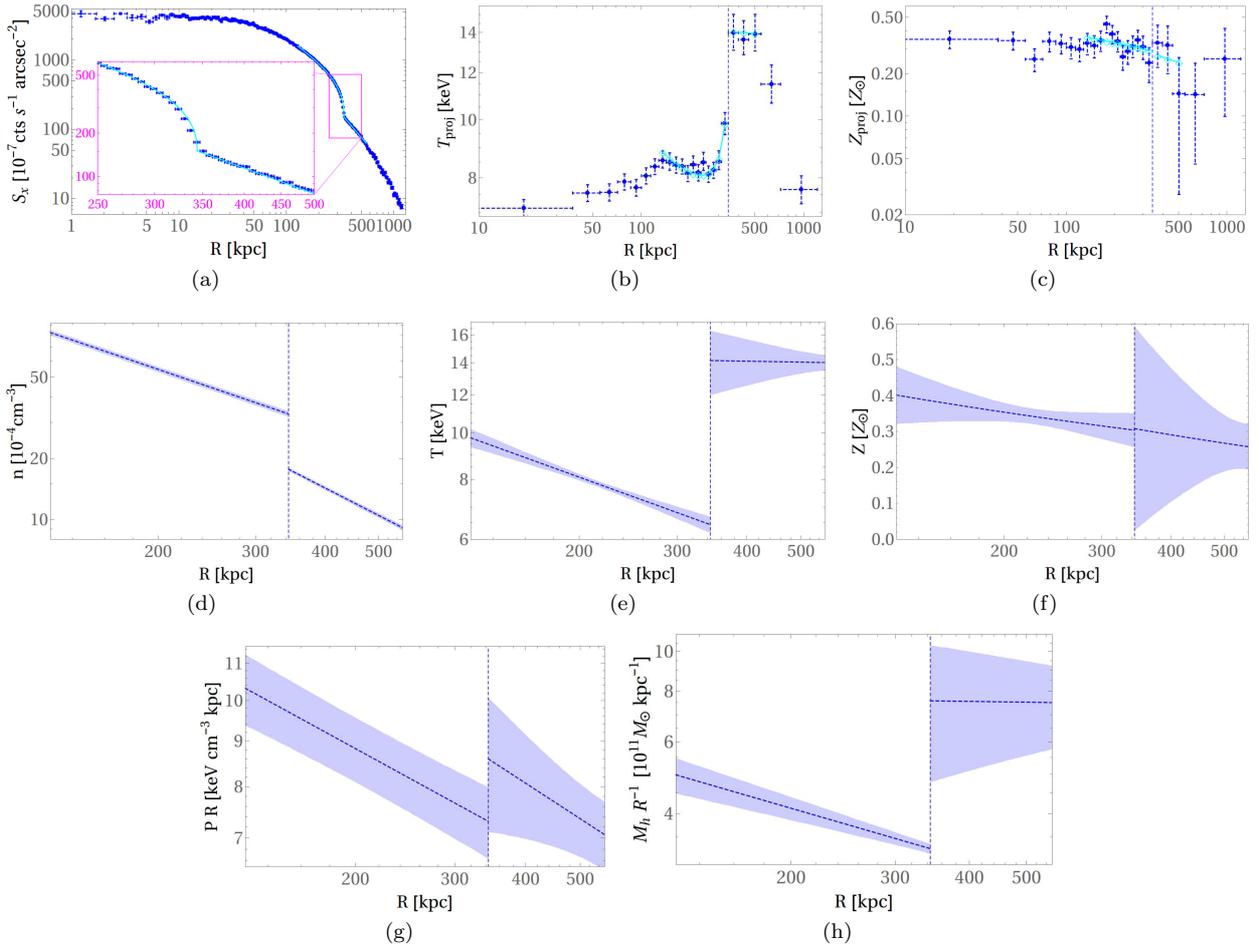

\centering
\subfigure[]
{
\DrawFig{\includegraphics[width=2.095in]{\myfig{A2142Sx.eps}}}
\label{fig:A2142Sx}
}
\subfigure[]
{
	\DrawFig{\includegraphics[width=2.015in]{\myfig{A2142T2d.eps}}}
	\label{fig:A2142ProjT}
}
\subfigure[]
{
	\DrawFig{\includegraphics[width=2.075in]{\myfig{A2142Z2d.eps}}}
	\label{fig:A2142ProjZ}
}\\
\subfigure[]
{
	\DrawFig{\includegraphics[width=2.065in]{\myfig{A2142n1z_newncf.eps}}}
	\label{fig:A2142n}
}
\subfigure[]
{
	\DrawFig{\includegraphics[width=2.065in]{\myfig{A2142T.eps}}}
	\label{fig:A2142T}
}
\subfigure[]
{
	\DrawFig{\includegraphics[width=2.065in]{\myfig{A2142Z.eps}}}
	\label{fig:A2142Z}
}\\
\subfigure[]
{
	\DrawFig{\includegraphics[width=2.1in]{\myfig{A2142rP_newncf.eps}}}
	\label{fig:A2142rP}
}
\subfigure[]
{
\DrawFig{\includegraphics[width=2.2in]{\myfig{A2142rM_CFside1.eps}}}
\label{fig:A2142rM}
}
\caption{A2142 - radial thermodynamic properties on the plane of the sky. Same as \MyApJ{Figures}\MyMNRAS{Figs.}~\ref{fig:A2029Sx}--\ref{fig:A2029rM} for the NW (dashed blue) sector of A2142 (see \MyApJ{Figure}\MyMNRAS{Fig.}~\ref{fig:A2142Figs}). The best-fit models of $\TtwoD$ and $\ZtwoD$ at each bin are presented by a cyan rhombus. The best-fit model of $\Sx$ is presented by a cyan line.}\label{fig:A2142Thermal}
\end{figure*}}

\MyMNRAS{\begin{figure*}
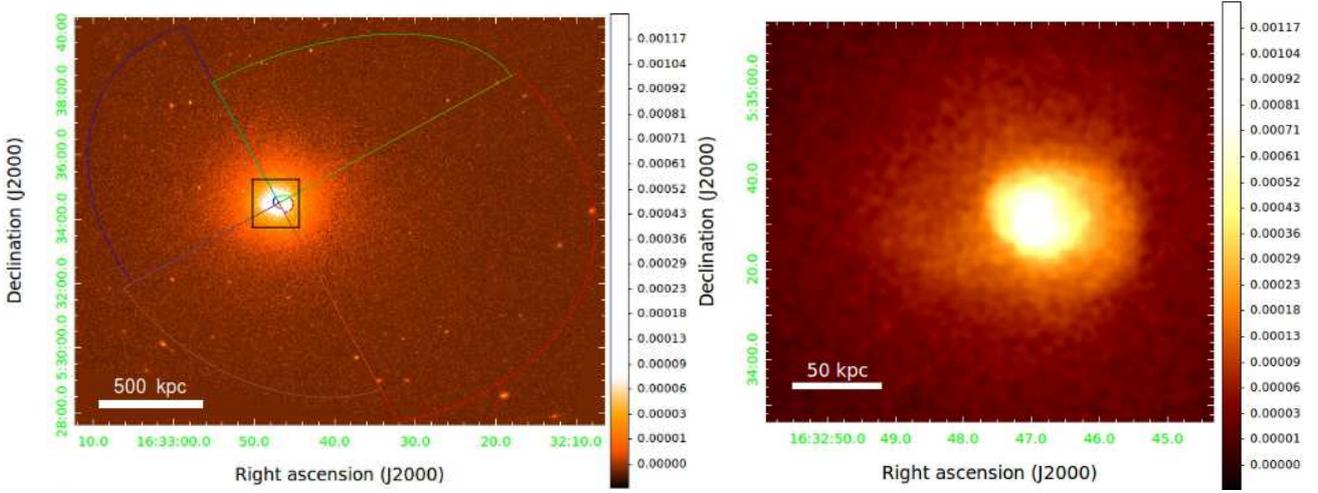

	\centering
	\DrawFig{
	\includegraphics[width=9cm]{\myfig{A2204B.eps}}
	\includegraphics[width=8cm]{\myfig{A2204_ZoomIn.eps}}
	}
\caption{\emph{Chandra} surface brightness image of A2204 in the $0.8-7.0$ keV band. The image was exposure corrected, background subtracted, and Gaussian smoothed with a $3''$ radius. In the image to the left, we divided the GC to four regions and fit a different ellipse to each of the CFs (dashed lines), visible at the west, east, north and south sectors.
Notations are as in Figure \ref{fig:A2029Figs}.}
\label{fig:A2204Figs}
\end{figure*}}

\MyMNRAS{\begin{figure*}
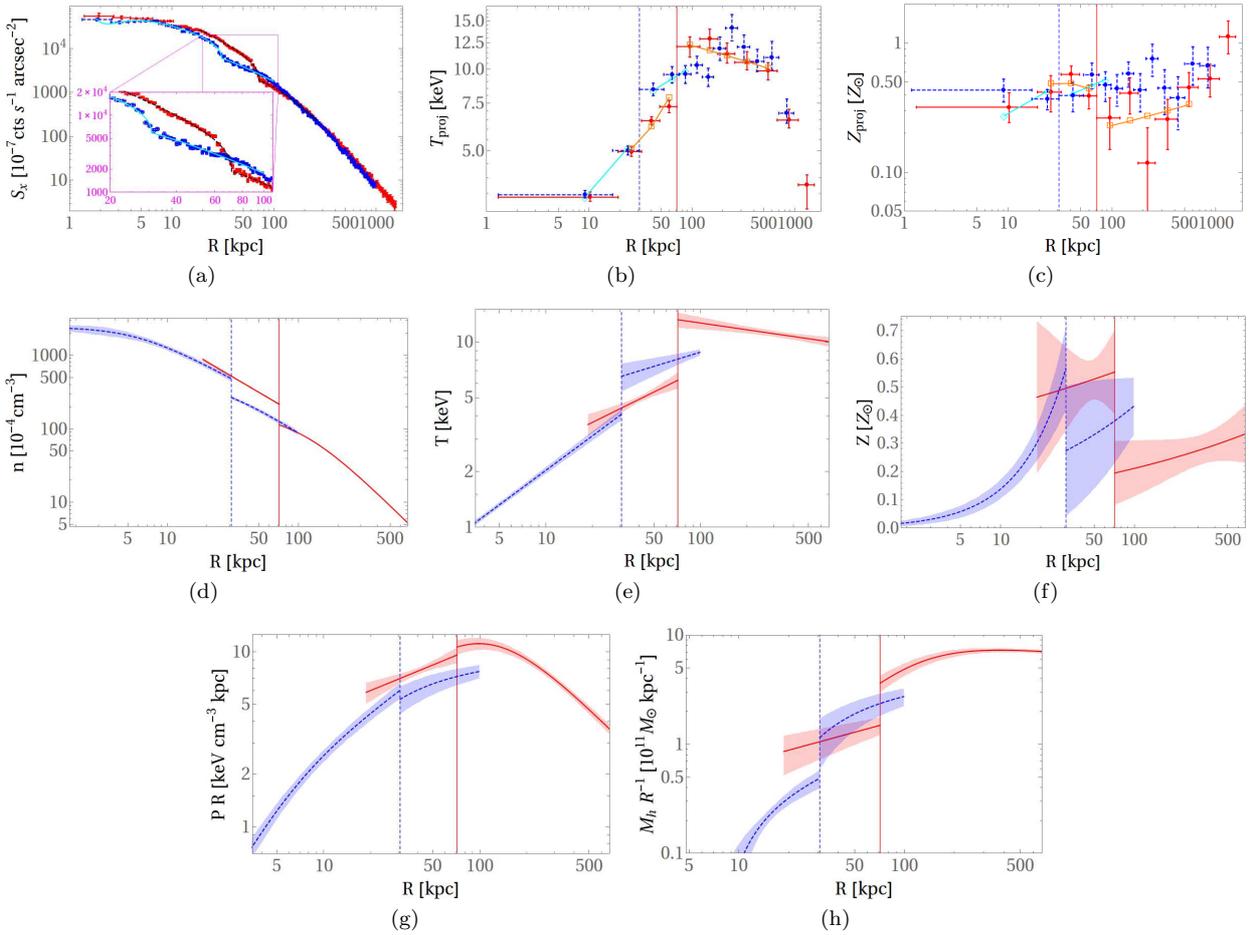

\centering
\subfigure[]
{
\DrawFig{\includegraphics[width=2.065in]{\myfig{A2204SxWE.eps}}}
\label{fig:A2204SxWE}
}
\subfigure[]
{
\DrawFig{\includegraphics[width=2.045in]{\myfig{A2204T2dWE.eps}}}
\label{fig:A2204ProjTWE}
}
\subfigure[]
{
\DrawFig{\includegraphics[width=2.065in]{\myfig{A2204Z2dWE.eps}}}
\label{fig:A2204ProjZWE}
}\\
\subfigure[]
{
\DrawFig{\includegraphics[width=2.095in]{\myfig{A2204nWE1z_newncf.eps}}}
\label{fig:A2204nWE}
}
\subfigure[]
{
\DrawFig{\includegraphics[width=2.065in]{\myfig{A2204TWE.eps}}}
\label{fig:A2204TWE}
}
\subfigure[]
{
\DrawFig{\includegraphics[width=2.025in]{\myfig{A2204ZWE.eps}}}
\label{fig:A2204ZWE}
}\\
\subfigure[]
{
\DrawFig{\includegraphics[width=2.1in]{\myfig{A2204rPWE_newncf.eps}}}
\label{fig:A2204rPWE}
}
\subfigure[]
{
\DrawFig{\includegraphics[width=2.12in]{\myfig{A2204rM1_WE.eps}}}
\label{fig:A2204rMWE}
}
\caption{A2204 - radial thermodynamic properties on the plane of the sky. Same as \MyApJ{Figures}\MyMNRAS{Figs.}~\ref{fig:A2029Sx}--\ref{fig:A2029rM} for the west (red) and east (dashed blue) sectors of A2204 (see \MyApJ{Figure}\MyMNRAS{Fig.}~\ref{fig:A2204Figs}). The best-fit models of $\TtwoD$ and $\ZtwoD$ at each bin are presented by a cyan rhombus (orange square) for the eastern (western) sector. The best-fit model of $\Sx$ is presented by a cyan (dashed black) line for the eastern (western) sector.}\label{fig:A2204WEThermal}
\end{figure*}}

\MyMNRAS{\begin{figure*}
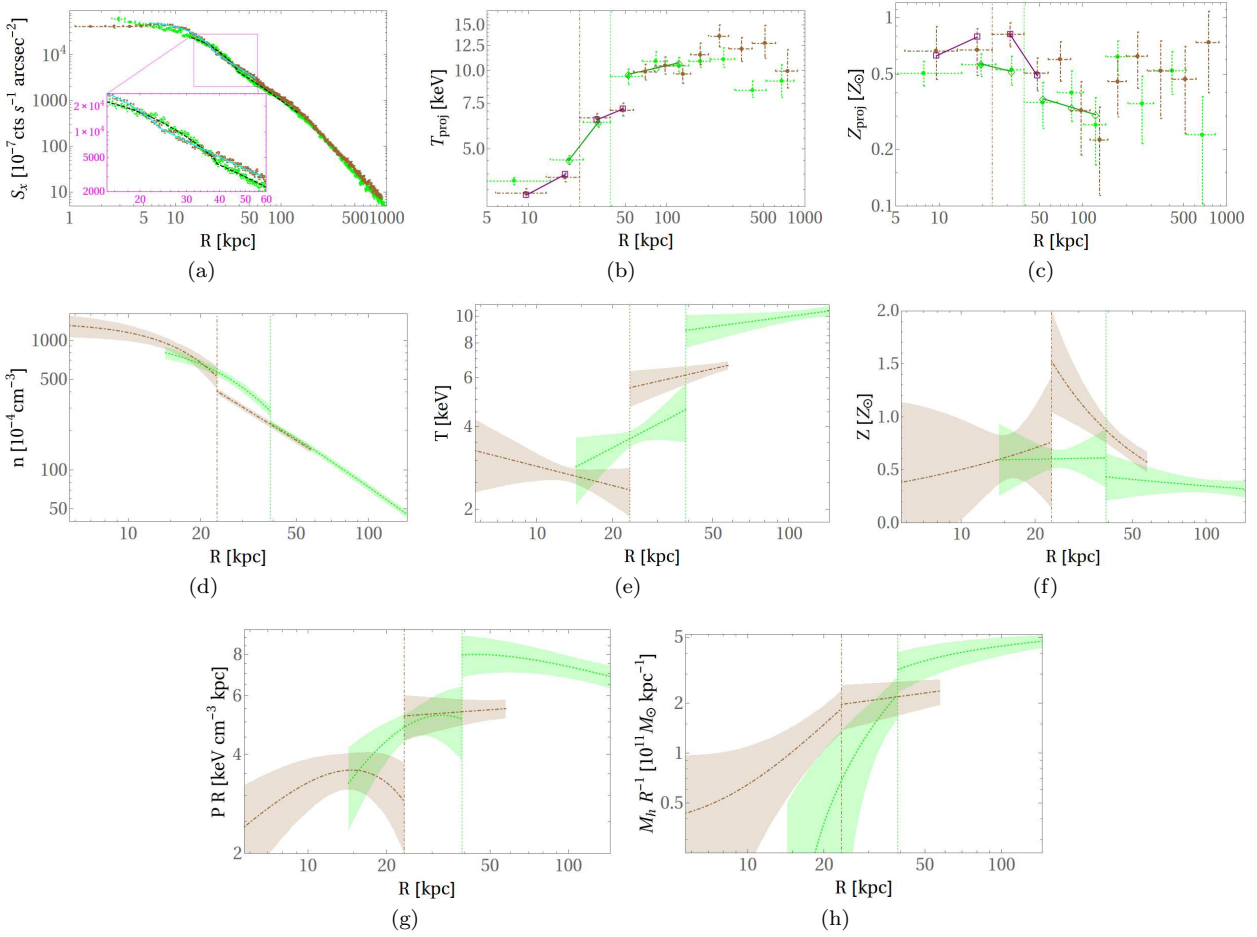

\centering
\subfigure[]
{
\DrawFig{\includegraphics[width=2.065in]{\myfig{A2204SxNS.eps}}}
\label{fig:A2204SxNS}
}
\subfigure[]
{
\DrawFig{\includegraphics[width=2.045in]{\myfig{A2204T2dNS.eps}}}
\label{fig:A2204ProjTNS}
}
\subfigure[]
{
\DrawFig{\includegraphics[width=2.065in]{\myfig{A2204Z2dNS.eps}}}
\label{fig:A2204ProjZNS}
}\\
\subfigure[]
{
\DrawFig{\includegraphics[width=2.095in]{\myfig{A2204nNS1z_newncf.eps}}}
\label{fig:A2204nNS}
}
\subfigure[]
{
\DrawFig{\includegraphics[width=2.065in]{\myfig{A2204TNS.eps}}}
\label{fig:A2204TNS}
}
\subfigure[]
{
\DrawFig{\includegraphics[width=2.025in]{\myfig{A2204ZNS.eps}}}
\label{fig:A2204ZNS}
}\\
\subfigure[]
{
\DrawFig{\includegraphics[width=2.1in]{\myfig{A2204rPNS_newncf.eps}}}
\label{fig:A2204rPNS}
}
\subfigure[]
{
\DrawFig{\includegraphics[width=2.12in]{\myfig{A2204rM1_NS.eps}}}
\label{fig:A2204rMNS}
}
\caption{A2204 - radial thermodynamic properties on the plane of the sky. Same as \MyApJ{Figures}\MyMNRAS{Figs.}~\ref{fig:A2029Sx}--\ref{fig:A2029rM} for the north (dotted green) and south (dot-dashed brown) sectors of A2204 (see \MyApJ{Figure}\MyMNRAS{Fig.}~\ref{fig:A2204Figs}). The best-fit models of $\TtwoD$ and $\ZtwoD$ at each bin are presented by a a green rhombus (purple square) for the southern (northern) sector. The best-fit model of $\Sx$ is presented by a cyan (dashed black) line for the southern (northern) sector.}\label{fig:A2204NSThermal}
\end{figure*}}

\MyMNRAS{\begin{figure*}
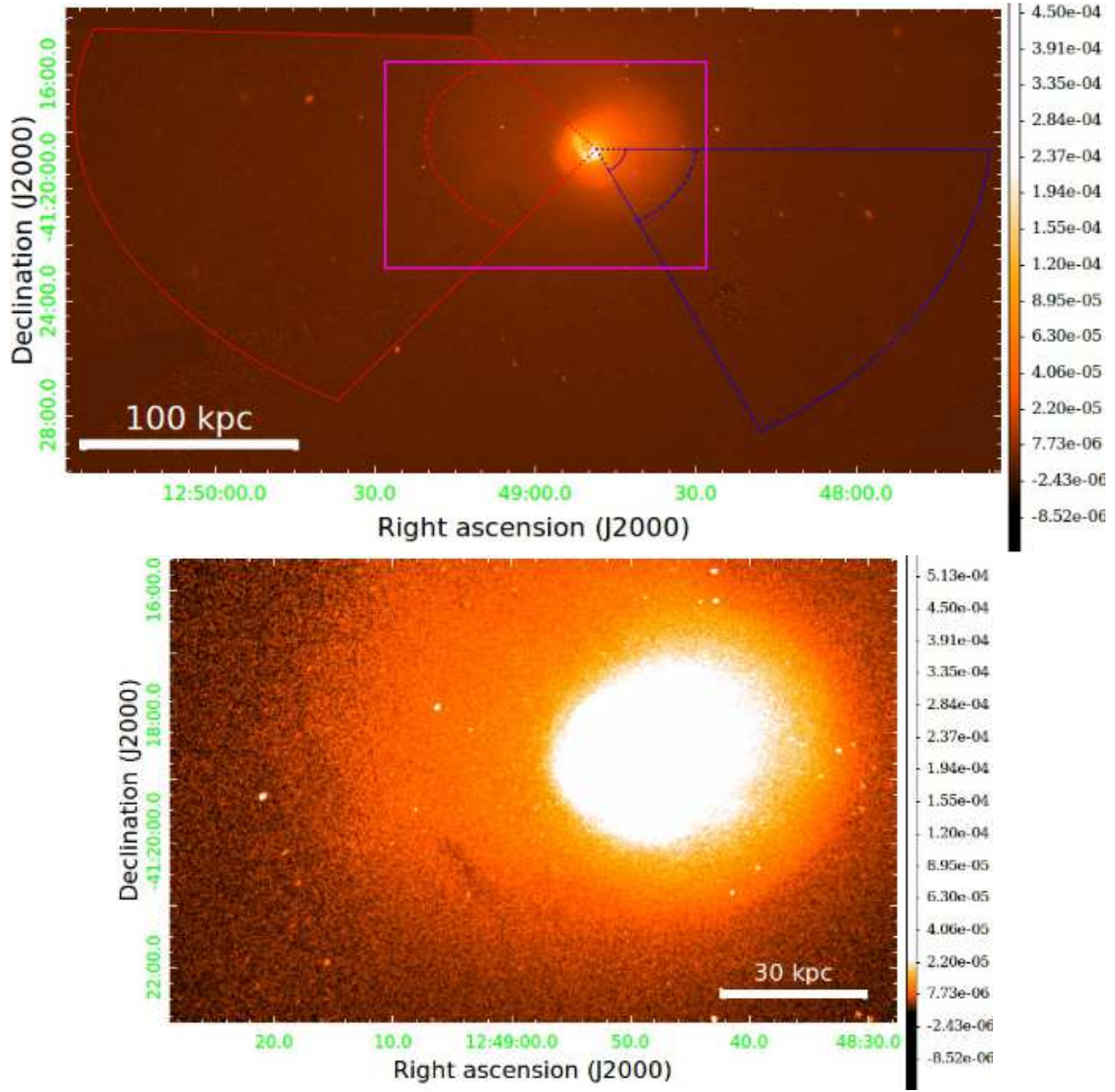

\centering
\DrawFig{
\includegraphics[width=15cm]{\myfig{A3526B.eps}}
\includegraphics[width=12cm]{\myfig{A3526_ZoomIn.eps}}
}
\caption{\emph{Chandra} surface brightness image of Centaurus in the $0.8-7.0$ keV band. The image was exposure corrected, background subtracted, and Gaussian smoothed with a $3''$ radius. Two CFs are visible to the west and two to the east of the GC \MyMNRAS{centre}\MyApJ{center}. The distant CFs from the \MyMNRAS{centre}\MyApJ{center} are those that were \MyMNRAS{analysed}\MyApJ{analyzed}. Notations are as in Figure \ref{fig:A2029Figs}.}
\label{fig:A3526Figs}
\end{figure*}}

\MyMNRAS{\begin{figure*}
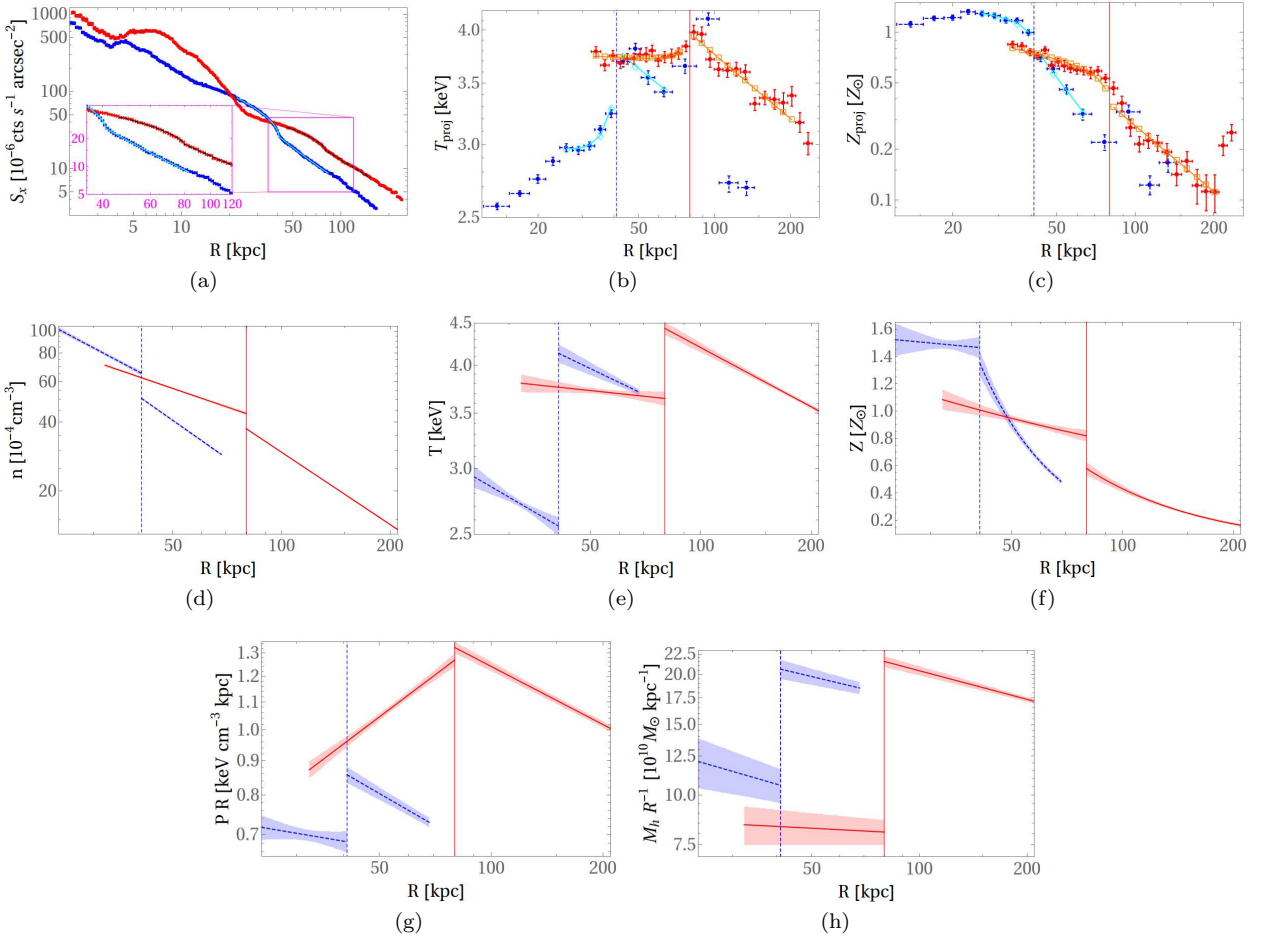

\centering
\subfigure[]
{
\DrawFig{\includegraphics[width=2.095in]{\myfig{A3526Sx.eps}}}
\label{fig:A3526Sx}
}
\subfigure[]
{
	\DrawFig{\includegraphics[width=2.015in]{\myfig{A3526T2d.eps}}}
	\label{fig:A3526ProjT}
}
\subfigure[]
{
	\DrawFig{\includegraphics[width=2.075in]{\myfig{A3526Z2d.eps}}}
	\label{fig:A3526ProjZ}
}\\
\subfigure[]
{
	\DrawFig{\includegraphics[width=2.065in]{\myfig{A3526n1z_newncf.eps}}}
	\label{fig:A3526n}
}
\subfigure[]
{
	\DrawFig{\includegraphics[width=2.065in]{\myfig{A3526T.eps}}}
	\label{fig:A3526T}
}
\subfigure[]
{
	\DrawFig{\includegraphics[width=2.065in]{\myfig{A3526Z.eps}}}
	\label{fig:A3526Z}
}\\
\subfigure[]
{
	\DrawFig{\includegraphics[width=2.1in]{\myfig{A3526rP_newncf.eps}}}
	\label{fig:A3526rP}
}
\subfigure[]
{
	\DrawFig{\includegraphics[width=2.08in]{\myfig{A3526rM1.eps}}}
	\label{fig:A3526rM}
}
\caption{Centaurus -
radial thermodynamic properties, on the plane of the sky. Same as \MyApJ{Figures}\MyMNRAS{Figs.}~\ref{fig:A2029Sx}--\ref{fig:A2029rM} for the western (dashed blue) and eastern (red) sectors of Centaurus (see \MyApJ{Figure}\MyMNRAS{Fig.}~\ref{fig:A3526Figs}). The best-fit models of $\TtwoD$ and $\ZtwoD$ at each bin are presented by a a cyan rhombus (orange square) for the western (eastern) sector. The best-fit model of $\Sx$ is presented by a cyan (dashed black) line for the south-western (eastern) sector.}\label{fig:A3526Thermal}
\end{figure*}}

\section{Thermal property derivation}\label{sec:Spectral Analysis}

In each GC, we extract the spatial profiles of the electron number density, $n$, electron temperature, $T$, and metallicity, $Z$, above and below the identified CFs.
To do so, we use the projected, binned profiles of $\Sx$, the temperature, and the metallicity.

In addition, in order to increase our sample of deprojected CFs, and thus investigate the thermal properties across CFs on a global extent, we extract deprojected thermal profiles across previously deprojected CFs in the literature.

\subsection{Deprojection procedure}\label{subsec:ProjectedProfiles}

In order to derive the deprojected thermodynamic properties $n$, $T$, and $Z$, we need three equations that relate these profiles to the observed, projected data.
The value of $\Sx$ may be related to the X-ray emissivity $j_x(\vect{r})$, which strongly depends on $n(\vect{r})$, integrated over the line of sight $\bm{l}$ (henceforth LOS),
\begin{equation}\label{eq:Sx}
\Sx=\int j_x\, dl\fin
\end{equation}
Standard packages (for example XSPEC\footnote{https://heasarc.gsfc.nasa.gov/xanadu/xspec}) use spectral fitting to provide the projected values of the thermal properties $A\in\{T,Z\}$ in a given angular bin $\check{b}$.
The value of $A$ thus fitted may be related, under certain assumptions, to the weighted average of $A$ over the volume $V$ corresponding to $\check{b}$ and $l$,
\begin{equation}\label{eq:A_EW}
A_{proj}(\check{b})\simeq \frac{\int W_A A \,dV}{\int W_A\,dV}\coma
\end{equation}
where $W_A$ is the corresponding weight.
Given expressions for the weights $j_x$, $W_T$ and $W_Z$, some assumptions on the spatial distribution of the gas, and the projected profiles $\Sx$, $\TtwoD$, and $\ZtwoD$, one can carry out the above three integrals, and solve for the spatial distributions of $n$, $T$, and $Z$.

\subsection{Projected quantities}\label{subsec:ProjectedQuantities}

To derive the projected profiles, $\Sx$, $\TtwoD$, and $\ZtwoD$, we first define the sectors for the analysis. Subsequently, we derive the finely-binned $\Sx$ profile, and the coarsely-binned $\TtwoD$ and $\ZtwoD$ profiles.

\subsubsection{CF sectors}\label{subsubsec:sectors}\label{subsec:sectors}

To simplify the analysis of the quasi-spiral CF observed, we divide the CF into sectors, and impose a common centre to all sectors, coincident with the GC centre.
This cannot be done with a circular fit to the CF, so for each sector we find the ellipse that best-fits the CF.
The region of interest inside (outside) the CF is bounded by this ellipse and a similar ellipse, rescaled by a small (large) number.
The inner and outer radii of each region of interest are determined by data availability.
As the CF ellipse is chosen by eye, its parameters carry some errors.
However, the radial position of the CF along the sector is derived more carefully (see \S\ref{subsec:FitProcedure}).

The \MyMNRAS{centre}\MyApJ{center} of the CF ellipse is chosen at the position of the GC \MyMNRAS{centre}\MyApJ{center}, defined as the X-ray peak. For the \MyMNRAS{analysed}\MyApJ{analyzed} GCs, this coincides with the position of the cD galaxy to within $\sim10\kpc$ for A2142 and $\sim1\kpc$ for the rest of the GCs.
This choice is justified for regions where the gas is relaxed \citep{MarkevitchVikhlinin2007}.
It is a necessary choice in some of our profiles (in A2029 and A2204), in which the inner region of interest extends all the way to the GC \MyMNRAS{centre}\MyApJ{center}.

In each region of interest, we then define a sector, by placing angular cuts emanating from the GC \MyMNRAS{centre}\MyApJ{center}.
The angular size of a sector is limited by the location where the CF either loses its coherence, or is no longer fit by a simple ellipse.
For simplicity, we associate the radial thermal profile in a given sector with its angular bisector.
As an example, see the A2029 CF sector in \MyApJ{Figure}\MyMNRAS{Fig.}~\ref{fig:A2029Figs}.

As the CFs are non-spherical, the angle $\alpha$ between the discontinuity and a coincident sphere around the GC centre is in general non-zero.
We quantify this deviation by the angle $\alpha_{\tiny\mbox{proj}}$ between the projected CF and its fitted ellipse, measured at the bisector; the results are presented in Table~\ref{tab:parameters}.
Six out of our eight newly deprojected CFs are approximately circular, and accordingly, $\alpha_{\tiny\mbox{proj}}\simeq10^\circ$ is small.
For the remaining two CFs, in the A2029SW and A2204N sectors, we find that $\alpha_{\tiny\mbox{proj}}\simeq30^\circ\mbox{--}40^\circ$.
These larger values reflect an elliptical GC morphology (in A2029) or an inclined spiral CF plane (see \S\ref{subsubsec:shearflows}), with a bisector that is not aligned with the axes of the ellipse.

\subsubsection{Surface brightness}\label{subsubsec:Sx}

The $\Sx$ profile along the sector is extracted in radially logarithmic, elliptical annuli, according to the properties of the aforementioned ellipse.
We convert the distribution of $\Sx$ among the pixels in the sector into a radial profile, by assigning each bin with its average $\Sx$ and the range of projected radius $R$ values it spans along the angle bisector.
For example, \MyApJ{Figure}\MyMNRAS{Fig.}~\ref{fig:A2029Sx} shows the $\Sx(R)$ profile in the A2029 sector of  \MyApJ{Figure}\MyMNRAS{Fig.}~\ref{fig:A2029Figs}.

There are some CFs that are seen in the  $\Sx$ maps, but we do not \MyMNRAS{analyse}\MyApJ{analyze}.
In Centaurus, four CFs are visible in \MyApJ{Figure}\MyMNRAS{Fig.}~\ref{fig:A3526Figs}: two to the east of the \MyMNRAS{centre}\MyApJ{center}, and two to the west.
However, we \MyMNRAS{analyse}\MyApJ{analyze} only the two peripheral CFs, as the regions below the more central CFs are obscured by substructure.
Notice that the peripheral CF to the east is reported here for the first time.
According to \citet{PaternoMahlerEtAl2013}, additional discontinuities lie above both CFs A2204E and A2204S, when subtracting a model from the $\Sx$ maps; this is also faintly visible in \MyApJ{Figure}\MyMNRAS{Fig.}~\ref{fig:A2204Figs}.
However, these additional discontinuities, if real, are not sufficiently significant in the $\Sx$ profiles, and so are disregarded.
In A2142, substructure prevents us from \MyMNRAS{analysing}\MyApJ{analyzing} the south-eastern CF.
In A2029, statistics are insufficient to \MyMNRAS{analyse}\MyApJ{analyze} the region below the north-eastern CF.

Just above the A3526W CF we find a substructure that is starched from the west to the north (see Fig.~\ref{fig:A3526Figs}). Therefore, we focus on the south-western side of the edge (A3526SW).

\subsubsection{Projected temperature and metallicity profiles}\label{subsubsec:TZ}

We extract the $\TtwoD$ and $\ZtwoD$ profiles above and below each CF, as follows.
We divide each sector into radial bins, to be used for both $\TtwoD$ and $\ZtwoD$.
The bins are defined such that each bin contains approximately the same number of counts, as listed in Table~\ref{tab:EventsBin}.
This number of counts is chosen large enough to ensure sufficient statistics for the $\TtwoD$ and $\ZtwoD$ derivation, and small enough to produce at least two bins both above and below the CF.
The radial range associated with each bin is determined along the angle bisector, as for $\Sx$.

The bins are aligned such that the CF lies at the boundary between two bins, in order to make the implicit assumption of binning --- an approximately uniform thermal distribution within each bin --- more plausible.
The by-eye tracing of the CF is thought to capture its shape reasonably well, but often introduces a substantial error in the determination of $\rcf$. Such an error has a minor effect on the  $\Sx$ profile, which is both continuous and finely binned.
This is not the case, however, for the sparsely-binned, discontinuous $\TtwoD$ and $\ZtwoD$ profiles, which are sensitive to errors in $\rcf$. The latter are therefore rebinned after $\rcf$ is more precisely determined; see \S\ref{subsec:FitProcedure}.

\MyApJ{\begin{table}[h]}
\MyMNRAS{\begin{table}}
	\caption{Events in each temperature bin} 
	\centering 
	\setlength{\tabcolsep}{0.5em} 
	{\renewcommand{\arraystretch}{1.5}%
		\begin{tabular}{| c| c |c | c|} 
			\hline 
			GC& $\#$ of events ($10^3$)& Circle radii& $N_H\left(10^{20}\mbox{cm}^{-2}\right)$\\
			(1) & (2) &(3)&(4)\\ 
			\hline 
			\multirow{ 1}{*}{A2029}& $8.5$& $3'$&$4.01\pm0.02$\\
			\hline 
				A2142& $25$& $3'$&$5.16\pm0.03$\\
			\hline 
			\multirow{1}{*}{A2204}& $9$& $3'$&$6.6\pm0.4$\\
			\hline 
			\multirow{1}{*}{Centaurus} & $60$& $3'.5$&$10.21\pm0.05$\\
			\hline
		\end{tabular}}\label{tab:EventsBin}\vspace{0.2cm}
		\begin{tablenotes}
			\item	Columns: (1) The GC; (2) The number of counts in each radial bin of $\TtwoD$ and $\ZtwoD$ profiles; (3) The radii of the circle in which $N_H$ was measured; (4) The absorption column value.
		\end{tablenotes}\vspace{0.2cm}
	\end{table}

The values of $\TtwoD$ and $\ZtwoD$ in each bin are derived with the spectral fitting package XSPEC, using an APEC*WABS model.
The required co-added instrument response files of the different ObsIDs are generated as described in \citet{VikhlininEtAl2005}.
Before fitting the APEC*WABS model to the data, we freeze the redshift at the mean GC value (see Table~\ref{tab:Obsids}).
We freeze the absorption column density $N_H$ at the value obtained when fitting the co-added GC spectra with a single temperature. The fitted region here is the largest disk, \MyMNRAS{centred}\MyApJ{centered} upon the X-ray peak, still confined within the chips. The disk radii and the corresponding $N_H$ values in each GC are given in Table~\ref{tab:EventsBin}.
While the simplifying assumption of an isothermal GC introduces some systematic errors, the resulting $N_H$ values are in good agreement with the expected results \citep[\eg][]{KalberlaEtAl2005}.

\MyApJ{Figures}\MyMNRAS{Figs.}~\ref{fig:A2142ProjT} and \ref{fig:A2142ProjZ} demonstrate the projected $\TtwoD$ and $\ZtwoD$ profiles, respectively, in A2142.

\subsection{Deprojection}

To model the deprojected radial thermal profiles, $n$, $T$, and $Z$, in a sector, we first assign a $3$D parametric model to each of the profiles. We then determine the weights $j_x$, $W_T$, and $W_Z$. Finally, we determine $\rcf$, rebin the $\TtwoD$ and $\ZtwoD$ profiles, and derive the best-fit parameters of the $\{n,T,Z\}$ models.

\subsubsection{Parametric models}\label{subsec:ModParameters}

Deprojecting a CF requires some assumptions on the underlying 3D plasma distribution.
Observations indicate that the gas distribution in a GC can be described as a triaxial spheroid \citep[\eg][]{Cooray2000,DeFilippisETal2005,PazEtal2006,SerenoEtal2006,Kawahara2010,PlanckCollaboration2016,SerenoEtAl2017}.
Simulations indicate that a prolate symmetry provides a good approximation \citep[\eg][]{JingSuto2002,AllgoodEtAl2006,MaccioEtal2007,MunozEtAl2011,DespaliEtAl2014}.
Hence, we model the $3$D gas distribution in each CF sector as a prolate spheroid.
For simplicity, we assume that the large axis of the spheroid lies on the plane of the sky.
Therefore, we infer the three axes of the spheroid from the CF ellipse; in particular, the axis along the LOS is identified with the minor axis of the ellipse.
The thermal properties are thus taken as functions of a single 3D parameter, namely the spheroidal radius $r$, identified on the plane of the sky along the bisector with the 2D coordinate $R$.

To model the thermal profiles while taking into account the presence of the CF at spheroidal radius $r=\rcf$, and the effect of the CF on the gas distribution, we adopt, for each CF, different models for the gas inside ($r<\rcf$) and outside ($r>\rcf$) the CF.

For the density profile, we first attempt a $\beta$-model \citep{CavaliereFuscoFemiano1976,CavaliereFuscoFemiano1978},
\begin{equation}\label{eq:nBetaModel}
n(r)=n_j\left(\frac{r_{c,j}^2+r^2}{r_{c,j}^2+\rcf^2}\right)^{-\frac{3}{2}\beta_j}\coma
\end{equation}
where the index $j\in\{i,o\}$ denotes regions inside, outside the CF.
Here, $\beta_j$ is the slope parameter, and the \MyMNRAS{normalisation}\MyApJ{normalization} $n_j$ is the density extrapolated (from each side) to $\rcf$.
The parameter $r_{c,j}$, representing the core radius, is allowed to take different values on each side of the CF.
At least on one side of the CF, $r_c$ cannot be assumed to correspond to any physical core scale.
If the $\beta$-model on either side of the CF yields a negligibly small $r_{c}$ (with respect to the radial data range), the model is replaced by a power-law,
\begin{equation}\label{eq:n_PLaw}
n_j(r) = n_j\left(\frac{r}{\rcf}\right)^{\alpha_{n,j}}\fin
\end{equation}

For $T(r)$ and $Z(r)$, we simply adopt power-law profiles,
\begin{equation}\label{eq:T_PLaw}
T(r)=T_j\left(\frac{r}{\rcf}\right)^{\alpha_{T,j}}\coma
\end{equation}
and
\begin{equation}\label{eq:Z_PLaw}
Z(r)=Z_j\left(\frac{r}{\rcf}\right)^{\alpha_{Z,j}}\coma
\end{equation}
respectively.
Here, the \MyMNRAS{normalisations}\MyApJ{normalizations} $T_j$ and $Z_j$ are the temperature and metallicity extrapolated to the CF radius, whereas $\alpha_T$ and $\alpha_Z$ are the corresponding profile slopes.

The various models described above are not expected to hold over a wide radial range, especially if more than one CF is present in the the modelled sector.
We therefore restrict the radial range of each model to the CF region, using data at radii reaching up to a factor of $\sim 3$ both above and below the CF radius.
The use of a multiplicative radial factor here is justified by the nearly logarithmic CF spirals inferred from observations and simulations (\UKspirals), and we choose the factor to be $\sim3$ based on some multi-CF sectors, for example in Centaurus \citep[\eg][]{SandersEtAl2016} and in A2029 \citep[\eg][]{PaternoMahlerEtAl2013}.
In some sectors, an evident change in the slope of $\Sx$, $\TtwoD$, or $\ZtwoD$ can be seen even closer to the CF; in such cases we accordingly limit the modelled radial range.
Our results depend somewhat on the precise locations of these radial cutoffs; we discuss this sensitivity in \S\ref{sec:Discussion}.

\subsubsection{Weight functions}
\label{subsubsec:Weights}

Once the projected profiles are derived and the 3$D$ parametric models are chosen, we only need to determine the weights $j_x$, $W_T$, and $W_Z$ in order to derive the deprojected thermal profiles (according to Eqs.~\ref{eq:Sx} and \ref{eq:A_EW}).

The emissivity $j_x$ is parameterised as $j_x=\Lambda n^2$.
For a given GC redshift $z$, observed photon energy band $\{\epsilon_a,\epsilon_b\}$, and given gas parameters $T$ and $Z$, XSPEC provides\footnote{When using the XSPEC function ``flux'' (with variable ``norm'' defined as unity, and with the energy range defined in the observer frame), we use $\Lambda=10^{-14}(1+z)\tilde{F}\mbox{ photons}\cm^3\se^{-1}$, where $\tilde{F}$ is the first output of the function; notice the factor $(1+z)$, absent from the documentation.
The redshift factor leads to a small multiplicative correction to $n_i$ and $n_o$, which does not affect the contrast; in any case this effect is modest in our low-redshift GCs.}
the cooling function $\Lambda$.
We model $\Lambda$ by interpolating tabular data from XSPEC, for the given GC redshift and the observed energy band, \ie $\epsilon_a=0.8\,\keV$ and $\epsilon_b=7.0\,\keV$.

The emissivities of different slices along the LOS have an additive effect on $\Sx$. This justifies writing $\Sx$ (in Eq.~\ref{eq:Sx}) as the integral over $j_x=n^2\Lambda$ (which can be seen as the mean of $n^2$ weighted by $\Lambda$), inasmuch as the model for $\Lambda(T,Z)$ can be trusted and the local (3D) temperature and metallicity can be determined.
In contrast, it is not necessarily true that the projected temperature and metallicity depend linearly on the contribution of each slice, because $\TtwoD$ and $\ZtwoD$ are determined from a spectral fit which is inherently non-linear. As the weights $W_T$ and $W_Z$ are thus ill-defined, there is no obvious way to choose them, and different expressions were explored in the literature. It is important to verify that the results of a deprojection are qualitatively insensitive to the precise choice (within reason) of these weights.

Among the expressions adopted in the literature for $W_T$, notable are the emissivity weight $W_T=n^2 \Lambda$, the emission measure weight $W_T=n^2$ (\ie approximating $\Lambda=\const$), and a weight based on a spectral fit,  $W_T=n^2 T^{-3/4}$ \citep[$\Lambda\propto T^{-3/4}$; see][and references therein]{MazzottaEtAl2004}.
As our nominal choice, we adopt the emissivity weight, $W_T=n^2 \Lambda$, which takes into account both temperature and metallicity variations, and can be justified for the simplified case where the spectral fit is additive. In \S\ref{app:deproj_models}, we examine different choices of the weight $W_T$, and quantify the sensitivity of the results to this choice.

The metallicity weight $W_Z$, defined through Eq.~(\ref{eq:A_EW}), received less attention in the literature, probably because measuring $Z$ is more difficult, and is not strictly necessary if $\Lambda$ and $W_T$ are approximated as $Z$-independent.
Several studies \citep[\eg][]{EttoriEtAl2002,PizzolatoEtAl2003,SiegelEtAl2018} adopted the emissivity as the metallicity weight, $W_Z=n^2\Lambda$, albeit not (as far as we know) in the context of CF analyses.
The metallicity in CF sectors was, however, non-parametrically deprojected: \citet{SandersEtAl2005} used the XSPEC deprojection routine PROJCT (see Appendix~\ref{app:deproj_models}), and \citet{SandersEtAl2016} used the deprojection routine DSDEPROJ \citep[\eg][]{SandersFabian2007,FabianEtAl2008}.

We adopt the emissivity as our nominal weight, $W_Z=n^2\Lambda$, for the same reasons outlined above for $W_T$; the sensitivity of our results to this choice are discussed in \S\ref{sec:Discussion}.

\subsubsection{Fit procedure}\label{subsec:FitProcedure}

As mentioned above, the by-eye determination of $\rcf$ is prone to error.
Therefore, unlike the parameters describing the shape of the CF, $\rcf$ is now treated as a free parameter.
This introduces some difficulty, because the bins already introduced into the analysis may now become misaligned with the CF.
As explained above, the results are not sensitive to a misaligned $\Sx$ binning, because $\Sx$ is continuous and finely binned, with bins near the CF thinner than $5\,\kpc$ in A2142, and $<1\,\kpc$ in the rest of the \MyMNRAS{analysed}\MyApJ{analyzed} GCs.
The results can however become significantly distorted by  a misaligned $\TtwoD$ or $\ZtwoD$ binning, because a thermal discontinuity may thus become lodged deep within a large bin.
This severely complicates the analysis: binning correctly requires an accurate determination of $\rcf$, but this determination is skewed by the incorrect binning.

To break the degeneracy, and self-consistently both determine $\rcf$ and accordingly bin the data, we use an iterative approach. First, we consider a model which does not depend on $T$ and $Z$.
This is achieved by fitting only the density profile, based on the binned $\Sx$ data interpreted using  Eq.~(\ref{eq:Sx}) with a simplified, $j_x\propto n^2$ emissivity that is independent of $T$ and $Z$.
The resulting best-fit value of $\rcf$ is then used to rebin the data, with an improved bin--CF alignment.
The rebinned $\TtwoD$ and $\ZtwoD$ profiles, along with the $\Sx$ profile, can now be used to fit the three thermal profiles, $n$, $T$, and $Z$, using the nominal ($T$- and $Z$-dependent) weights.
This step --- rebinning the data and refitting the nominal model --- can be repeated until it converges. In practice, we find that a single iteration is sufficient, leading to an $\rcf$ determination accurate to within the $\Sx$ bin resolution.

The outcome of this process is the determination of the values of the model parameters in Eqs.~(\ref{eq:nBetaModel}--\ref{eq:Z_PLaw}) that, when projected according to Eqs.~(\ref{eq:Sx}) and (\ref{eq:A_EW}) with our nominal weights, provide the best fit the projected data.
This includes $\rcf$, the temperature profile parameters $T_j$ and $\alpha_{T,j}$, the metallicity profile parameters $Z_j$ and $\alpha_{Z,j}$, and the density profile parameters $n_j$ and either $\{r_{c,j},\beta_j\}$ or $\alpha_{n,j}$.

\subsection{Previously deprojected CFs}\label{subsec:DeptojCFsLit}

In order to improve the statistics of our CF analysis, we supplement our CF sample by \MyMNRAS{analysing}\MyApJ{analyzing} previously deprojected CFs from the literature. We \MyMNRAS{reanalyse}\MyApJ{reanalyze}, in particular, part of  the CF sample discussed in {\RK}, as detailed below.
A thorough discussion of systematic effects associated with the analysis of data from the literature is given by {\RK}.

Some CF analyses in the literature show only one thermal bin on one side of of the CF, either above or below it. This is the case, for example, in three out of the 17 CFs in the sample of {\RK}. \MyMNRAS{Modelling}\MyApJ{Modeling} the thermal profile on one side of the CF with a single data point requires strong assumptions on the profile; one typically assumes that the plasma here is isothermal.
To avoid such unjustified assumptions and the corresponding large systematic errors, we exclude these CFs from our analysis.

We find $13$ CFs in $10$ GCs which are usable here, namely were reported in the literature with more than one thermal bin on each side of the CF.
The CFs in this previously-deprojected CF sample are found in the following GCs (one CF in a GC, unless otherwise stated): A133 \citep[two CFs;][]{RandallEtAl2010}, A496 \citep{TanakaEtAl2006}, A1644 \citep{JohnsonEtAl2010}, A1795 \cite{MarkevitchEtAl2001}, A2052 \citep{dePlaaEtAl2010},  A2199 \citep{NulsenEtAl2013}, A3158 \citep{WangEtAl2010}, RXJ1532 \citep{HlavacekLarrondoEtAl2013}, Virgo \citep{UrbanEtAl2011}, and 2A0335 \citep[three CFs;][]{SandersEtAl2009}.

Our sample of previously deprojected CFs includes three CFs, found in the GCs A3158, A2199, and RXJ1532, that were not \MyMNRAS{analysed}\MyApJ{analyzed} by {\RK}.
The remaining 10 CFs, already studied by {\RK}, are \MyMNRAS{reanalysed}\MyApJ{reanalyzed} here starting from the published, binned thermal profiles.
The seven CFs studied by {\RK} and excluded from the present analysis include the three aforementioned CFs with a single temperature bin below the CF (in RXJ1347.5 and the inner CF in A1644), three CFs in A2204 which we \MyMNRAS{reanalyse}\MyApJ{reanalyze} from the raw data, and one CF in A1664 which was not deprojected.

Results from the 13 previously deprojected CFs are presented in the relevant figures below. However, only eight of these CFs are actually included in our following statistical computations.
The remaining five CFs suffer from analysis complications.
The CF in Virgo is based on a very narrow sector, leading to strong fluctuations; see \S\ref{sec:Introduction}.
The outer CF in A1644 is suspected as being affected by a merger; see discussion in {\RK}.
The thermal binnings used in A133 (the innermost CF), RXJ1532, and A2199 appear to be misaligned with the CF; here we ignore the bin \MyMNRAS{harbouring}\MyApJ{harboring} the CF, and use the bins above and below this problematic bin.
As these five CFs provide interesting information but differ from the other CFs and carry potentially large systematic errors, they are only shown in figures, but not incorporated in the statistical analysis.

\section{Deprojection results}\label{sec:DeproProp}

The best-fit model parameters in Eqs.~(\ref{eq:nBetaModel}--\ref{eq:Z_PLaw}) are presented in Tables~\ref{tab:jumps} and \ref{tab:parameters}, for each of the \MyMNRAS{analysed}\MyApJ{analyzed} sectors in each of the GCs.
Since we are interested in the fractional variations in the thermal parameters across a CF, we \MyMNRAS{analyse}\MyApJ{analyze} the thermal profiles in log space, implicitly assuming that the statistical errors are distributed symmetrically in their respective log space; the sensitivity to this assumption is discussed in \S\ref{sec:PjumpsSec}.
The projected and deprojected thermal profiles across the different \MyMNRAS{analysed}\MyApJ{analyzed} sectors are shown in
\MyApJ{Figure}\MyMNRAS{Fig.}~\ref{fig:A2029Thermal} for A2029, \MyApJ{Figure}\MyMNRAS{Fig.}~\ref{fig:A2142Thermal} for A2142, \MyApJ{Figures}\MyMNRAS{Figs.}~\ref{fig:A2204WEThermal} and \ref{fig:A2204NSThermal} for A2204, and \MyApJ{Figure}\MyMNRAS{Fig.}~\ref{fig:A3526Thermal} for Centaurus.

\subsection{Density drops and temperature jumps}

As Table~\ref{tab:jumps} shows, across the \MyMNRAS{analysed}\MyApJ{analyzed} CFs, the density sharply drops (with increasing radius) and the temperature sharply jumps, as expected.
The fractional density drops are of order $\njump\equiv n_i/n_o\sim1.6$, whereas the fractional temperature jumps are of order $\Tjump\equiv T_o/T_i\sim2$ (see detailed discussion below).
Here and below, we define contrasts, such as $\njump$ and $\Tjump$, such that their value is positive for a typical CF.
Comparing these typical $\njump$ and $\Tjump$ values already hints at a non-isobaric transition, as we discuss later in \S\ref{sec:PjumpsSec}.

In some CFs, it is possible to compare our results with previous deprojections.
We confirm that in A2142NW, A2204E, and A2204W, our $\njump$ and $\Tjump$ values agree (within $\lesssim1\sigma$) with previous studies that employed different deprojection methods \citep{SandersEtAl2005, ChenEtAl2017, WangEtAl2018}.
The same holds in A3526SW for $\njump$, but not for $\Tjump$, because the previous deprojection \citep{SandersEtAl2016} used a smaller range of radii outside the CF.

The mean, uncertainty-weighted values of $\njump$ and $\Tjump$, averaged over all the newly deprojected CFs, are $\barnjump=1.32\pm0.01$ and $\barTjump=1.38\pm0.03$, respectively.
However, the dispersion among these CFs is substantial, as we discuss below; indeed, the reduced chi-squared values \citep[$\chi^2$ per degree of freedom; DOF;][]{Pearson1900} of fitting universal contrasts are $\chi_\nu^2\equiv \chi^2/\nu\simeq79$ for $\barnjump$, and $\chi_\nu^2\simeq7$ for $\barTjump$.
Here, $\nu = N_{\tiny{\mbox{CF}}}-1=7$ is the number of DOF, and $N_{\tiny{\mbox{CF}}}=8$ is the number of CFs.

The dispersion among CFs in different GCs is especially notable.
Without the CFs in Centaurus, which dominate the results due to their relatively small uncertainties, the mean-weighted values become $\barnjump=1.78\pm0.03$ ($\chi_\nu^2\simeq16$) and $\barTjump=2.0\pm0.2$ ($\chi_\nu^2\simeq0.4$).
Fitting only the two CFs in Centaurus with uniform contrast values yields $\barnjump=1.19\pm0.01$ ($\chi_\nu^2\simeq20$) and $\barTjump=1.34\pm0.03$ ($\chi_\nu^2\simeq28$).
Fitting the four CFs in A2204, we find $\barnjump=1.81\pm0.04$ ($\chi_\nu^2\simeq9.5$) and $\barTjump=2.0\pm0.2$ ($\chi_\nu^2\simeq0.5$).
In the other GCs, A2029 and A2142, we analysed only one CF sector.
Thus, while the dispersion within a GC can be substantial (as in Centaurus), the overall dispersion is dominated by the variation between different GCs.

The dispersion in contrast values among different CFs in different GCs exceeds the statistical uncertainties.
The measured contrast values $\njump$ and $\Tjump$, and their considerable dispersion among CFs in different GCs, are consistent with previous contrast values inferred from the literature \citep[\eg][]{MarkevitchEtAl2000,MarkevitchEtAl2001,SandersEtAl2005,SandersEtAl2009,	
OgreanEtAl2014,ChenEtAl2017}.
It is interesting to examine the origins of this dispersion, and to quantify the variations in contrast within a single GC.

As shown in \S\ref{sec:PjumpsSec}, the parameters $\njump$ and $\Tjump$ are closely related, as the CFs are not far from being isobaric.
We may thus reduce the statistical noise by examining the distribution of the mean CF contrast, $q\equiv(\njump\Tjump)^{1/2}$, averaging the density and temperature contrasts.
We find that the dispersion in $q$ within a given GC is relatively small, \ie the CFs in the GC have a roughly constant contrast; this is shown in \S\ref{subsec:qTr500}, using both newly and  previously (see \S\ref{subsec:DeptojCFsLit}) deprojected CFs.
This conclusion suggests that the CF contrast is a characteristic of the GC, and as such, may scale with some other GC property --- for example, the GC mass --- leading to the aforementioned dispersion in contrast among different GCs; this possibility is examined in \S\ref{subsec:qTM200}.

\subsubsection{The CF contrast is nearly constant inside a GC}\label{subsec:qTr500}

To examine the variation in contrast $q$ within a given GC, we focus on GCs that \MyMNRAS{harbour}\MyApJ{harbor} more than one deprojected CF sector.
We begin with the newly deprojected, multiple-CF GCs, namely A2204 and  Centaurus.
Next, we consider GCs with multiple deprojected CF sectors in the literature, namely A133 and 2A0335.
Finally, we study the joint population of these two GC samples.

In Centaurus, the contrast value $q=1.44\pm0.03$ across the A3526SW CF, is only slightly higher than the $q=1.19\pm0.02$ value across the A3526E CF.
Yet, due to the exquisite statistics in this GC, the two contrast values are inconsistent with each other at the $\sim6.6\sigma$ confidence level.
Among the four CFs in A2204, those with the highest contrast
($q=2.0\pm0.02$ across the A2204W CF) and with the lowest contrast ($q=1.5^{+0.3}_{-0.2}$ across the A2204N CF) are consistent with each other at the $\sim1.7\sigma$ level.

More quantitatively,
\MyApJ{Figure}\MyMNRAS{Fig.}~\ref{fig:qnTr500} presents the \MyMNRAS{normalised}\MyApJ{normalized} contrast $q/\bar{q}$, plotted against the \MyMNRAS{normalised}\MyApJ{normalized} CF distance $\RCF/\RCFbar$ from the GC \MyMNRAS{centre}\MyApJ{center}.
Here, $\bar{q}$ and $\bar{R}_{cf}$ are the mean values of $q$ and $\RCF$, respectively, among all the CFs in the same GC.
The mean variation in $q/\bar{q}$ among the new sample of six CFs (two in Centaurus and four in A2204), weighted by the statistical errors, gives  $\langle \delta q/\bar{q}\rangle = 0.007\pm0.017$, with $\chi^2_{\nu}\simeq6.7$, where we defined $\delta q\equiv q-\bar{q}$.
The relatively large $\chi^2_{\nu}$ value is mostly due to the good statistics in Centaurus.
The $\langle \delta q/\bar{q}\rangle\simeq0$ result support the hypothesis of a nearly constant contrast inside a given GC.

We repeat this analysis for the sample of multiple CFs previously deprojected in the same GC.
In order to extract the parameters of a CF from the literature (listed in \S\ref{subsec:DeptojCFsLit}), we extrapolate the reported density and temperature profiles on each side of the CF to the CF radius.
First, we fit power-law models for the $n(r)$ and $T(r)$ profiles, each on both sides of the CF.
Next, we use these fits to extrapolate each quantity on each side of the CF, to the CF radius.
The mean CF contrast $q$ is then estimated as $q=[n_iT_o/(n_oT_i)]^{1/2}$.

We find two GCs which harbor multiple previously deprojected CFs: two CFs in A133 and three in 2A0335.
However, as mentioned in \S\ref{subsec:DeptojCFsLit}, one of the CFs in A133 is excluded from our study.
In the remaining GC, 2A0335, $\langle \delta q/\bar{q}\rangle = 0.01\pm0.07$ ($\chi^2_{\nu}\simeq0.7$).
For the joint sample, of nine newly and previously deprojected CFs, $\langle\delta q/\bar{q}\rangle=0.007\pm0.016$, with $\chi^2_{\nu}\simeq4.3$. This result further support a nearly constant contrast inside a GC.

To test for a possible subtle trend in CF contrast within a GC, we examine the Pearson linear correlation $\corl$ \citep{Pearson1895} between $\ln(q/\bar{q})$ and $\ln(\RCF/\RCFbar)$.
In order to incorporate the statistical errors, we Monte-Carlo simulate the distribution of mock CFs corresponding to each data point, and use this distribution to evaluate $\corl$ and its confidence interval.
As expected, the Fisher z-transformation \citep[][]{Fisher1921}, $\zeta=(1/2)\ln[(1+\corl)/(1-\corl)]$, leads to an approximately normal distribution in $\zeta$-space, providing a useful measure of the mean correlation $\bar{\zeta}$ and its standard deviation $\sigma_{\zeta}$. We use $\sigma_\zeta$ to quantify the $1\sigma$ confidence interval on $\corl$, in the form $\corl=\bar{\corl}_{-\delta\corl_-}^{+\delta\corl_+}$, where $\bar{\corl}\equiv\corl(\bar{\zeta})$ and $\delta\corl_\pm\equiv \pm\corl(\bar{\zeta}\pm\sigma_{\zeta})\mp\bar{\corl}$;
see discussion in \citet{Fisher1921}.

In this method, we parameterise the correlation between the contrast values and radii of CFs within a GC in terms of both Fisher $\zeta$ and Pearson $\corl$.
For the sub-sample of six newly-deprojected CFs in Centaurus and A2204, $\bar{\zeta}=0.0\pm0.4$ and $\corl=0.0^{+0.4}_{-0.4}$.
For the joint sub-sample of nine CFs, $\bar{\zeta}=0.1\pm0.3$ and $\corl=0.1^{+0.3}_{-0.3}$.
These results are consistent with no correlation between $q$ and $R$, and with nearly constant contrast values inside each GC.

It is useful to test for a subtle radial trend in contrast, $q=q(R)$, also by fitting a model of the form
\begin{equation} \label{eq:ModelA}
(q/\bar{q})=a (\RCF/\RCFbar)^b \fin
\end{equation}
This gives $a=1.01\pm0.02$, $b=-0.25\pm0.05$ with $\chi^2_{\nu}\simeq2.1$ for the newly deprojected sub-sample, and $a=1.01\pm0.02$, $b=-0.23\pm0.05$ ($\chi^2_{\nu}\simeq1.8$) for the nine-CF joint sample.
Excluding the two CFs in Centaurus (which dominate the results due to their relatively small uncertainties) yields $a=1.01\pm0.05$ and $b=0.1\pm0.1$ ($\chi^2_{\nu}\simeq0.4$).
The best fit models are presented in \MyApJ{Figure}\MyMNRAS{Fig.}~\ref{fig:qnTr500}.
The results of the model fits, like those of the correlation tests, are consistent with an approximately constant $q$ value inside each GC.

The model (\ref{eq:ModelA}) fits the data substantially better than the null hypothesis (labelled with subscript '$-$'),
\begin{equation} \label{eq:ModelB}
q=q_- \equiv a_- (\RCF/\RCFbar)^{b_-} \coma
\end{equation}
that does not invoke a fixed contrast inside a GC.
For example, for the joint nine CF sample, the best fit here is $a_-=1.30\pm0.02$ and $b_-=0.20\pm0.04$ ($\chi^2_{\nu-}\simeq8.7$).

We may use the TS test \citep{Wilks1938, MattoxEtAl1996} to quantify how much better model (\ref{eq:ModelA}) performs with respect to model (\ref{eq:ModelB}), \ie determine the confidence level at which a uniform CF contrast value can be assigned to each GC.
To this end, we supplement model (\ref{eq:ModelA}) with a control parameter, arriving at the refined model (labelled with subscript ``$+$'')
\begin{equation} \label{eq:ModelC}
q=q_+ \equiv a_+ (\RCF/\RCFbar)^{b_+}\bar{q}^{\,c_+} \fin
\end{equation}
Here, for the joint sample $a_+=1.01\pm0.02$, $b_+=-0.23\pm0.05$, and $c_+=1.1\pm0.2$ ($\chi^2_{\nu+}\simeq2$), and the implied $\mbox{TS}\equiv\chi^2_--\chi^2_+\simeq 49$ approximately follows a $\chi^2$ distribution with one degree of freedom.
We conclude that model (\ref{eq:ModelC}), which invokes a nearly constant contrast within each GC, is favored over the null hypothesis at the $\gtrsim6.9\sigma$ confidence level.

We conclude that although the sample of GCs with multiple CFs is small, and the data is noisy, there is evidence that variations in contrast are modest inside each GC, compared to the dispersion among different GCs.
Next, we examine how the mean contrast $\bar{q}$ inside each GC depends on other GC properties.

\MyMNRAS{\begin{figure}
\centering
\DrawFig{\includegraphics[width=9.5cm]{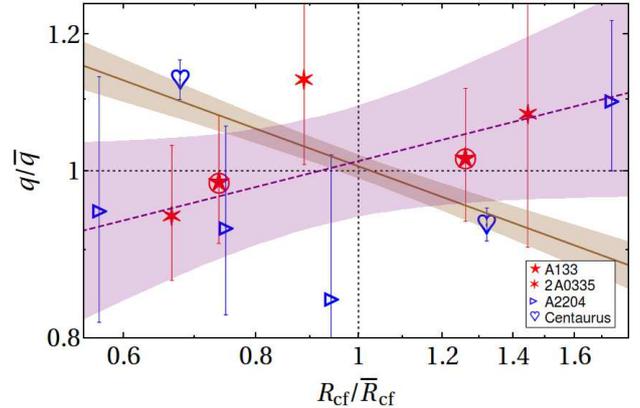}}
\caption{The CF contrast $q=(\njump \Tjump)^{1/2}=[(n_i T_o)/(n_o T_i)]^{1/2}$ plotted against its radius $\RCF$, after \MyMNRAS{normalising}\MyApJ{normalizing} both $q$ and $\RCF$ by their mean values $\bar{q}$ and $\RCFbar$ averaged over all CFs in their respective GC. Only GCs with multiple \MyMNRAS{analysed}\MyApJ{analyzed} CFs are shown.
Empty blue symbols (filled red symbols) represent the newly (previously) deprojected CFs (see legend).
The best fit (\ref{eq:ModelA}) is shown for the joint sample, both with (solid brown curve) and without (dashed purple) the two dominant CFs in Centaurus.
In both cases, we show (shaded region) the $1\sigma$ confidence interval, and exclude from the fit the two problematic CFs in A133 (marked by a circle; see text).
}
\label{fig:qnTr500}
\end{figure}}

\subsubsection{The CF contrast scales with the GC mass}\label{subsec:qTM200}

Having identified the CF contrast as a global property of the GC, it is natural to expect it to correlate with other GC properties.
In particular, some correlation is anticipated with the GC mass, which itself correlates with GC properties such as the mean temperature \citep[\eg][]{HornerEtAl1999, VikhlininEtAl2006, MantzEtAl2010} and the characteristic radii $R_{\kappa}$.
Here, $R_{\kappa}$ is defined as the radius enclosing a mean density that is $\kappa$ times the critical density of the Universe; in particular, we \MyMNRAS{utilise}\MyApJ{utilize} the scale $R_{500}$ below.

The relation between the CF contrast parameter $q$ and the host GC mass is shown in  \MyApJ{Figure}\MyMNRAS{Fig.}~\ref{fig:qnTM200}.
As a proxy for the GC mass, we use $M_{200}$, the mass within $R_{200}$.
Measurements of $M_{200}$ are based on weak lensing for A2029, A2142, and A2204.
In Centaurus, no weak lensing estimate is available, to our knowledge, so we adopt an estimate of $M_{200}$ derived from X-ray observations assuming hydrostatic equilibrium.
In A2029, A2142, and A2204, the masses inferred from weak lensing and from X-rays are found to be consistent with each other, so using the X-ray-based mass in Centaurus seems reasonable.
For the previously deprojected CFs, we found a  weak-lensing measurement of $M_{200}$ only for RXJ1532; for the rest of the GCs in this sample we use $M_{200}$ estimates based on combining X-ray observations with the assumption of hydrostatic equilibrium  \citep[see][]{ReiprichBohringer2002,MartinetEtal2017,SimionescuEtAl2017}.

\MyMNRAS{\begin{figure}
\centering
\DrawFig{\includegraphics[width=8.45cm]{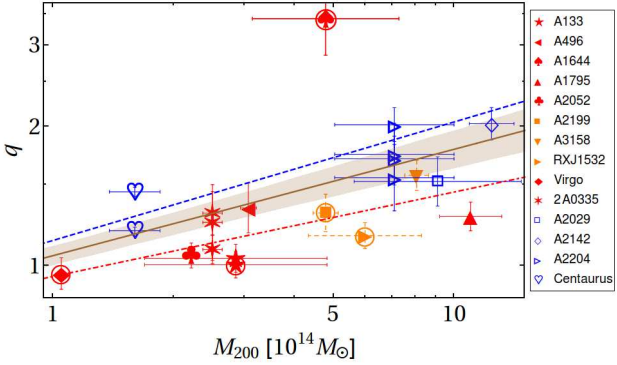}}
\caption{
The CF contrast $q=(\njump\Tjump)^{1/2}$ plotted against host GC mass $M_{200}$.
CF symbols (see legend) are the same as in Fig.~\ref{fig:qnTr500}, but here we include also single-CF GCs from the literature (filled symbols; red if found in {\RK}, orange with dashed error bars if not).
The best-fit (\ref{eq:qTM200}) is shown for the subsample of newly deprojected CFs (dashed blue curve), previously deprojected CFs (dot-dashed red), and the joint sample (solid brown, with shaded region showing the $1\sigma$ confidence level).
Problematic CFs excluded from the fit are marked by a circle (see text).
}
\label{fig:qnTM200}
\end{figure}}

We examine the $q$--$M_{200}$ relation in both the newly and the previously deprojected CF samples.
First, we use Pearson's correlation test to estimate the correlations between the contrast and mass in each GC, in terms of $\corl$ and its $\zeta$-space value.
Next, we fit the data with a power-law model,
\begin{equation} \label{eq:qTM200}
q \propto M_{200}^b \fin
\end{equation}
Applying these tests to our CF sub-samples shows a consistent positive, sub-linear relation between $q$ and $M_{200}$, as follows.

For the newly deprojected CFs, the correlation coefficient between $\ln (q)$ and $\ln (M_{200})$ is found to be $\bar{\zeta}=0.46\pm0.07$ in $\zeta$-space ($\sim5.7\sigma$), and $\corl=0.43^{+0.06}_{-0.06}$.
When fitting the data to the model (\ref{eq:qTM200}), we obtain the best-fit slope $b=0.26\pm0.04$, with $\chi^2_{\nu}\simeq2.3$.
Both of these tests suggest a strong positive correlation between $q$ and $M_{200}$.

Repeating the above tests for the sample of previously deprojected CFs (see \S\ref{subsec:DeptojCFsLit}), we again find a rather strong positive correlation, $\bar{\zeta}=0.24\pm0.05$ ($\sim4.7\sigma$), $\corl=0.23^{+0.05}_{-0.05}$, and $b=0.18\pm0.06$ ($\chi^2_{\nu}\simeq1.6$). In a joint analysis of both newly and previously deprojected CFs, we obtain $\bar{\zeta}=0.46\pm0.07$ (a $\sim6.4\sigma$ effect), $\corl=0.43^{+0.06}_{-0.06}$, and $b=0.23\pm0.04$ ($\chi^2_{\nu}\simeq4.5$).
Notice that the previously deprojected CF sample indicates, on its own accord, a similar scaling $b$ of $q$ with $M_{200}$ as found from the new CFs.

To conclude, our results are consistent with the CF  contrast being a property of the GC, showing little variability inside the GC but positively correlating with the GC mass.
This behavior is found in both sub-samples, as well as in the joint sample.
The correlation between $q$ and $M_{200}$ appears to be positive but sublinear, best-fit by $q\propto M_{200}^{0.23\pm0.04}$.

\subsection{Metallicity drops}
\label{subsec:ZDrops}

Previously \MyMNRAS{analysed}\MyApJ{analyzed} metallicity profiles, based on deprojected \citep[][]{SandersEtAl2005}, but more often projected \citep[\eg][]{GhizzardiEtAl2007,RossettiEtal2007,SimionescuEtal2010,FabianEtal2011,RussellEtAl2012,GhizzardiEtAl2013,WalkerEtAl2017}, data, suggested a sharp drop in $Z$ across CFs, $\Zjump\equiv Z_i/Z_o>1$.
The only metallicity profiles that were, as far as we know, previously deprojected across a CF \citep[][]{SandersEtAl2005,SandersEtAl2016}, suggested such a drop at only marginal, $\lesssim 2\sigma$ confidence levels.
We examine the deprojected metallicity profiles in our sample of eight CFs (see Table~\ref{tab:jumps}), as shown in \MyApJ{Figure}\MyMNRAS{Fig.}~\ref{fig:qzr500} by plotting $\Zjump$ against $\RCF/R_{500}$.

For the CFs in A2204E, A2204W, and A3526SW, it is possible to compare our $\Zjump$ values with previous deprojections \citep{SandersEtAl2005,ChenEtAl2017,SandersEtAl2016}. We confirm that the agreement is good (within $\lesssim1\sigma$).

Across the A3526E CF, we find a contrast $\Zjump=1.4^{+0.2}_{-0.1}$, inconsistent with an equal metallicity on both sides of the CF at the $\sim2.9\sigma$ confidence level.
The contrast values across the other seven CFs are also consistent with a metallicity drop, $\barZjump=1.1\pm0.1$, but this is only a marginally-significant, $\sim1.4\sigma$ effect in this subsample.
The contrast value of the combined sample of eight CFs is  $\barZjump=1.25^{+0.09}_{-0.08}$, inconsistent with a constant metallicity across the CF at the $\sim2.9\sigma$ confidence level.

To test if our analysis --- in particular the binning, fit, and extrapolation to the CF radius --- introduces systematic errors that may bias the $\Zjump$ measurement, we compute $\Zjump$ values across a control CF sample, composed of bins that do not \MyMNRAS{harbour}\MyApJ{harbor} a CF, in a method similar to that used in \RK.
Namely, we place fictitious CFs between consecutive metallicity (and so, also temperature) bins in the same sectors \MyMNRAS{harbouring}\MyApJ{harboring} the real CFs, but at radii sufficiently far removed from these CFs.
In order to produce useful mock profiles and minimise the influence of nearby real CFs, we place the mock CFs only above the known CFs, with at least two CF-free bins with $Z$ measurements both above and below each mock CF.
The metallicity contrasts of the 50 mock CFs in the resulting control sample, shown in \MyApJ{Figure}\MyMNRAS{Fig.}~\ref{fig:mockqz}, have a weighted (by statistical errors) mean metallicity contrast $\barZjump=1.05\pm0.06$, showing that any positive bias due to our analysis method is not large, and is unlikely to account for the mean $\Zjump\sim1.25$ drop across the real CFs.

\MyApJ{\begin{figure}[h]}
\MyMNRAS{\begin{figure}}
	\centering
	\DrawFig{\includegraphics[width=8.9cm]{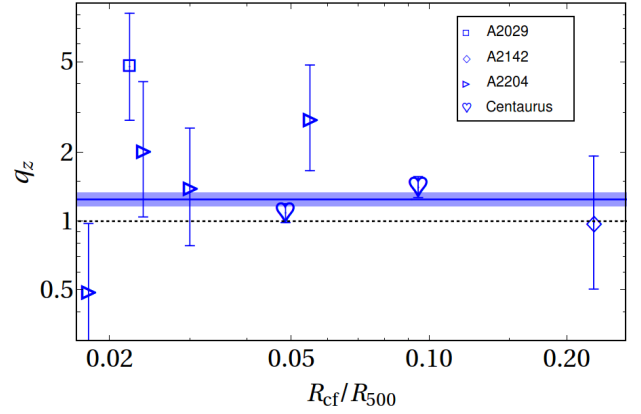}}
	\caption{The deprojected metallicity ratio $\Zjump=Z_i/Z_o$ across the newly deprojected CFs (symbols; see legend and Table~\ref{tab:jumps}), plotted against their \MyMNRAS{normalised}\MyApJ{normalized} radii $\RCF/R_{500}$  \citep[$R_{500}$ values taken from][]{ReiprichBohringer2002}.
    The weighted-mean value of $\Zjump$ (solid blue line with $1\sigma$ confidence interval as shaded region) lies above the constant metallicity line (dotted black).
    }
	\label{fig:qzr500}
\end{figure}

\MyApJ{\begin{figure}[h]}
\MyMNRAS{\begin{figure}}
	\centering
	\DrawFig{\includegraphics[width=9.85cm]{\myfig{mock_qz_rcfr500EP2.eps}}}
	\caption{The deprojected metallicity ratio $\Zjump$ across control (fictitious) CFs, plotted against their \MyMNRAS{normalised}\MyApJ{normalized} radii $\RCF/R_{500}$ in the control sample. The symbols and lines are the same as in \MyApJ{Figure}\MyMNRAS{Fig.}~\ref{fig:qzr500}.
	}
	\label{fig:mockqz}
\end{figure}

\MyApJ{\begin{table*}[h]}
\MyMNRAS{\begin{table*}}
\caption{The thermal properties ratio/difference across the CFs} 
\centering 
\setlength{\tabcolsep}{0.01em} 
{\renewcommand{\arraystretch}{1.5}%
\begin{tabular}{| c| c |c| c |c |c|c |c|c|} 
\hline 
GC& CF & $\rcf$[kpc]&$\njump$& $\Tjump$&$\Zjump$&$\Pjump$&$\Mydelta$&$R_{500}[\mbox{kpc}]$\\
(1) & (2) & (3) & (4) & (5)& (6)& (7)&(8)&(9)\\ 
\hline 
\multirow{1}{*}{A2029}& SW &$31.8\pm0.9$&$1.23\left(^{+0.07}_{-0.06}\right)\left[^{+0.26}_{-0.22}\right]$&$1.9\left(^{+0.5}_{-0.4}\right)\left[^{+1.9}_{-0.9}\right]$&$5\left(^{+3}_{-2}\right)\left[^{+13}_{-3}\right]$&$1.5\left(^{+0.4}_{-0.3}\right)\left[^{+1.6}_{-0.8}\right]$&$0.2\pm(0.1)[0.6]$&$1436^{+79}_{-86}$\\
\hline 
\multirow{1}{*}{A2142}& NW&$343.8\pm0.3$ &$1.87\left(^{+0.04}_{-0.04}\right)\left[^{+0.17}_{-0.15}\right]$&$2.2\left(^{+0.4}_{-0.3}\right)\left[^{+1.4}_{-0.9}\right]$&$1.0\left(^{+1.0}_{-0.5}\right)\left[^{+3.6}_{-0.8}\right]$&$1.18\left(^{+0.2}_{-0.2}\right)\left[^{+0.8}_{-0.5}\right]$&$1.1\pm(0.7)[2.6]$&$1500^{+121}_{-100}$\\
\hline 
\multirow{4}{*}{A2204}
& S&$23.4\pm0.8$&$1.3\left(^{+0.2}_{-0.2}\right)\left[^{+0.9}_{-0.5}\right]$&$2.3\left(^{+0.7}_{-0.5}\right)\left[^{+2.8}_{-1.3}\right]$&$0.5\left(^{+0.5}_{-0.2}\right)\left[^{+1.8}_{-0.4}\right]$&$1.8\left(^{+0.7}_{-0.5}\right)\left[^{+2.7}_{-1.1}\right]$&$0.1\pm(0.8)[2.9]$&\multirow{4}{*}{$1300^{+36}_{-29}$}\\
& E&$30.8\pm0.6$&$1.8\left(^{+0.1}_{-0.1}\right)\left[^{+0.5}_{-0.4}\right]$&$1.6\left(^{+0.4}_{-0.3}\right)\left[^{+1.4}_{-0.7}\right]$&$2\left(^{+2}_{-1}\right)\left[^{+8}_{-2}\right]$&$0.9\left(^{+0.2}_{-0.2}\right)\left[^{+0.8}_{-0.4}\right]$&$0.3\pm(0.3)[1.0]$&\\
& N&$39.0\pm1.0$&$1.2\left(^{+0.1}_{-0.1}\right)\left[^{+0.4}_{-0.3}\right]$&$1.9\left(^{+0.7}_{-0.5}\right)\left[^{+2.6}_{-1.1}\right]$&$1.4\left(^{+1.1}_{-0.6}\right)\left[^{+4.6}_{-1.1}\right]$&$1.6\left(^{+0.5}_{-0.4}\right)\left[^{+2.1}_{-0.9}\right]$&$0.3\pm(0.4)[1.5]$&\\
& W &$71.4\pm1.0$&$1.91\left(^{+0.06}_{-0.05}\right)\left[^{+0.22}_{-0.19}\right]$&$2.1\left(^{+0.4}_{-0.3}\right)\left[^{+1.4}_{-0.8}\right]$&$3\left(^{+2}_{-1}\right)\left[^{+8}_{-2}\right]$&$1.1\left(^{+0.2}_{-0.2}\right)\left[^{+0.7}_{-0.4}\right]$&$0.5\pm(0.2)[0.7]$&\\
\hline 
\multirow{2}{*}{Centaurus}& SW&$40.9\pm0.1$&$1.29\left(^{+0.03}_{-0.02}\right)\left[^{+0.09}_{-0.09}\right]$&$1.62\left(^{+0.07}_{-0.07}\right)\left[^{+0.26}_{-0.23}\right]$&$1.1\left(^{+0.1}_{-0.1}\right)\left[^{+0.4}_{-0.3}\right]$&$1.26\left(^{+0.06}_{-0.06}\right)\left[^{+0.24}_{-0.20}\right]$&$0.62\pm(0.08)[0.29]$&\multirow{2}{*}{$843^{+14}_{-14}$}\\
& E&$79.8\pm0.4$&$1.16\left(^{+0.01}_{-0.01}\right)\left[^{+0.05}_{-0.05}\right]$&$1.22\left(^{+0.04}_{-0.04}\right)\left[^{+0.14}_{-0.13}\right]$&$1.4\left(^{+0.2}_{-0.1}\right)\left[^{+0.6}_{-0.4}\right]$&$1.04\left(^{+0.03}_{-0.03}\right)\left[^{+0.12}_{-0.11}\right]$&$0.59\pm(0.03)[0.12]$&\\
\hline 
\multicolumn{3}{|c|}{Weighted-mean value}&$1.32\left(^{+0.01}_{-0.01}\right)\left[^{+0.04}_{-0.04}\right]$&$1.38\left(^{+0.03}_{-0.03}\right)\left[^{+0.13}_{-0.12}\right]$&$1.25\left(^{+0.09}_{-0.08}\right)\left[^{+0.34}_{-0.26}\right]$&$1.11\left(^{+0.03}_{-0.03}\right)\left[^{+0.11}_{-0.10}\right]$&$0.57\pm(0.03)[0.11]$&$-$\\
\hline 
\end{tabular}}\label{tab:jumps}\vspace{0.2cm}
\begin{tablenotes}
\item	Columns: (1) The GC; (2) The position of the CF at that \MyMNRAS{analysed}\MyApJ{analyzed} sector, see \MyApJ{Figures}\MyMNRAS{Figs.}~\ref{fig:A2029Figs}, \ref{fig:A2142Figs}, \ref{fig:A2204Figs}, and \ref{fig:A3526Figs}; (3) The CF distance from the \MyMNRAS{centre}\MyApJ{center}; (4) The \MyMNRAS{modelled}\MyApJ{modeled} electron number density ratio at the CF, $\njump$ (5) The \MyMNRAS{modelled}\MyApJ{modeled} temperature ratio at the CF, $\Tjump$; (6) The \MyMNRAS{modelled}\MyApJ{modeled} metallicity ratio at the CF, $\Zjump$; (7) The thermal pressure jumps at the CF, $\Pjump$; (8) The shear parameter; (9) the radius enclosing a mean density $500$ times the critical density of the Universe \citep[values taken from][]{ReiprichBohringer2002}. In round (square) brackets are the $1\sigma$ single (multi-) parameter confidence levels. The statistical errors are assumed to have symmetrical distribution in log space.
\end{tablenotes}
\end{table*}

\section{CF thermal pressure jumps}
\label{sec:PjumpsSec}

We now turn our attention to the thermal pressure profiles across CFs, examining if a pressure discontinuity can be robustly identified.
As reviewed in \S\ref{sec:Introduction}, thermal pressure jumps were previously discovered across deprojected CFs by stacking their pressure profiles (\RK).
Our analysis includes eight newly deprojected CFs and 13 previously deprojected CFs, eight of which are considered high quality (one of these eight literature CFs was not included in the {\RK} analysis).
We find that the present CF sample is sufficiently sensitive for critically testing the nature, magnitude, and significance of the pressure transition across CFs.
We \MyMNRAS{analyse}\MyApJ{analyze} our newly deprojected CFs and the deprojected CFs from the literature, both separately and in a joint analysis, thus generalising, testing, and correcting the stacking analysis of {\RK}.

\subsection{Pressure jumps: newly deprojected CFs}\label{sec:PjumpsUs}

Using the deprojected $n$ and $T$ profiles derived in \S\ref{sec:Spectral Analysis}, we compute the deprojected profile of the electron thermal pressure, $\PTh=nk_BT$, across the eight newly deprojected CFs. Here, $k_B$ is the Boltzmann constant.
The thermal pressure profiles in the different \MyMNRAS{analysed}\MyApJ{analyzed} sectors are shown in \MyApJ{Figure}\MyMNRAS{Fig.}~\ref{fig:A2029rP} for A2029, \MyApJ{Figure}\MyMNRAS{Fig.}~\ref{fig:A2142rP} for A2142, \MyApJ{Figures}\MyMNRAS{Figs.}~\ref{fig:A2204rPWE} and \ref{fig:A2204rPNS} for A2204, and \MyApJ{Figure}\MyMNRAS{Fig.}~\ref{fig:A3526rP} for Centaurus.
In order to better show deviations of the pressure from its radially-declining mean, in these figures we plot $R\,\PTh(R)$ instead of $\PTh(R)$.

The pressure contrast values inferred across these CFs are provided in Table~\ref{tab:jumps}, and are plotted against the normalised CF radius, $\RCF/R_{500}$, in \MyApJ{Figure}\MyMNRAS{Fig.}~\ref{fig:xir500}.
In our nominal analysis, statistical errors are assumed to be symmetrically distributed in $\ln(\Pjump)$; alternative assumptions on the error are explored in \S\ref{sec:PjumpsLit}.
For the A3526SW CF, we find a significant, $\sim4.3\sigma$ pressure jump, $\Pjump=1.26\pm0.06$.
The seven other newly-deprojected CFs too suggest a pressure jump, albeit at a low significance level for each individual CF.
For six of these CFs, the best fit gives $1.04\lesssim\Pjump\lesssim1.8$ in each, only marginally suggesting pressure jumps (at confidence levels ranging between $0.7\sigma$ and $1.7\sigma$); in the remaining CF, $\Pjump=0.9\pm0.2$.
The error-weighted mean value of these seven CFs is $\barPjump=1.06\pm0.03$, hinting at a pressure jump at the $\sim1.9\sigma$ confidence level.
Combining all eight newly-deprojected CFs gives $\barPjump=1.11\pm0.03$, indicating a pressure jump at the $\sim3.8\sigma$ confidence level.
As shown below, our nominal method is rather conservative; by replacing $\Pjump$ with $\xi$ (defined above as $\Pjump^{-1}$), for example, these eight CFs would indicate a $4.9\sigma$ pressure jump.

\MyApJ{\begin{figure}[h]}
\MyMNRAS{\begin{figure}}
	\centering
	\DrawFig{\includegraphics[width=8.35cm]{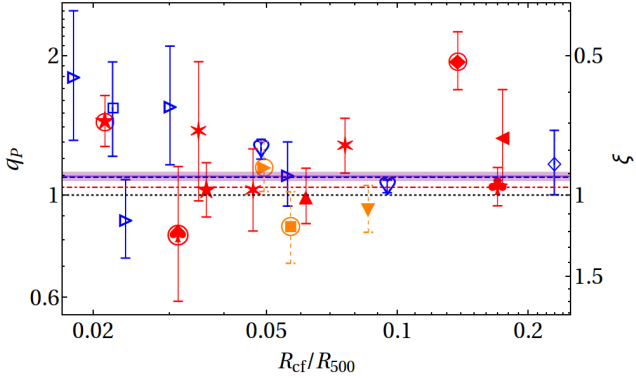}}
	\caption{The CF thermal-pressure ratio $\Pjump=\PTho/\PThi$, plotted against the normalised radius $\RCF/R_{500}$ \citep[$R_{500}$ values taken from][]{ReiprichBohringer2002,EhlertEtAl2013,SimionescuEtAl2017}. Empty symbols (filled symbols) represent the newly (previously) deprojected CFs. The symbols are the same as in \MyApJ{Figure}\MyMNRAS{Fig.}~\ref{fig:qnTM200}.
    The weighted-mean value of $\Pjump$ is shown for the subsample of newly deprojected CFs (dashed blue line), previously deprojected CFs (dot-dashed red), and the joint sample (solid purple, with shaded region showing the $1\sigma$ confidence level).
	 Problematic CFs excluded from the fit are marked by a circle (see text).
}
	\label{fig:xir500}
\end{figure}

The most significant pressure jump is found across a CF with nearly the best photon statistics, suggesting that our results are not an \MyMNRAS{artefact}\MyApJ{artifact} of poor statistics.
We examine different methods of deprojection, indicating that the results are unlikely to be deprojection \MyMNRAS{artefacts}\MyApJ{artifacts}; see discussion in Appendix~\ref{app:deproj_models}.
To test if our analysis is dominated by systematic errors, we compute $\Pjump$ values across the same control sample used in \S\ref{subsec:ZDrops}.
The weighted mean value of the control sample is $\barPjump=0.98\pm0.02$, consistent with no pressure jumps.
This result disfavours any putative bias of our analysis toward the $\Pjump>1$ values measured across real CFs.

\MyApJ{\begin{figure}[h]}
\MyMNRAS{\begin{figure}}
	\centering
	\DrawFig{\includegraphics[width=9.8cm]{\myfig{mock_qP_rcfr500EP2.eps}}}
	\caption{The deprojected thermal-pressure ratio $\Pjump$ across control (fictitious) CFs, plotted against their \MyMNRAS{normalised}\MyApJ{normalized} radii $\RCF/R_{500}$ in the control sample. The symbols and lines are the same as in \MyApJ{Figure}\MyMNRAS{Fig.}~\ref{fig:xir500}.
	}
	\label{fig:mockxi}
\end{figure}

\subsection{Including previously deprojected CFs}\label{sec:PjumpsLit}

The jump in $\PTh$ across each of the previously deprojected CFs (listed in \S\ref{subsec:DeptojCFsLit}) is estimated as $\Pjump=\Tjump/\njump=n_oT_o/(n_iT_i)$, and presented in \MyApJ{Figure}\MyMNRAS{Fig.}~\ref{fig:xir500}.
Here we use the same extrapolation procedure for $n_i$, $n_o$, $T_i$, and $T_o$, as discussed in \S\ref{subsec:qTr500}.
The results are not sensitive to the details of the extrapolation, as we show below.

To compute the weighted-mean $\barPjump$, we first assume that $\ln(\Pjump)$ is normally distributed.
We find that for the high-quality sample of eight CFs from the literature (shown in the figure as symbols without a circle; see \S\ref{subsec:DeptojCFsLit}), $\barPjump=1.06\pm0.06$, consistent with an isobaric CF at the $\sim1\sigma$ confidence level.
The weighted mean of the joint CF sample, including both newly and high-quality previously deprojected CFs, is $\barPjump=1.10\pm0.03$, inconsistent with an isobaric CF at the $\sim3.9\sigma$ confidence level.

To test if the results are sensitive to the extrapolation to the CF radius, we test the literature CFs with several other fitting models.
Namely, we replace the power-law fits of $n(r)$ and $T(r)$ with linear fits (giving $\barPjump=1.03\pm0.05$), exponential fits ($\barPjump=1.07\pm0.06$), or logarithmic fits ($\barPjump=1.04\pm0.05$).
These results are similar to the nominal, $\barPjump=1.06\pm0.06$, indicating that our approach is reasonable and perhaps slightly conservative.

The results weakly depend on the number of bins taken on each side of the CF. If we restrict the data to two bins on each side of the CF, the results become $1.07\pm0.07$ (nominal method), $1.13\pm0.09$ (linear fits), $1.10\pm0.07$ (exponential fits), and $1.09\pm0.07$ (logarithmic fits).

If one assumes that $\Pjump$, rather than $\ln(\Pjump)$, is distributed normally, the inferred pressure jumps change only slightly.
Here, we obtain
$\barPjump=1.09\pm0.03$ for the eight newly deprojected CFs,
$\barPjump=1.04\pm0.06$ for the eight previously deprojected CFs,
and $\barPjump=1.08\pm0.02$ for the joint sample.
These results are inconsistent with an isobaric transition at the $\sim3.3\sigma$, $\sim0.7\sigma$, and $\sim3.3\sigma$ confidence levels, respectively.
If one assumes, instead, that $\xi\equiv\Pjump^{-1}$ is normally distributed, the tension with an isobaric transition increases.
Here, we obtain
$\bar{\xi}=0.89\pm0.02$ for the newly deprojected CFs,
$\bar{\xi}=0.93\pm0.05$ for the literature sample,
and $\bar{\xi}=0.89\pm0.02$ for the joint sample, inconsistent with an isobaric transition at the $\sim4.9\sigma$, $\sim1.4\sigma$, and $\sim5.0\sigma$ confidence levels, respectively.
Note that the result remains fairly strong ($\sim2.7\sigma$) even if A3526SW is excluded from the joint sample.

We deduce that the results are not highly sensitive to the error distribution, that a $\sim 10\%$ pressure jump is favoured, and that an isobaric CF is ruled out at the $(3\mbox{--}5)\sigma$ level.
These results are based on a conservative selection of CFs, excluding potentially problematic profiles as discussed in \S\ref{subsec:DeptojCFsLit}.
However, it should be noted that incorporating such problematic profiles in our sample would not alter our conclusions.
In fact, including the CFs in A133 and Virgo would only raise the significance of our results.
The mean pressure jumps are nearly independent of the CFs in A1644 and RXJ1532, as their analyses carry considerable errors; the above results are obtained whether or not these CFs are included in the sample.
Including the A2199 CF in our sample would have a negligible effect on the overall sample, although the significance of the pressure jump in the literature-only sample would somewhat decrease.
Including all problematic CF profiles would lead to a more significant mean pressure jump than in the nominal method.

As a consistency test, we repeat the analysis for the {\RK} CF sample, adopting their assumption of normally-distributed $\xi$. For the full {\RK} sample of 17 CFs (including, in addition to our literature sample, also Virgo, A133, A1644, A1664, A2204, and RXJ1347, but excluding A3158), we obtain $\bar{\xi}=0.77\pm0.04$, inconsistent with an isobaric CF at the $\sim5.8\sigma$ confidence level. This significance is quite high, although somewhat lower than reported in {\RK}, due to a correction to their error propagation routine. For the nominal 14 CF sample of {\RK} (excluding also Virgo, A1644, and RXJ1347), we obtain $\bar{\xi}=0.83\pm0.05$, inconsistent with an isobaric transition at the $\sim 3.7\sigma$ level; again, this significance is fairly high, albeit lower than reported in {\RK}, for the same reasons.

\subsection{
Approximately uniform magnetisation}\label{subsec:PjumpsPic}

The above discussion indicates that $\Pjump>1$ thermal pressure jumps are robustly found across spiral CFs.
Depending on the sample selection, we find $10\%$ to $30\%$ pressure jumps, at fairly high (stacked: typically $>3\sigma$, and up to $5\sigma$) significance levels.
These jumps do not seem to be highly sensitive to photon statistics, deprojection bias, or assumptions on the underlying error distribution, and do not appear in control samples.
A significant pressure jump is found even without stacking in one case, presenting with $\Pjump=1.26\pm0.06$ ($4.3\sigma$) in the A3526SW CF.

Having identified CFs as long-lived discontinuity surfaces, the total pressure across them should be continuous.
Our results then indicate that an inhomogeneous, nonthermal pressure component $P_{nt}$ must exist, carrying at least $10\%\mbox{--}30\%$ of the thermal pressure just below the CF, consistent with {\RK}.
This conclusion alone does not uniquely identify the nature of the nonthermal component, nor does it determine whether this component is confined below the CF or simply discontinuously stronger below it.
Nevertheless, the sharpness of the jump indicates that $P_{nt}$ is predominantly magnetic,
\begin{equation}
P_B\equiv \frac{B^2}{8\pi}\gtrsim \frac{1}{2}P_{nt} \simeq 0.1 \PTh \, ;
\end{equation}
see discussion in {\RK}.

Inspecting Fig.~\ref{fig:xir500}, it is interesting to examine the variation of $\Pjump$ with CF radius.
We thus model our various samples of CFs with $\Pjump\propto (\RCF/R_{500})^b$.
The best-fit power-law index is found to be $b=-0.17\pm0.06$ ($\chi^2_{\nu}\simeq1.4$) for the newly deprojected CFs sample,
$b=0.01\pm0.10$ ($\chi^2_{\nu}\simeq0.7$) for the sample of deprojected CFs from the literature, and $b=-0.12\pm0.05$ ($\chi^2_{\nu}\simeq1.1$) for the joint sample.
It is unclear if the slightly negative values of $b$ in the new and joint samples reflect a physical trend.
These values are entirely governed by the high statistics in Centaurus; without either of the CFs in this cluster, $b$ would become consistent with zero at the $<1\sigma$ level ($b=-0.07\pm0.09$ and $b=-0.03\pm0.06$ without A3526SW, and $b=-0.03\pm0.10$ and $b=-0.07\pm0.07$ without A3526E).
Moreover, the control sample of mock CFs shown in Fig.~\ref{fig:mockxi} yields a similar, $b=-0.07\pm0.07$.
Pearson
correlation tests between $\ln (\Pjump)$ and $\ln (\RCF/R_{500})$ give $\bar{\corl}=-0.3\pm0.3$ for the newly deprojected CFs sample, whether or not the CFs in Centaurus are included.
The joint sample yields $\bar{\corl}=-0.3^{+0.3}_{-0.2}$ ($\bar{\corl}=-0.1\pm0.2$) if we include (exclude) the two CFs in Centaurus.

We conclude that the results are consistent with a roughly uniform value of $\Pjump$ among different CFs in different GCs; the statistics are insufficient to identify any trend here.
The preceding discussion thus indicates a roughly constant magnetic equipartition value, $\eta_{\tiny \mbox{B}}\equiv P_B/\PTh\simeq 0.1$, below all CFs.
Such a magnetisation level could arise if, for example, shear magnetisation becomes saturated below CFs, as suggested in {\RK} based on the strength of the field; see discussion in \S\ref{sec:Discussion}.

\section{
Projection hides pressure drops}\label{sec:PjumpsProj}

Several studies \citep[\eg][]{VikhlininEtAl2001,MarkevitchVikhlinin2007,MachacekEtAl2011,GastaldelloEtAl2013,LalEtAl2013,GhizzardiEtAl2014} have estimated the $\PTh$ transitions across CFs without fully deprojecting the data.
To critically test this simpler approach and its systematics, we \MyMNRAS{analyse}\MyApJ{analyze} similarly projected pressure profiles across CFs in our sample and from the literature.
Using the same methods applied above to the deprojected profiles, here we compute the projected pressure transitions $\PjumpP$ across CFs, and compare them with the corresponding deprojected jumps $\Pjump$ inferred in \S\ref{sec:PjumpsSec}.

We follow the standard approach in the literature for studying the projected pressure, by combining the projected temperature with the deprojected density.
Namely, we compute the deprojected density contrast $\njump\equiv n_i/n_o$ and the projected temperature contrast $\TjumpP\equiv \TtwoDo/\TtwoDi$, and define the projected pressure contrast as
\begin{equation}\label{eq:qPProj}
  \PjumpP \equiv \TjumpP/\njump \, ;
\end{equation}
We compute $\TjumpP$ by \MyMNRAS{modelling}\MyApJ{modeling} $\TtwoD(R)$ with a (different) power-law profile on each side of the CF, as in Eq.~(\ref{eq:T_PLaw}), and extrapolating the inside and outside profiles to the CF position to obtain $\TtwoDi$ and $\TtwoDo$.

We apply this procedure to our sample of eight newly deprojected CFs, as well as to CFs from the literature.
The resulting  $\PjumpP$ profiles are plotted against $\RCF/R_{500}$ in \MyApJ{Figure}\MyMNRAS{Fig.}~\ref{fig:xiproj}.
The literature sample used here differs from that of \S\ref{sec:PjumpsSec}, because there we required the deprojected, rather than projected, temperature profiles.
Consequently, here we are able to include also CFs in galaxy groups (GGs), and not only in GCs.

As in \S\ref{sec:PjumpsSec}, here too we only use profiles with data available for at least two bins on each side of the CF. This leads to the following literature sample of seven CFs in six GCs and seven CFs in four GGs (one CF per object, unless otherwise stated).
The GCs include  A133 \citep[two CFs;][]{RandallEtAl2010},  A496 \citep{TanakaEtAl2006}, A1664 \citep[][]{KirkpatrickEtAl2009}, A3158 \citep{WangEtAl2010}, RXJ1532 \citep{HlavacekLarrondoEtAl2013}, and Virgo \citep{UrbanEtAl2011}.
The GGs include IC1860 \citep{GastaldelloEtAl2013}, NGC5098 \citep[two CFs;][]{RandallEtAl2009},  NGC5846 \citep[two CFs;][]{MachacekEtAl2011}, and 3C449 \citep[two CFs;][]{LalEtAl2013}.

The projected profiles in A133 and RXJ1532 suffer from problems similar to those pertaining to the corresponding deprojected profiles (see \S\ref{subsec:DeptojCFsLit}).
Moreover, in A133, both CFs (and not only the inner one) are misaligned with the projected bins.
In NGC5098, the projected profiles across both CFs suffer from the same problem as in A133 and RXJ1532: in all cases the CF is misaligned with the projected bins.
The six potentially problematic profiles found in these GCs and in Virgo are shown circled in \MyApJ{Figure}\MyMNRAS{Fig.}~\ref{fig:xiproj}, but are not used in our \MyMNRAS{modelling}\MyApJ{modeling}.
Our high quality projected-CF sample from the literature, after excluding these CFs, consists of three CFs in three GCs and five CFs in four GGs.

As \MyApJ{Figure}\MyMNRAS{Fig.}~\ref{fig:xiproj} shows, the $\PjumpP$ value across each of the projected CFs, in both the new sample and the high-quality literature sample, is consistent with a $\PjumpP<1$ pressure drop, rather than a pressure jump.
This is not the case for the CFs in A133 and in Virgo, both of which are problematic and thus excluded from the statistical analysis, as mentioned above.
The weighted-mean pressure contrast is $\PjumpPbar=0.90\pm0.01$ for the newly analysed CFs, $\PjumpPbar=0.77\pm0.04$ for the literature sample, and $\PjumpPbar=0.89\pm0.01$ for the joint sample.
These results strongly favour a projected pressure drop, rather than a jump, and  are inconsistent with an isobaric ($\PjumpP=1$) projected CF at respectively the $\sim8\sigma$, $\sim5.9\sigma$, and $\sim9.2\sigma$ confidence levels.
We conclude that in a typical CF, $\PjumpPbar<1<\barPjump$.

The $\PjumpPbar<\barPjump$ behaviour is anticipated from projection effects (\RK).
The projected $\TtwoD$ in a radial bin is affected by all the bins above it.
However, as the density profile is steep, so is the emission measure, and the effect is typically modest and short range.
The effect of projection on a certain bin is thus dominated by the adjacent bin from above, and can be noticeable if the temperature contrast between the two is large.
In a CF, $T_o$ is larger than $T_i$, by a factor of $\sim2$ or so, so the bin $i$ immediately inside the CF is affected by the first bin $o$ outside the CF, leading to $\TtwoDi>T_i$.
The temperature above the CF changes slowly with radius, so $\TtwoDo\simeq T_o$ is not sensitive to projection.
Combining these results, projection leads to $\TtwoDo/\TtwoDi=\TjumpP<\Tjump=T_o/T_i$.
As both projected and deprojected analyses employ the deprojected density, it follows that $\PjumpP<\Pjump$.
Indeed, we verify that $\PjumpP<\Pjump$ holds not only on average, but also on a CF-by-CF basis, for the CF sub-sample that has both projected and deprojected profiles.

To examine if our results, $\barPjump/\PjumpPbar\simeq 1.1/0.9\simeq 1.2$, are self-consistent, consider a simple projection model.
Assume, for simplicity, a spherical geometry, $\Tjump=2$, and a $\Lambda\propto1$ emission-measure cooling function.
For the typical radial bins, logarithmically separated with factor $\sim 1.3$ increments, projection of the bin just outside the CF gives in this model $P_{i,proj}\simeq 1.4 P_i$. Consequently, $\Pjump/\PjumpP=(P_{i,proj}P_o)/(P_{o,proj}P_i)\simeq 1.4$, consistent with our results.
Namely, projection has a large, $\sim 40\%$ on the estimated pressure jump under these conditions, so deprojection is essential for uncovering the nature of the transition.

\MyApJ{\begin{figure}[h]}
\MyMNRAS{\begin{figure}}
	\centering
	\DrawFig{\includegraphics[width=8.6cm]{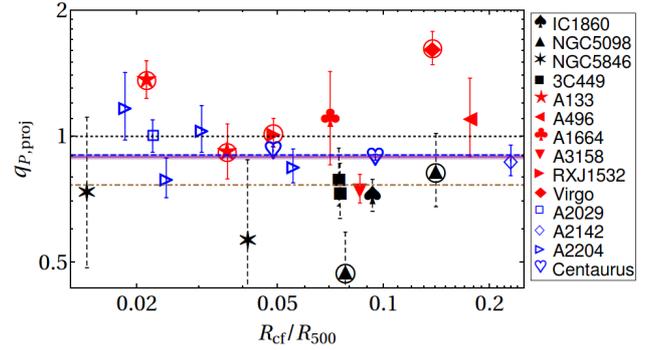}}
	\caption{
	The projected thermal-pressure ratio, $\PjumpP=(n_o\TtwoDo)/(n_i\TtwoDi)$, plotted against the \MyMNRAS{normalised}\MyApJ{normalized} CF radius $\RCF/R_{500}$, including both new (empty symbols) and literature (filled symbols) CFs (see text and legend).
Literature CFs colored red (black) with solid (dashed) error bars are found in GCs (GGs).
	Weighted mean values of $\PjumpP$ are shown for the new CFs (dashed blue line), the literature CFs (dot-dashed brown), and the joint sample (solid purple, with shaded region showing the $1\sigma$ confidence level).
	Problematic CFs excluded from the fit are marked by a circle (see text). Data references for $R_{500}$: \citet{ReiprichBohringer2002,GastaldelloEtAl2007,SunEtAl2009,EhlertEtAl2013,ZhuEtAl2016}.}
	\label{fig:xiproj}
\end{figure}

Note that while our projected pressure profiles are, as usual, based on the product of the projected temperature and the deprojected density (see Eq.~\ref{eq:qPProj}), our nominal density profiles themselves are derived using the \emph{deprojected} temperature (see \S\ref{sec:Spectral Analysis}).
Density deprojection is quite robust; nevertheless, we repeat the analysis using a modified deprojected density obtained in Appendix~\ref{app:deproj_models} with the emission-measure model, independent of temperature.
This method gives $\njump$ values slightly greater than our nominal values, by $\sim5\%$, resulting in slightly more substantial projected pressure drops than inferred in the nominal method.

\section{Shear flows along cold fronts}\label{sec:delta}

Spiral-type CFs typically show a strong, order unity discontinuity in the hydrostatic mass $\MHE$.
Namely, just above the CF one infers an $\MHE$ value considerably larger than $\MHE$ just below the CF \citep[see \eg][and \UK]{MarkevitchEtAl2001}.
Such a discontinuity is naturally interpreted as arising from a tangential bulk flow along, and below, the CF (\UK).
We examine such hydrostatic mass discontinuities, across both newly deprojected and literature CFs, and estimate the associated shear.

We assume that the gravitating mass has a spherically symmetric distribution, and define $M(r)$ as the mass enclosed inside $r$.
For simplicity, we assume that the pressure is predominantly thermal (as suggested by our above results; see \S\ref{sec:PjumpsSec}), and that its distribution is approximately spherically symmetric, so $P_{tot}\simeq \PTh(r)$.
Notice that this assumption is consistent with our deprojection procedure (which implicitly assumes a constant pressure in bins aligned with the CF and not, in general, spherical; see \S\ref{sec:Spectral Analysis}) only in the limit of a nearly spherical CF (like in A2204) or a sufficiently narrow sector.
But in any case, the deprojection geometries used here (a prolate spheroid elongated in the plane of the sky; see \S\ref{subsec:ModParameters}) and in other studies are not fully consistent with a 3D quasi-spiral CF manifold.

In \S\ref{subsubsec:ThermalEqb}, we present a simplified analysis of the $\MHE$ jumps assuming a non-rotating, spherical CF surface.
The resulting $\MHE$ jumps are presented and analysed in \S\ref{subsubsec:deltaResults}.
We generalise the analysis to the more realistic picture of a rotating, quasi-spiral CF manifold, in \S\ref{subsubsec:shearflows}, showing that our conclusions are not substantially modified with respect to the spherical picture.

\subsection{Spherically symmetric limit}\label{subsubsec:ThermalEqb}

For simplicity, let us first approximate the gas distribution as spherically symmetric, $n=n(r)$, and assume that the GC is non-rotating.
The hydrostatic mass $\MHE$, defined as the gravitating mass in hydrostatic equilibrium, is then directly determined from the density and the radial derivative of the total pressure, as inferred from the Euler equation for a steady flow,
\begin{align}\label{eq:EulerStatic}
&0=\unit{r}\cdot\frac{d\vect{u}}{dt}=-\frac{1}{\mu n}\frac{\partial \PTh}{\partial r}-\frac{G \MHE}{r^2}\fin
\end{align}
Here, $\{r,\theta,\phi\}$ are spherical coordinates, where $r$ is the radius from the GC centre, $\theta$ is the polar angle, and $\phi$ is the azimuth.
The radial component of the velocity field $\vect{u}$ is denoted $\myu_r$, $\mu\simeq0.6m_p$ is the average particle mass, $m_p$ is the proton mass, and $G$ is the gravitational constant.

Under the above assumption of a spherical, non-rotating gas distribution, one can measure $\MHE(r)$ using the deprojected $n(r)$ and $\PTh(r)$ profiles along the bisector, as $\MHE\simeq -r^2\PTh'(r)/(\mu n G)$.
The radial $\MHE(R)$ profiles on the plane of the sky are shown in \MyApJ{Figure}\MyMNRAS{Fig.}~\ref{fig:A2029rM} for A2029, \MyApJ{Figure}\MyMNRAS{Fig.}~\ref{fig:A2142rM} for A2142, \MyApJ{Figures}\MyMNRAS{Figs.}~\ref{fig:A2204rMWE} and \ref{fig:A2204rMNS} for A2204, and \MyApJ{Figure}\MyMNRAS{Fig.}~\ref{fig:A3526rM} for Centaurus, for each of the eight newly analysed sectors.
In order to better show deviations of the hydrostatic mass from its radially-increasing mean, in these figures we plot $R^{-1}\,\MHE(R)$ instead of $\MHE(R)$.

These figures suggest that $\MHE$ discontinuously jumps, $\MHEi<\MHEo$, across each of the eight newly deprojected CFs.
As the gravitating mass is continuous, such discontinuities indicate that the above assumptions break down.
A natural interpretation is that the ICM near the CF is not in hydrostatic equilibrium, such that the acceleration term in Eq.~(\ref{eq:EulerStatic}) does not vanish.
As there can be no flow across a tangential discontinuity, the measured $\MHE$ jump thus implies the presence of a tangential bulk flow, at least below the CF.
As long as the CF can be approximated as spherical, and assuming that the gas is not rapidly rotating, one can then measure a dimensionless shear parameter quantifying the normalised centripetal-acceleration difference across the CF (\UK),
\begin{equation}\label{eq:deltaUK}
\frac{\myu_o^2-\myu_i^2}{c_{s,i}^2}\simeq \delta\equiv\frac{\RCF}{\gamma\, k_BT_i}\left(\frac{1}{n_o}\frac{\partial\PTho}{\partial R}-\frac{1}{n_i}\frac{\partial\PThi}{\partial R}\right)\fin
\end{equation}
Here, $\myu$ is velocity of the flow along the CF, $c_{s}=(\gamma k_B T/\mu)^{1/2}$ is the speed of sound, $\gamma=5/3$ is the adiabatic index, and we define the shorthand notation $\partial_R \ldots\equiv(\partial_R \ldots)_{\varphi,l}$, where $\varphi$ is the projected azimuthal angle, and $l$ is (as defined above) the distance along the LOS.
It is convenient to define
\begin{equation}\label{eq:Upsilon}
\Upsilon\equiv-\delta=\frac{\RCF}{\gamma\, k_BT_i}\left[\frac{\partial_R \PTh}{n}\right]_o^i \simeq \frac{[\myu^2]_o^i}{c_{s,i}^{2}} \coma
\end{equation}
which is typically positive.
Here, $[F]_o^i\equiv F_i-F_o$ for any quantity $F$.

Alternative interpretations appear to be far less plausible, as discussed in {\UK}.
For example, one may try to attribute the change in the pressure slope to the presence of some nonthermal pressure component.
However, the large $\MHE$ contrast consistently observed requires the implied nonthermal pressure gradient, $\partial_rP_{nt}$, to be at least comparable to its thermal counterpart, $\partial_rP_{t}$, below the CF. Such a strong gradient would lead to a large nonthermal contribution to the pressure below the CF, in contrast to the small nonthermal component, $\eta_{\tiny \mbox{B}}\simeq0.1$, we derived in \S\ref{sec:PjumpsSec}.

Additional, competing effects are similarly implausible.
For example, a viscous acceleration term,
\begin{equation}\label{eq:acc_visc}
a_{\mbox{\tiny visc}}=\frac{1}{\mu n}\bm{\nabla}\cdot\left\{\eta\left[\bm{\nabla}\vect{u}+\left(\bm{\nabla}\vect{u}\right)^\mathsf{T}\right]\right\} \coma
\end{equation}
where $\eta$ is the dynamic viscosity, was neglected in Eq.~(\ref{eq:EulerStatic}) with respect to the pressure gradient term, $a_{\mbox{\tiny p}}=-\bm{\nabla}\PTh/(\mu n)$; namely, the Reynolds number $\mbox{Re}$ was assumed large.
(This remains the relevant criterion also when considering the difference in the Euler equation between the two sides of the CF, as the fractional jump in $a_{\mbox{\tiny p}}$ is found to be substantial.)
In terms of the Spitzer viscosity, $\eta_{\tiny\mbox{S}}\simeq1.4\times10^3T_5^{5/2}\erg\se\cm^{-3}$ \citep{Braginskii1965,spitzer2006}, this assumption becomes
\begin{equation}\label{eq:eta_no_B}
1\gg \frac{a_{\mbox{\tiny visc}}}{a_{\mbox{\tiny p}}} \simeq\mbox{Re}^{-1}\simeq 0.3\left(\frac{\eta}{\eta_{\tiny\mbox{s}}}\right)\left(\frac{T_5^{2}}{\MyMach\,n_3\,\Delta_{10}}\right)\coma
\end{equation}
where we defined $k_B T\equiv5T_5\keV$, $n\equiv 10^{-3}n_3\cm^{-3}$, and assumed a flow of Mach number $\MyMach$ confined to a layer of width $\Delta\equiv10\Delta_{10}\kpc$.
This criterion is satisfied because the viscosity is thought to be suppressed near CFs \citep[\eg][]{RoedigerEtAl2013,ZuHoneEtAl2015,ZuhoneRoediger2016} by a factor $\eta/\eta_{\tiny\mbox{s}}\sim0.1$.
Indeed, the shear flow inferred below the CF is accompanied by tangential magnetization, shown in {\RK} and in \S\ref{subsec:PjumpsPic} to be strong, capable of suppressing viscosity by orders of magnitude \citep{kaufman1960}.

Next, we derive the value of $\Mydelta$ for the newly (\S\ref{sec:Data Reduction}) and previously (\S\ref{subsec:DeptojCFsLit}) deprojected CFs, discussed above.

\subsection{Shear flows across CFs: results}\label{subsubsec:deltaResults}

We derive the value of the shear parameter, $\Mydelta$, for each of the newly and previously deprojected CFs according to Eq.~(\ref{eq:Upsilon}).
The resulting $\Mydelta$ values are plotted against $\RCF/R_{500}$ in Fig.~\ref{fig:deltar500}; the results for the newly deprojected CFs are presented in Table~\ref{tab:jumps}.
As Fig.~\ref{fig:deltar500} shows, all (21 out of 21) of our analysed CFs show $\Mydelta>0$, indicating bulk flows below all these CFs.
This result, valid also for quasi-spiral patterns (see \S\ref{subsubsec:shearflows}), testifies to the ubiquity of the phenomenon.

The weighted-mean value of the eight newly deprojected CFs is $\barMydelta=0.57\pm0.03$, inconsistent with a stationary fluid at the $\sim19\sigma$ confidence level.
To test if our analysis introduces systematic errors, we compute mock $\Mydelta$ values across the same control sample used in \S\ref{subsec:ZDrops}; see Fig.~\ref{fig:mockdelta}.
The weighted mean value of the control sample is $\barMydelta=0.01\pm0.02$, consistent with a stationary fluid, disfavouring any putative bias in our analysis.
We confirm that our $\Mydelta$ values in the A2204E and A2204W CFs agree (within $<1\sigma$) with those of \citet{ChenEtAl2017}, which were deduced using a different deprojection method.

The weighted-mean value of the eight previously deprojected CFs is
$\barMydelta=0.32\pm0.10$, inconsistent with a stationary fluid at the $\sim3.3\sigma$ confidence level.
For the joint-sample of both newly and previously deprojected CFs the weighted-mean value is $\barMydelta=0.55\pm0.03$ ($\sim20\sigma$).
The latter value is strongly dominated by the newly deprojected CFs, due to their superior statistics.

This mean $\barMydelta$ value is more significant, but somewhat lower, than the $\barMydelta=0.78\pm0.14$ ($5.5\sigma$) value reported by {\UK}.
The difference is mostly due to our conservative exclusion of CFs suspected of large systematic uncertainties, in particular RXJ1720 (where only one bin is available below the CF), the outer CF in A1644 (where a merger might distort the result), and in 2A0335 (where the profile analyzed by {\UK} was modified by an additional CF reported later by {\RK}).

The {\UK} analysis also included deprojection results superseded by improved ones; in particular, the first deprojections of the A2204E and A2204W CFs \citep{SandersEtAl2005} were noisier and gave much larger $\Mydelta$ values than obtained here \citep[and in][]{ChenEtAl2017}.
As a consistency check, we confirm that in CFs where the present analysis can be directly compared with {\UK} (in A1644, A1795, and A496), the results are in good agreement.

When interpreted in the spherical, non-rotating gas picture, these results indicate a nearly sonic flow below the CF. One should, however, consider the results in a more realistic picture, in which the ICM may be rotating, the CFs have a quasi-spiral structure, and the morphology depends on the unknown inclination angle between the LOS and the spiral's symmetry axis. These issues are considered in \S\ref{subsubsec:shearflows} below. As we show, the interpretation of the results does not change considerably with respect to the simple, spherical picture.

\begin{figure}
\vspace{0.2cm}
\centering
\DrawFig{\includegraphics[width=8.4cm]{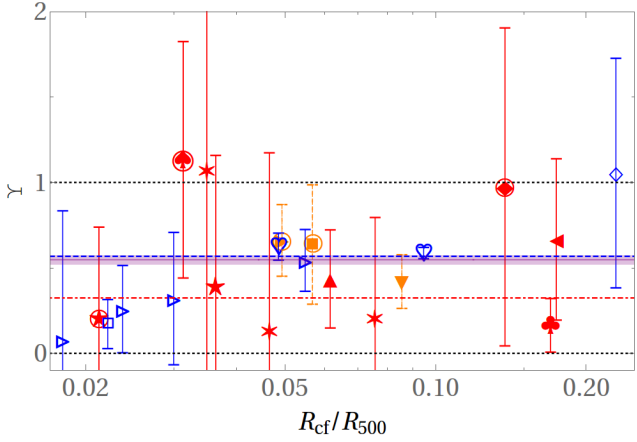}}
\caption{The shear parameter, $\Mydelta$, plotted against the normalised CF radius $\RCF/R_{500}$. Empty symbols (filled symbols) represent the newly (previously) deprojected CFs. The symbols are the same as in \MyApJ{Figure}\MyMNRAS{Fig.}~\ref{fig:qnTM200}.
The weighted-mean value of $\Mydelta$ is shown for the subsample of newly deprojected CFs (dashed blue line), previously deprojected CFs (dot-dashed red), and the joint sample (solid purple, with shaded region showing the $1\sigma$ confidence level).
The region between the dotted black, $\Mydelta =0,1$ lines is broadly consistent with subsonic flows below the CF.
}
\label{fig:deltar500}
\end{figure}

\begin{figure}
	\vspace{0.2cm}
	\centering
	\DrawFig{\includegraphics[width=9.5cm]{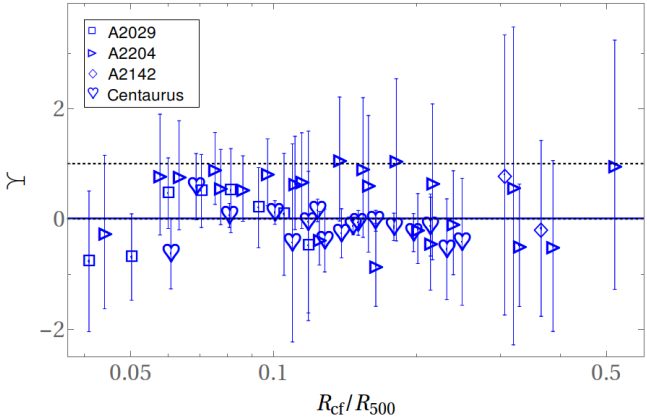}}
	\caption{The shear parameter, $\Mydelta$, plotted against the normalised CF radius $\RCF/R_{500}$, in the control samples.
	The symbols and lines are the same as in Fig.~\ref{fig:deltar500}.
	}
	\label{fig:mockdelta}
\end{figure}

\subsection{Quasi-spiral structure}\label{subsubsec:shearflows}

In practice, deviations from spherical symmetry are evident in all well-observed GCs.
We therefore generalise the discussion to incorporate the quasi-spiral pattern typically observed.
For simplicity, we retain the spherical definition of $M_h$ in Eq.~(\ref{eq:EulerStatic}).
It is necessary, however, to generalize the interpretation of $\Mydelta$ on the RHS of Eq.~(\ref{eq:Upsilon}), taking into account projection effects and the 3D spiral structure. The analysis largely follows \UKspirals, focusing on the semi-circular variant of the discontinuity manifold therein.

The observation of a quasi-spiral CF pattern in a large fraction of well-observed GCs suggests that there exists an optimal viewing angle in which a putative observer would see the projected CF pattern as a continuous, monotonic, and undistorted planar spiral.
Let us assume that such a unique viewing angle exists.
Define the plane harboring the GC centre and perpendicular to this viewing angle as the equatorial plane.
Denote the normal to this plane as the axis of symmetry as $\unit{\myz}$, and a line of sight parallel (perpendicular) to this axis as face on (edge on) viewing.
We assume that the centre of the spiral pattern in the equatorial plane coincides with the GC centre.
In the analysis below, we adopt both cylindrical $\{\varrho,\phi,\myz\}$ and spherical $\{r,\theta,\phi\}$ coordinate systems, both centred on the GC centre, with $\theta=0$ identified as the $+\myz$ direction and $\varrho$ as the distance from the $\myz$-axis, such that $r^2=\varrho^2+\myz^2$.

Adiabatic numerical simulations of sloshing in a GC \citep[\eg][]{AscasibarMarkevitch2006,RoedigerEtal2011}
suggest that in a constant $\phi$ cut (edge-on viewing), the CF discontinuity would appear semi-circular, $\rcf(\theta,\phi)\simeq \rcf(\phi)$.
Hence, the CF radius of curvature in the polar direction is approximately fixed by the CF radius, $R_\theta(\phi) \simeq \rcf(\phi)$.
Equivalently, the projected CF radius would be nearly constant, $\RCF \simeq \const\simeq \rcf(\phi)$, in edge-on viewing.
For simplicity, we adopt such a geometry here, although simplistic analytical models of spiral flows that take into account radiative cooling \citep[\UKspirals;][]{ReissKeshet2015} suggest that $R_\theta>\rcf$.

For this simple CF geometry in the polar direction, $\rcf(\theta,\phi)\simeq \rcf(\phi)\simeq R_\theta(\phi)$, the two-dimensional discontinuity surface may be compactly parameterised in 3D as
\begin{equation}\label{eq:Spiral_Sf}
0=S(\bm{r})=S(r,\theta,\phi)\simeq S(r,\phi) =r-f(\phi)\coma
\end{equation}
where $f(\phi)$ is a monotonic function, chosen, without loss of generality, to be monotonically increasing.

In projection, the CF is observable at viewing angles for which the line of sight $\bm{l}$ is tangent to the discontinuity surface,
\begin{equation}\label{eq:projdmnd}
  \unit{l}\cdot\grad S=0\fin
\end{equation}
The precise appearance of the projected CF curve depends on the inclination angle, $\incAng$, defined as the angle between $\bm{l}$ and the $\mathsf{z}$-axis.
The measured values of $\alpha_{\tiny\mbox{proj}}$ indicate that $f$ varies slowly, $f'(\phi)/f(\phi)\lesssim \tan(\alpha_{\tiny\mbox{proj}})\simeq 0.2$, although this depends somewhat on projection effects.

We may now relate $\Mydelta$ to the dynamics.
First, generalise the spherical momentum Eq.~(\ref{eq:EulerStatic}) to the general Euler equation
\begin{equation}\label{eq:Euler}
\frac{d\vect{u}}{dt}  =\frac{\partial \vect{u}}{\partial t}+\left(\vect{u}\cdot\bm{\nabla}\right)\vect{u}
 = -\frac{\bm{\nabla}\PTh}{\mu n}-\frac{G M}{r^2}\unit{r}
\fin
\end{equation}
Under the present assumptions, the flow is driven by central forces only, implying a planar fluid motion.
Namely, the path function of any fluid element is confined to a plane that contains the GC centre.

Next, we assume that at least locally, there exists a frame of reference in which the CF pattern is stationary.
This frame may be rotating about the $\myz$-axis at some angular frequency $\omega$, which could vary slowly in space and time; the derivatives of $\omega$ are assumed negligible in what follows.
By symmetry, the angular velocity is chosen in the $\myz$ direction, $\bm{\Omega}=\omega\unit{\myz}$.
In our notations, a trailing spiral pattern corresponds to $\omega<0$.
For simplicity, we assume a steady flow in this rotating frame.
Under these assumptions, in the vicinity of the CF, the Euler equation (\ref{eq:Euler}) becomes
\begin{equation}\label{eq:Euller2}
\left(\vect{u}\cdot\bm{\nabla}\right)\vect{u} \simeq -\frac{\bm{\nabla}\PTh}{\mu n}-\frac{G M}{r^2}\unit{r}- 2\bm{\Omega}\times\vect{u}-\bm{\Omega}\times(\bm{\Omega}\times\vect{r})\fin
\end{equation}

The radial component of Eq.~(\ref{eq:Euller2}) may be written as
\begin{equation}\label{eq:NavierStokesSS}
\frac{  {\breve{\myu}}_{\phi}^2+\myu_\theta^2}{r} - \dot{\myu}_r \simeq \frac{\PTh'(r)}{\mu n} + \frac{G M}{r^2}
\coma
\end{equation}
where we denote $\dot{F}\equiv \left(\vect{u}\cdot\bm{\nabla}\right)F$
as the derivative along the flow,
and a breve designates the non-rotating frame; in particular,
$\breve{\myu}_{\phi}\equiv\myu_\phi+\omega\varrho$.
Taking the difference of this equation between the two sides of the CF, and noting that $\PTh'(r) = (r/R)\partial_R \PTh$, we may relate the measured $\Mydelta$
to the 3D flow,
\begin{eqnarray} \label{eq:Upsilon_spirals}
\Mydelta & \equiv & \frac{\RCF}{\gamma\, k_BT_i}\left[\frac{\partial_R \PTh}{n}\right]_o^i
= \frac{\RCF^2/\rcf}{c_{s,i}^2} \left[\frac{\PTh'(r)}{\mu n}\right]_o^i  \nonumber \\
& \simeq &
\left(\frac{\RCF}{\rcf}\right)^2\left(\frac{\left[{\breve{\myu}}_{\phi}^2+\myu_\theta^2\right]_o^i}{c_{s,i}^2}  - \frac{\rcf\left[\dot{\myu}_r\right]_o^i}{c_{s,i}^2}\right) \coma
\end{eqnarray}
to be evaluated where the LOS meets the tangent to the CF.

In the rotating frame of reference, where the flow is assumed steady, the velocity component normal to the CF must vanish in its vicinity, by the definition of the CF as a tangential discontinuity.
The measured large, near-unity values of $\Mydelta$, and the constraint $\RCF\leq\rcf$, thus indicate that the $\dot{\myu}_r$ term in Eq.~(\ref{eq:Upsilon_spirals}) is subdominant.
Otherwise, the radial velocity $u_r$ would quickly become large, and even supersonic, in contradiction to the slow velocity  anticipated from the nearly spherical CF surface.

We therefore see that $\Mydelta\sim 0.6$ provides an approximate measure of the difference in the squared non-radial velocity in the inertial frame, $\left[{\breve{\myu}}_{\phi}^2+\myu_\theta^2\right]_o^i=\left[{\breve{\myu}}^2-\myu_r^2\right]_o^i$.
Furthermore, as $\Mydelta$ is large, we deduce that the flow is considerably faster inside, \ie below, the CF.
Up to the correction factor $(\RCF/\rcf)^2$, then, $\Mydelta\simeq \MyMach_i^2\equiv [({\breve{\myu}}_{\phi}^2+\myu_\theta^2)/c_{s}^2]_i$ approximately measures the square of the Mach number $\MyMach_i$ associated in the inertial frame with the non-radial velocity inside the CF.

The implied Mach number, $\bar{\MyMach}_i\simeq \langle\Mydelta^{1/2}\rangle=0.76\pm0.02$ for the joint CF sample, somewhat underestimates the inner flow tangential to the CF, because \emph{(i)} the latter also has some radial component; \emph{(ii)} the factor $(\RCF/\rcf)$ is smaller than unity; and \emph{(iii)} the fluid above the CF is not precisely stationary.
However, the spiral CFs observed are quite tight, so the radial component of the flow is small and the factor $\RCF/\rcf$ is close to unity.
Moreover, while a fast flow above the CF or an extreme projection with a small $(\RCF/\rcf)$ could render $\MyMach_i\gg\Mydelta^{1/2}$, this seems unlikely as the inner flow would become highly supersonic.
We conclude that our results robustly indicate that the tangential flow below the CF is nearly sonic, in agreement with {\UK}.

In the limit of a tight spiral, where the CF surface approaches a sphere, the interpretation (\ref{eq:Upsilon_spirals}) of $\Mydelta$ reduces to the spherical picture (\ref{eq:Upsilon}) in the inertial frame, regardless of the rotation frequency $\omega$.
The analysis similarly reduces to that of the spherical case, regardless of $\omega$ and the spiral function $f(\phi)$, if the flow is polar in the inertial frame, \ie if $\bm{\breve{\myu}}=\breve{\myu}\,\unit{\theta}$; however, such a flow pattern seems rather unlikely.

\section{Summary and Discussion}\label{sec:Discussion}

We perform an in-depth study of the accessible thermal transitions across CFs, comparing different deprojection and projection methods.
In addition to an analysis of all deprojected CFs from the literature (see \S\ref{subsec:DeptojCFsLit}), we process raw \emph{Chandra} data from the GCs A2029, A2142, A2204, and Centaurus.
The $\Sx$ maps of these four GCs (see \S\ref{sec:Data Reduction}) show sharp edges, which we confirm as CFs.
We spectrally analyse the sectors above and below these CFs (see \S\ref{sec:Spectral Analysis}) and derive the projected and deprojected thermal profiles in these sectors;
see Figs.~\ref{fig:A2029Figs}--\ref{fig:A2029Thermal} for A2029, Figs.~\ref{fig:A2142Figs}--\ref{fig:A2142Thermal} for A2142, Figs.~\ref{fig:A2204Figs}--\ref{fig:A2204NSThermal} for A2204, and Figs.~\ref{fig:A3526Figs}--\ref{fig:A3526Thermal} for Centaurus.

In our spectral analysis, we first derive the projected $\Sx$, $\TtwoD$, and $\ZtwoD$ profiles in radially logarithmic bins, aligned with each CF and centred on the X-ray peak.
The data are then fit by projected 3D models.
We adopt separate $n$ ($\beta$-model or power law), $T$ (power-law), and $Z$ (power-law) profiles above and below each CF (see Eqs.~\ref{eq:nBetaModel}--\ref{eq:Z_PLaw}).
Our nominal analysis projects these quantities, weighted (see Eqs.~\ref{eq:Sx} and \ref{eq:A_EW}) by the emissivity $j_x\simeq n^2 \Lambda(T,Z;\epsilon_a,\epsilon_b,z)$, and assuming an underlying prolate spheroidal geometry; the resulting model parameters are given in Tables~\ref{tab:jumps} and  \ref{tab:parameters}.
Other, simplified deprojection models
are examined in Appendix~\S\ref{app:deproj_models}, in order to test the robustness of our nominal results.
Overall, the results are found to be robust to the analysis method; the main differences lie in the inferred confidence levels.
We confirm that our results are consistent with previous deprojection studies of the A2204E, A2204W, A2142NW, and A3526SW sectors \citep[][]{SandersEtAl2005, SandersEtAl2016, ChenEtAl2017, WangEtAl2018}, which computed some of the CF parameters $\njump$, $\Tjump$, $\Zjump$, $\Pjump$, or $\Mydelta$.

As expected, as one crosses outside any of these CFs, the density abruptly drops while the temperature jumps (see Table~\ref{tab:jumps}).
The average density and temperature contrast values across the eight newly deprojected CFs are $\barnjump\equiv\langle n_i/n_o\rangle\sim 1.3$ and $\barTjump\equiv\langle T_o/T_i\rangle\sim1.4$, but these means are dominated by the superior  statistics in Centaurus.
We find large dispersions in $\njump$ and in $\Tjump$ among the different CFs in our sample.
A similarly large dispersion is found in the mean $q\equiv(\njump\Tjump)^{1/2}$, which provides a more robust estimate of the contrast for nearly-isobaric (see \S\ref{sec:PjumpsSec}) CFs.
(Note that in simplified spiral flow models \citep[{\UKspirals} and][]{ReissKeshet2015}, $\njump$ and $\Tjump$ are expected to differ and show some opposite radial trends.)
In contrast, the dispersion in $q$ within each GC is small (see \MyApJ{Figure}\MyMNRAS{Fig.}~\ref{fig:qnTr500}), with no evidence for a radial trend.
Statistical tests (Pearson's correlation test, TS test, and chi-squared test; see \S\ref{subsec:qTr500}) are indeed consistent with a null hypothesis of a constant contrast among all CFs inside each GC.
We thus propose that the CF contrast $q$ is a global property of each GC.
This GC contrast is shown (see \S\ref{subsec:qTM200}) to strongly ($>5\sigma$, see \MyApJ{Figure}\MyMNRAS{Fig.}~\ref{fig:qnTM200}) correlate with the cluster mass, with a power-law fit $q\propto M_{200}^{0.23\pm0.04}$.

We identify metallicity drops across CFs (see \S\ref{subsec:ZDrops}), established at a high ($\sim3\sigma$) significance for the first time (see Table~\ref{tab:jumps} and \MyApJ{Figure}\MyMNRAS{Fig.}~\ref{fig:qzr500}).
Across the A3526E CF, we find a highly significant, $\Zjump\equiv Z_i/Z_o=1.4_{-0.1}^{+0.2}$ metallicity drop, inconsistent with a continuous metallicity transition at the $\sim2.9\sigma$ confidence level.
The remaining seven newly deprojected CFs also suggest a metallicity drop, with an average value $\barZjump=1.1\pm0.1$, albeit at lower $\sim1.4\sigma$ confidence level.
The method is tested using a  control sample of mock CFs (see \MyApJ{Figure}\MyMNRAS{Fig.}~\ref{fig:mockqz}), found to be consistent with continuous metallicity profiles, revealing no bias in the measurement of $\Zjump$ and thus supporting our conclusions.

An extended spiral CF is thought to arise from large-scale flows in the GC, following a major perturbation such as a merger.
Assuming that the metallicity in a parcel of gas changes slowly, on a timescale longer than the dynamical timescale, the metallicity drop then indicates that the gas below (above) the CF originates from radii smaller (larger) than the CF radius.
Namely, the metallicity distribution follows the spiral pattern of the CF \citep[as was found in the Fornax cluster; \eg][]{SuEtal2017}.
The motions involved span nearly an order of magnitude in $r$, if one assumes a typical, $Z\propto r^{-0.3}$ \citep[][]{SandersonEtAl2009} metallicity profile in a relaxed cluster.
One might then expect to find correlations between $\Zjump$ and measures of the velocity (such as the shear parameter $\Mydelta$, see \S\ref{sec:delta}) or of the CF strength (such as the contrast $\njump$).
As shown in Appendix~\ref{app:Corl}, we are unable to establish or rule out such correlations using the present data.
For an analysis of the imprint of a spiral flow on the metallicity and entropy discontinuities, see Naor et al. (in preparation).

We find that the eight newly deprojected CFs are all consistent with a pressure jump ($\Pjump>1$; see \S\ref{sec:PjumpsUs}) across the CF (see Table~\ref{tab:jumps} and \MyApJ{Figure}\MyMNRAS{Fig.}~\ref{fig:xir500}).
The A3526W CF shows a particularly significant, $\Pjump=1.26\pm0.06$ pressure jump, inconsistent with an isobaric transition at the $\sim4.3\sigma$ confidence level.
The remaining seven CFs show a mean $\barPjump=1.06\pm0.03$ ($\sim1.9\sigma$) pressure jump.
Our control sample of mock CFs (see \MyApJ{Figure}\MyMNRAS{Fig.}~\ref{fig:mockxi}) is found to be consistent with isobaric transitions, thus disfavouring any bias in our analysis.
Measuring the pressure behavior is subtle, so we examine variations to our nominal analysis by testing different deprojection models that were used in the literature (see Appendix~ \ref{app:deproj_models}).
While the pressure jumps depend somewhat on the deprojection model, with $\Pjump$ changing by $\lesssim 30\%$ in each CF, all models are consistent with an average deprojected pressure jump.
We conclude that the inferred pressure jumps are robust, and unlikely to be systematic \MyMNRAS{artefacts}\MyApJ{artifacts} of our analysis, nor statistical fluctuations.

Our new deprojections are supplemented by a reanalysis of the thermal pressure contrast across CFs that were previously deprojected in the literature (see \MyApJ{Figure}\MyMNRAS{Fig.}~\ref{fig:xir500}).
We reanalyse the 17 CF sample of {\RK}, confirming a significant ($\sim5\sigma$, high but slightly lower than reported in {\RK}; see discussion in \S\ref{sec:PjumpsLit}) stacked pressure jump when using their assumptions.
Our nominal analysis differs from that of {\RK} in two ways.
First, we impose more stringent constraints on the data and deprojection quality, leading to a smaller sample of high-quality CFs from the literature, composed of seven CFs from the {\RK} sample supplemented by one additional CF.
Second, our assumption of a normally distributed $\ln(\Pjump)$ is more conservative than the normal $q_p^{-1}$ distribution assumed by {\RK}.
Consequently, we find a weighted mean $\barPjump=1.06\pm0.06$ for previously deprojected CFs, suggesting a pressure jump, but alone inconsistent with an isobaric CF at the $\sim1\sigma$ confidence level.
The weighted mean value of the joint sample of both newly deprojected and high-quality literature deprojected CFs, is $\barPjump\sim1.10\pm0.03$ ($\sim3.9\sigma$).
Modifying our assumptions yields similar, but typically more significant, pressure jumps.

We perform several sensitivity tests to examine the robustness of the deprojected pressure jumps.
These include tests of the sensitivity to biases (using control samples; see \S\ref{sec:PjumpsUs}), alternative assumptions on the deprojection weights (see Appendix~\S\ref{app:deproj_models}), different models for the underlying 3D density, temperature and metallicity profiles (see Appendix~\S\ref{app:deproj_models}), variations in the assumed parameters --- redshift, column density, CF radius, CF morphology (including best-fit spheres or ellipses with an unconstrained center, and different radii of curvature along the LOS), numbers of bins on each side of the CF, and bin size --- around their nominal values, tests of the stacking sensitivity to the presence of a few dominant CFs, and different methods of extrapolation to the CF radius.
Our main results, in particular the presence of pressure jumps across CFs, are not sensitive to reasonable variations in these assumptions.
For example, we consider different assumptions for the distribution of pressure values within their inferred uncertainty, including normal distributions of either $\ln(\Pjump)$ (nominal), $\Pjump$, or $\xi\equiv\Pjump^{-1}$.
We show (in \S\ref{sec:PjumpsLit}) that the inferred pressure jumps depend only weakly on these assumptions, that the detection of the pressure jump is robust, and that our nominal assumption is conservative.

Previous studies of spiral-type CFs that were not properly deprojected, typically reported pressure drops ($\PjumpP<1$), as opposed to our deprojected pressure jumps ($\Pjump>1$).
We critically test the effect of projection (see \S\ref{sec:PjumpsProj}), and show that it usually mistakes an actual pressure jump for an apparent pressure drop (see \MyApJ{Figure}\MyMNRAS{Fig.}~\ref{fig:xiproj}).
This effect is tested both by projecting our newly-deprojected CFs, and by examining projected CFs from the literature.
We confirm that $\PjumpP<\Pjump$ for individual CFs, in the sub-sample where both projected and deprojected profiles are available.
We find that a CF typically shows a projected $\sim10\%$ pressure drop and an opposite, $\gtrsim 10\%$ deprojected pressure jump.
On average, we find $\PjumpPbar=0.89\pm0.01$ (inconsistent with an isobaric transition at the $\sim9\sigma$ confidence level) for our joint sample of projected CFs.
Projection can lead to a $\sim 40\%$ reduction in the value of $\Pjump$ (see discussion in \S\ref{sec:PjumpsProj}), so this projected average is consistent with the mean $\barPjump\sim1.10\pm0.03$ obtained after deprojection.

Our deprojected $1.1\lesssim\Pjump\lesssim1.3$ values indicate an excess of non-thermal pressure just below spiral CF discontinuities, at a level $\sim 10\%-30\%$ of equipartition (see \S\ref{subsec:PjumpsPic}).
As discussed in {\RK}, some of this non-thermal pressure could possibly be attributed to poorly-constrained components of the ICM, such as turbulence and perhaps cosmic-rays.
However, the non-thermal pressure must be predominantly magnetic, in order to maintain the CF thinness \citep[much narrower than the Coulomb mean free path; see][]{VikhlininEtAl2001} and its general stability against Kelvin-Helmholtz instabilities.
Such magnetisation in indeed expected to arise from tangential shear flows, which were inferred from hydrostatic mass jumps across spiral CFs ({\UK}).

We compute the shear parameter $\Mydelta$ for each of the $21$ deprojected CFs in our sample (see \S\ref{subsubsec:deltaResults}), presented in Fig.~\ref{fig:deltar500}.
This parameter, defined in Eq.~(\ref{eq:Upsilon}), is closely related to the discontinuity in the power-law index $d\ln \PTh/d\ln R$ of the radial pressure profile.
All analysed CFs are found to be consistent with a jump in hydrostatic mass across the CF, $\MHEo>\MHEi$, such that $\Mydelta>0$.
On average, we obtain mean $\Mydelta$ values of $0.57\pm0.03$ (for our eight newly deprojected CFs), $0.32\pm0.10$ (for the eight high-quality CFs out of the 13 deprojected CF from the literature; see \S\ref{subsec:DeptojCFsLit}), and $0.55\pm0.03$ (for the joint sample), all highly inconsistent with a continuous $M_h$.
Control samples of mock CFs (see \MyApJ{Figure}\MyMNRAS{Fig.}~\ref{fig:mockdelta}) indicates that the measurement is robust against a putative bias in our $\Mydelta$ measurement.

The large $\Mydelta$ values we infer can be naturally explained only if a nearly sonic, tangential bulk flow is present below each CF, with a mean Mach number $\bar{\MyMach}_i\simeq \langle\Mydelta^{1/2}\rangle= 0.76\pm0.02$, consistent with the conclusions of \UK.
Such a picture directly derives from a spherical CF approximation (see \S\ref{subsubsec:ThermalEqb}), but remains the most plausible interpretation also for the more realistic, rotating spiral CF manifold inferred from simulations (see \S\ref{subsubsec:shearflows}), provided that $\MyMach$ is measured in the inertial frame.
To this end, we model the three-dimensional CF discontinuity as $r(\theta,\phi)=f(\phi)$, and use the CF morphology to show that the spiral is tight, $f'(\phi)/f\lesssim \tan(\alpha_{\tiny\mbox{proj}})\simeq 0.2\ll1$; see Table~\ref{tab:parameters}.

A fast flow below the CF can quickly reach high, near equipartition magnetisation levels.
To quantify this, consider a simple model with a cylindrical CF and an azimuthal velocity profile $\vect{u}=\omega(\varrho)\varrho\,\unit{\phi}$.
Assume some preexisting, weak magnetic field $\tilde{\vect{B}}$, with a radial component $\tilde{B}_\varrho$ corresponding to an equipartition fraction $\tilde{\eta}_B$.
Equations (25) and (26) of \citet{NaorKeshet2015} directly yield the field after time $t$,
\begin{equation}\label{eq:Bcf}
\bm{B}=\tilde{\bm{B}}+\tilde{B}_\varrho\, \omega'(\varrho) \varrho\,t\,\unit{\phi}\, \fin
\end{equation}
Assuming that the flow is confined to a layer of width $\Delta<\varrho$ below the CF, the initially radial component is amplified by the shear, giving a dominant azimuthal component,
\begin{equation}\label{eq:Bcf_deltarho}
B_\phi\sim\tilde{B}_\varrho\frac{\myu_i  t}{\Delta}\fin
\end{equation}
The time required to stretch the field to the equipartition value $\eta_B\simeq (\Pjump-1)/\Pjump$ inferred from the CF analysis may then be estimated as
\begin{eqnarray}
t & \sim & \frac{\Delta}{\myu_i}\frac{B_\phi}{\tilde{B}}
\sim \frac{\Delta}{\myu_i}\sqrt{\frac{\Pjump-1}{\Pjump \tilde{\eta}_B}} \\
& \sim & 100\left(\frac{\Delta}{10\kpc}\right)\left(\frac{\tilde{\eta}_B}{10^{-3}}\right)^{-\frac{1}{2}}\Myr \coma \nonumber
\end{eqnarray}
where we adopted the typical values inferred above for $u_i$ and $\Pjump$.
This timescale, much shorter than the estimated $\gtrsim \Gyr$ age of the spiral structure, suggests that shear magnetization can rapidly generate the $\eta_B\simeq 0.1$ fields we infer below CFs, and that this magnetization is saturated.

A spiral flow could lead to correlations between the different measures of the deprojected CFs, namely $\njump$, $\Tjump$, $\Zjump$, $\Pjump$, and $\Mydelta$.
For instance, one might expect $\Mydelta$ and $\Pjump$ to be correlated, if the former gauges the flows giving rise to the nonthermal pressure measured by the latter.
The various available correlations are computed in Appendix~\ref{app:Corl} for our deprojected CF sample (see Table \ref{tab:nTZ_Crl_delta}), and calibrated against a control sample of mock clusters (see Table \ref{tab:nTZ_Crl_delta_mock}).
A strong correlation is found between $\njump$ and $\Tjump$, as expected in a CF, but no other significant correlation is identified, possibly due to the substantial statistical uncertainties.
Future observations, with better statistics, may be able to identify some underlying correlations.

In particular, we find no correlation between $\Pjump$ and the other thermal parameters, including the shear flow parameter $\Mydelta$ (see Appendix~\ref{app:Corl}).
We also find no significant radial dependence of the pressure jump, which appears to be consistent with a constant value, $\Pjump\sim1.1$, among all CFs (see \S\ref{subsec:PjumpsPic}).
These results, suggesting a roughly constant magnetic equipartition value, $\eta_{\tiny \mbox{B}}\simeq0.1$, below all CFs, are consistent with the aforementioned picture of saturated magnetisation.
Interestingly, magnetic saturation at such $\eta_{\tiny \mbox{B}}\simeq0.1$ levels is sometimes inferred in radio halos and relics in GCs \citep[][]{Keshet2010}, from observations of radio-bright regions in the ICM.

In summary, the evidence presented above supports the notion that GCs harbor extended, magnetised, spiral bulk flows.
Future, high resolution studies may be able to unravel the geometry of the CF discontinuity surface, and measure the orientation, width, and structure of the tangential flow layers.
However, our results suggest that systematic uncertainties at the $\sim5\%$ level (see Appendix \ref{app:deproj_models}), associated with the deprojection method, will limit such studies.
It may be necessary to develop a dedicated spiral deprojection routine in order to facilitate future progress.

\MyApJ{\acknowledgements}
	\MyMNRAS{\section*{Acknowledgements}}
We are greatly indebted to M. Markevitch for his extensive help, patience, and hospitality.
We thank K. Arnaud for helpful discussions. This research was supported by the Israel Science Foundation (grant No. 1769/15), has received funding from the IAEC-UPBC joint research foundation (grants No. 257/14 and 300/18) and from the European Union Seventh Framework Programme (FP7/2007-2013) under grant agreement n\textordmasculine ~293975.

\MyMNRAS{
\begin{landscape}
\begin{table}
	\caption{Modelled and measured parameters.} 
	\centering 
	\setlength{\tabcolsep}{0.156em} 
	{\renewcommand{\arraystretch}{1.5}
		\begin{tabular}{| c |c |c| c| c| c| c| c| c| c| c|c|c| c |c |} 
			\hline 
			GC& Region & $n_o[10^{-3}\mbox{cm}^{-3}]$& $T_i[\mbox{keV}]$ &$Z_o[Z_{\sun}]$ & $\rcf[\mbox{kpc}]$&  $\beta_i$ (or $\alphani$)&$\beta_o$ (or $\alphano$)& $\rcorei[\mbox{kpc}]$ &$\rcoreo[\mbox{kpc}]$& $\alphaTi$ &$\alphaTo$& $\alphaZi$ &$\alphaZo$&$\alpha_{\tiny\mbox{proj}}$\\
			(1) & (2) & (3) & (4) & (5) & (6) & (7)  & (8)& (9)& (10)& (11)& (12)& (13)& (14)& (15)\\ 
			\hline 
			\multirow{ 1}{*}{A2029}& SW &$31.6\pm3$&$4\pm3$&$0.3^{+0.7}_{-0.3}$&$31.8\pm0.9$&($-0.6\pm0.3$)&$0.41\pm0.04$&$-$&$36\pm10$&$0.0\pm1.2$&$0.0\pm0.2$&$0.4\pm2.2$&$0.2\pm1.5$&$42^\circ$\\
			\hline 
				\multirow{ 1}{*}{A2142}& NW &$1.8\pm0.1$&$6.4\pm0.9$&$0.3^{+1.1}_{-0.3}$&$343.8\pm0.3$&($-0.93\pm0.05$)&$(-1.4\pm0.1)$&$-$&$-$&$-0.4\pm0.3$&$0.0\pm1.1$&$-0.3\pm1.2$&$-0.4\pm7.2$&$8^\circ$\\
			\hline 
			\multirow{ 4}{*}{A2204}
			& S &$40\pm6$&$2\pm2$&$2\pm2$&$23.4\pm0.8$&$7\pm162$&$(-1.2\pm0.1)$&$76^{+934}_{-76}$&$-$&$-0.2\pm1.3$&$0.2\pm0.6$&$0.5\pm7.4$&$-1\pm2$&$8^\circ$\\
			& E &$27\pm5$&$4\pm1$&$0.3^{+0.9}_{-0.3}$&$30.8\pm0.6$&$0.3\pm0.1$&$0.38\pm0.08$&$6^{+7}_{-6}$&$23^{+26}_{-23}$&$0.6\pm0.1$&$0.3\pm0.5$&$1\pm1$&$0.4\pm2.2$&$11^\circ$\\
			& N &$23\pm4$&$4^{+5}_{-4}$&$0.4^{+0.9}_{-0.4}$&$39.0\pm1.0$&$1\pm3$&$0.45\pm0.05$&$36^{+81}_{-36
		}$&$20^{+23}_{-20}$&$0.5\pm2.0$&$0.1\pm0.5$&$0\pm4$&$-0.2\pm2.0$&$31^\circ$\\
			& W &$11.3\pm0.7$&$6\pm2$&$0.2^{+0.4}_{-0.2}$&$71.4\pm1.0$&$(-1.0\pm0.2)$&$0.59\pm0.02$&$-$&$99\pm10$&$0.4\pm0.6$&$-0.1\pm0.2$&$0.1\pm2.3$&$0.2\pm1.4$&$10^\circ$\\
			\hline 
			\multirow{2}{*}{Centaurus} & SW&$5.1\pm0.2$&$2.6\pm0.2$&$1.3\pm0.3$&$40.9\pm0.1$&($-0.8\pm0.2$)&($-1.10\pm0.05$)&$-$&$-$&$-0.3\pm0.4$&$-0.2\pm0.2$&$-0.07\pm0.87$&$-2.0\pm0.6$&$12^\circ$\\
			& E&$3.74\pm0.08$&$3.6\pm0.3$&$0.6\pm0.2$&$79.8\pm0.4$&($-0.54\pm0.07$)&($-1.05\pm0.02$)&$-$&$-$&$-0.05\pm0.17$&$-0.24\pm0.08$&$-0.3\pm0.5$&$-1.3\pm0.5$&$13^\circ$\\
			\hline 
		\end{tabular}}\label{tab:parameters}\vspace{0.2cm}
		\begin{tablenotes}
			\item	Columns: (1) The GC name; (2) The \MyMNRAS{analysed}\MyApJ{analyzed} sector, see \MyApJ{Figures}\MyMNRAS{Figs.}~\ref{fig:A2029Figs}, \ref{fig:A2142Figs}, \ref{fig:A2204Figs}, and \ref{fig:A3526Figs}; (3) The electron number density just above the CF; (4) The temperature just below the CF; (5) The metallicity just above the CF; (6) The distance of the CF from the \MyMNRAS{centre}\MyApJ{center} of the sector; (7) The beta parameter of the $\beta$-model (or, if in brackets, the power-law slope) of the density profile below the CF; (8) Same as (7) but above the CF; (9) The $r_c$ core-radius parameter of the $\beta$-model below the CF; (10) Same as (9) but above the CF; (11) The power-law slope of the temperature below the CF; (12) The power-law slope of the temperature above the CF; (13) The power-law slope of the metallicity below the CF; (14) The power-law slope of the metallicity above the CF; (15) The angle between the CF and the azimuthal direction on the plane of the sky. The error bars are the $1\sigma$ multi-variant confidence levels.
		\end{tablenotes}
		\end{table}
	\end{landscape}}

\appendix

\section{Deprojection robustness}
\label{app:deproj_models}

Various sensitivity and robustness tests were outlined in \S\ref{sec:Discussion}, examining the validity of our nominal deprojection analysis method described in \S\ref{sec:Spectral Analysis}.
Here we demonstrate such tests by examining a few variations to the deprojection method which may be found in the literature.

First, we examine alternative choices of the deprojection weights, in particular replacing the interpolation over the tabulated cooling function $\Lambda$ used in the nominal method.
Simple variants tested include: (i) replacing $\Lambda$ by the emission measure cooling function $\Lambda=\const$ \citep[\eg][]{MarkevitchEtAl2001,VikhlininEtAl2001,HallmanMarkevitch2004,TanakaEtAl2006,JohnsonEtAl2010}; (ii) replacing $\Lambda$ by the spectroscopic-like cooling function $\Lambda=\Lambda(T)\propto T^{-3/4}$ \citep{MazzottaEtAl2004}; and (iii) using a mixed approach \citep[\eg][]{MazzottaEtAl2011,BourdinEtal2013,GiacintucciEtAl2017,WangEtAl2018}, where we use the emission measure cooling function for $\Sx$ (\ie $\Lambda=\const$ in $j_x$ in Eq.~\ref{eq:Sx}), and the spectroscopic-like cooling function for the projected temperature and metallicity (\ie $\Lambda\propto T^{-3/4}$ in $W_T$ and $W_Z$ in Eq.~\ref{eq:A_EW}).

Another common deprojection variant found in the literature is to ignore the metallicity drop at the CF.
We thus retain our nominal weights, but adopt a single beta-model for the metallicity profile across the CF \citep[\eg][]{PizzolatoEtAl2003,SiegelEtAl2018}, replacing Eq.~(\ref{eq:Z_PLaw}) by
\begin{equation}
Z(r)=Z_{cf}\left(\frac{r_z^2+r^2}{r_z^2+\rcf^2}\right)^{-\beta_z}\fin
\end{equation}
Here, $r_z$ and $\beta_z$ are free parameters analogous to $r_c$ and $\beta$.

\MyApJ{Figure}\MyMNRAS{Figure}~\ref{fig:AllModels} presents the thermal pressure contrast $\Pjump$, plotted against the shear parameter $\Mydelta$, for each of the eight newly deprojected CFs.
The results are shown both in the nominal method (large symbols) and in its variants outlined above (smaller symbols).
Overall, the different deprojection variants lead to small changes in $\Pjump$, and very small changes in $\Mydelta$.
Two variants show persistent trends: $\Pjump$ values slightly larger than those of the nominal model are obtained when using the spectroscopic-like $\Lambda$, and vice versa when using the mixed model.
The A3526SW and A2204S CFs are not shown for the $\Lambda\propto T^{-3/4}$ fits, as they are invalid for $T<3\keV$ regions \citep[see][]{MazzottaEtAl2004}.

\MyMNRAS{
\begin{figure}
\centering
\DrawFig{\includegraphics[width=9.1cm]{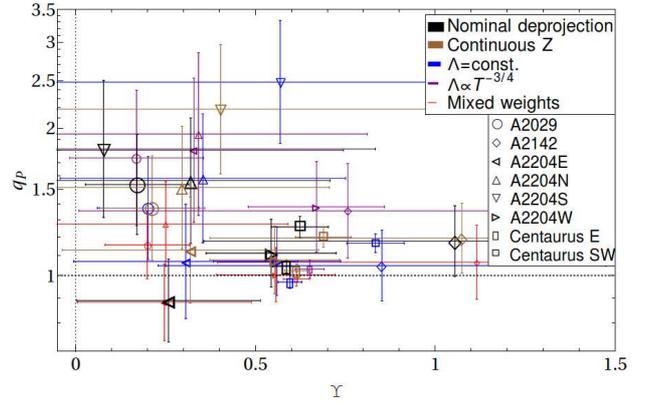}}
\caption{
The thermal pressure ratio, $\Pjump$, plotted against the shear parameter, $\Mydelta$, for the eight newly deprojected CFs, computed in different deprojection method variants.
Our nominal deprojection method (see \S\ref{sec:Spectral Analysis}) is shown in very large black symbols.
Other variants include a continuous metallicity model (large brown symbols), $\Lambda\propto T^{-3/4}$ (medium-large purple), emission-measure cooling function $\Lambda=\const$ (medium-small blue), and mixed weights ($j_x\propto T^0$ and $W_T=W_Z\propto T^{-3/4}$, small red symbols).
Most of the CFs suggest a pressure jump ($\Pjump>1$) and an inside flow ($\Mydelta>0$), delineated by dotted lines.
}
\label{fig:AllModels}
\end{figure}
}

As the figure shows, in all deprojection variants there is strong evidence for a fast inside flow ($\Mydelta>0$), and most of them indicate pressure jumps ($\Pjump>1$).
The weighted-average $\barPjump$ and $\barMydelta$ values for each of the four (non nominal) variants are $\barPjump=1.07\pm0.04$ and $\barMydelta=0.63\pm0.04$ for the spectroscopic-like model, $\barPjump=1.04\pm0.03$ and $\barMydelta=0.61\pm0.03$ for the emission-measure model, $\barPjump=1.00\pm0.03$ and $\barMydelta=0.57\pm0.03$ for the mixed approach, and $\barPjump=1.08\pm0.03$ and $\barMydelta=0.60\pm0.03$ for the continues $Z$ model.
These results should be compared to those of the nominal method, $\barPjump=1.11\pm0.03$ and $\barMydelta=0.57\pm0.03$.
We conclude that indeed, the shear parameter is not sensitive to the deprojection method and its errors.
The results indicate that the identification of pressure jumps across CFs does not crucially depend on the deprojection method.

Overall, we find that $\barPjump$ shows a weak dependence upon the model variant, changing the magnitude of the inferred pressure jump but not its overall presence.
Note that the variants of our nominal method seemingly yield smaller $\Pjump$ values, with the mixed approach showing no evidence for a pressure jump.
However, this behavior can be shown to arise in part due to the exclusion of the highly significant A3526SW CF from some variants, as mentioned above, and in part due to the less physical assumptions underlying these model variants.
We conclude that the identification of pressure jumps is not a deprojection \MyMNRAS{artefact}\MyApJ{artifact}, but the inferred magnitude of the jump, as a $\sim10\%$ effect, is sensitive to the underlying assumptions.

Finally, we test our nominal deprojection method by comparing it to the commonly used XSPEC deprojection routine PROJCT.
PROJCT does not attempt to fit the projected spectrum in each bin with some effective physical parameters $n$, $T$, and $Z$.
Instead, it produces a synthetic emissivity function for each 3D shell, given the mean physical parameters in that shell.
The projected spectrum in each bin is then fitted with the volume-weighted sum over all shells, by simultaneously adjusting the physical parameters in all shells.

PROJCT thus has the advantage of avoiding the contamination of spectra by inter-bin temperature variations along the line of sight, but has some disadvantages.
First, it does not utilise the superior angular resolution one can obtain in $\Sx$.
Second, for a multi-phase gas PROJCT is sensitive to changes in the radial binning (dropping bins from either ends of the region, altering bin size, etc.) and tends to produce spurious temperature oscillations \citep[\eg][]{FabianEtAl2006,SandersFabian2007,RussellEtal2008}.

\MyMNRAS{\begin{figure*}
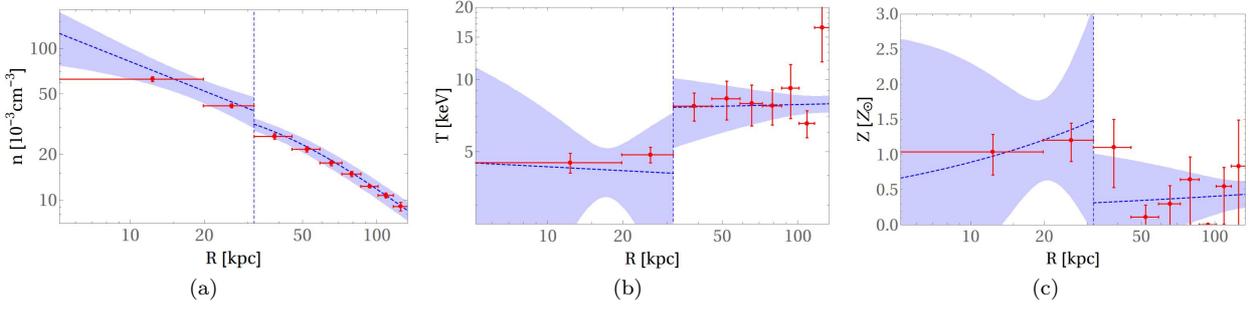

\centering
\subfigure[]
{
\DrawFig{\includegraphics[width=2.095in]{\myfig{A2029PROJCTn.eps}}}
\label{fig:A2029PROJCTn}
}
\subfigure[]
{
\DrawFig{\includegraphics[width=2.065in]{\myfig{A2029PROJCTT.eps}}}
\label{fig:A2029PROJCTT}
}
\subfigure[]
{
\DrawFig{\includegraphics[width=2.025in]{\myfig{A2029PROJCTZ.eps}}}
\label{fig:A2029PROJCTZ}
}
\caption{Radial profiles deprojected in the A2029 sector, both in the nominal method (dashed blue with shaded region as the $1\sigma$ confidence interval, identical to Fig.~\ref{fig:A2029Thermal}) and when using PROJCT (red disks with vertical error bars).
}\label{fig:A2029PROJCT}
\end{figure*}}

PROJCT should work well in relaxed GCs, where a single-phase approximation is good.
As A2029 is one of the most relaxed GCs known to date, we use it to compare the nominal deprojection method with PROJCT, utilising the same $\TtwoD$ and $\ZtwoD$ bins.
Figure \ref{fig:A2029PROJCT} compares the continuous $n$, $T$, and $Z$ profiles of our nominal deprojection (blue dashed lines) with the binned $n$, $T$, and $Z$ profiles derived using PROJCT.
The two methods are found to be in fairly good agreement, although some deviations are seen near the CF.
Extrapolating the PROJCT profiles to the CF radius, using the same models adopted in the nominal method, yields $\Pjump=1.3\pm0.2$ and $\Mydelta=0.3\pm0.3$.
These results, again suggesting a pressure jump and a fast flow below the CF, are consistent with the nominal results, $\Pjump=1.5^{+0.4}_{-0.3}$ and $\Mydelta=0.2\pm0.1$, obtained for this CF.

In the other GCs --- A2142, A2204, and Centaurus --- we obtain nonphysical results when using PROJCT.
In Centaurus, PROJCT produces large oscillations in both analysed sectors, giving poorly constrained CF contrast parameters.
In A2142, PROJCT yields profiles that are in reasonable agreement with our nominal profiles, but appears to overestimate the density just above the CF. This seemingly leads to a large pressure jump, which we dismiss because it corresponds to a nonphysical CF density jump, instead of the expected drop.
In A2204, PROJCT yields very low temperatures near the CF, in all four sectors, far colder than in the corresponding projected temperature. These low temperatures, found on both sides of the CFs, are related in all but the south sector with strong $T$ gradients in the CF vicinity, which seem non-physical.
Moreover, in the A2204W and A2204E sectors, PROJCT yields nonphysical temperature drops across the CF.

In conclusion, we find that special care must be taken when using PROJCT. In cases where PROJCT produces reasonable results, it is in good agreement with our nominal deprojection method.

\section{Cross Correlations}\label{app:Corl}

It is interesting to examine the cross-correlations between the different CF parameters derived in the analysis.
We thus consider the parameters $\ln(\njump)$, $\ln(\Tjump)$, $\ln(\Zjump)$, $\ln(\Pjump)$, and $\Mydelta$, the latter taken without a logarithm as it is the difference, rather than the ratio, between properties above and below the CF.
For each pair of parameters, we carry out two correlation tests: Pearson's correlation test and a linear fit.
Table~\ref{tab:nTZ_Crl_delta} presents the results for the newly deprojected, eight-CF sample.
For each pair of quantities, the table lists the Pearson's correlation coefficient $\bar{\corl}$, its value in Fisher's $\zeta$-space, $\bar{\zeta}$ (see \S\ref{subsec:qTr500}), the slope $b$ of the best linear fit, and its goodness of fit per DOF $\chi^2_{\nu}$; error estimates are computed as described in \S\ref{subsec:qTr500}.
In order to identify systematic biases, the same analysis is applied to the control, 50-CF sample (see \S\ref{subsec:ZDrops}); the results are displayed in Table~\ref{tab:nTZ_Crl_delta_mock}.

\MyApJ{\begin{table}[h]}
\MyMNRAS{\begin{table}}
	\caption{Correlations between the thermal transitions across the eight newly deprojected CFs}
	\centering 
	\setlength{\tabcolsep}{0.11em} 
	{\renewcommand{\arraystretch}{1.5}%
		\begin{tabular}{| c| c | c |c |c |c |} 
				\cline{3-6} 
			 \multicolumn{2}{c|}{}&$\ln(\njump)$ & $\ln(\Tjump)$ & $\ln(\Zjump)$& $\ln(\Pjump)$\\
			\hline 
			\multirow{4}{*}{$\Mydelta$}&$\bar{\corl}$&$0.24^{+0.29}_{-0.34}$ &$-0.05^{+0.40}_{-0.39}$&$-0.12^{+0.48}_{-0.43}$&$-0.22^{+0.45}_{-0.38}$\\
			&$\bar{\zeta}$& $0.25\pm0.34$ & $-0.05\pm0.42$ & $-0.12\pm0.50$ & $-0.22\pm0.46$\\
			&$\chi^2_{\nu}$& $2.2$& $1.5$ & $0.5$ & $1.3$\\
			&$b$ &$-4^{+9}_{-13}$& $-3^{+2}_{-2}$& $-3^{+2}_{-3}$&$-2^{+1}_{-2}$\\
			\hline
		     \multirow{4}{*}{$\ln(\Pjump)$}&$\bar{\corl}$&$-0.39^{+0.28}_{-0.22}$&$0.32^{+0.31}_{-0.39}$&$-0.15^{+0.45}_{-0.40}$\\
		    &$\bar{\zeta}$& $-0.41\pm0.30$& $0.33\pm0.40$ & $-0.15\pm0.46$ \\	
		    &$\chi^2_{\nu}$&$2.8$& $1.0$& $1.2$\\
		   &$b$&$4^{+5}_{-5}$&$2^{+1}_{-1}$&$-2^{+2}_{-2}$\\
		    \cline{1-5}
		   \multirow{4}{*}{$\ln(\Zjump)$}&$\bar{\corl}$&$0.09^{+0.34}_{-0.36}$&$-0.08^{+0.37}_{-0.35}$\\
		    &$\bar{\zeta}$& $0.09\pm0.36$ & $-0.08\pm0.38$\\
		    	&$\chi^2_{\nu}$ &$1.8$ &$1.6$ \\
		    &$b$&$-3^{+9}_{-10}$ & $-2^{+2}_{-2}$ \\
			\cline{1-4} 
			\multirow{4}{*}{$\ln(\Tjump)$}&$\bar{\corl}$&$0.32^{+0.28}_{-0.34}$\\
			&$\bar{\zeta}$& $0.33\pm0.36$\\
				&$\chi^2_{\nu}$ &$2.6$\\
			&$b$ &$0.7^{+0.2}_{-0.2}$\\
			\cline{1-3} 
		\end{tabular}}\label{tab:nTZ_Crl_delta}\vspace{0.2cm}
		\begin{tablenotes}
			\item	
	 For each of the quantity rows, we list different measures of correlation:
     \textbf{First sub-row} --- Pearson correlation;
	 \textbf{Second sub-row} --- the correlation coefficient in Fisher's $\zeta$-space;
	 \textbf{Third sub-row} --- the reduced chi square value of the best fit;
     \textbf{Fourth sub-row} --- the best fitted slope.
	 In the best fit the parameter to the left is regarded as the horizontal axis and the upper as the vertical axis.
	 \end{tablenotes}\vspace{0.2cm}
	\end{table}

\MyApJ{\begin{table}[h]}
\MyMNRAS{\begin{table}}
	\caption{Correlations between the thermal transitions across the $50$ CFs of the control sample}
	\centering 
	\setlength{\tabcolsep}{0.15em} 
	{\renewcommand{\arraystretch}{1.5}%
		\begin{tabular}{| c| c | c |c |c |c |} 
				\cline{3-6} 
			 \multicolumn{2}{c|}{}&$\ln(\njump)$ & $\ln(\Tjump)$ & $\ln(\Zjump)$& $\ln(\Pjump)$\\
			\hline 
			\multirow{4}{*}{$\Mydelta$}&$\bar{\corl}$&$-0.11^{+0.10}_{-0.10}$ &$-0.09^{+0.10}_{-0.10}$&$0.05^{+0.11}_{-0.11}$&$-0.03^{+0.10}_{-0.10}$\\
			&$\bar{\zeta}$& $-0.11\pm0.10$ & $-0.09\pm0.10$ & $0.05\pm0.11$ & $-0.03\pm0.10$\\
			&$\chi^2_{\nu}$& $4.2$ & $3.2$ & $2.8$ & $2.9$\\
			&$b$ &$-0.41^{+0.07}_{-0.07}$ & $0.10^{+0.06}_{-0.06}$& $0.69^{+0.25}_{-0.25}$&$0.11^{+0.02}_{-0.02}$\\
			\hline
	    \multirow{4}{*}{$\ln(\Pjump)$}&$\bar{\corl}$&$-0.41^{+0.10}_{-0.10}$&$0.53^{+0.14}_{-0.17}$&$0.11^{+0.10}_{-0.10}$\\
		    &$\bar{\zeta}$& $-0.43\pm0.12$ & $0.59\pm0.21$ & $0.11\pm0.10$\\
		    &$\chi^2_{\nu}$&$3.6$ & $1.2$ & $2.8$\\
		   &$b$ &$-0.43^{+0.02}_{-0.02}$ & $0.80^{+0.21}_{-0.21}$& $0.75^{+0.24}_{-0.24}$\\
		    \cline{1-5}
		  \multirow{4}{*}{$\ln(\Zjump)$}&$\bar{\corl}$&$-0.24^{+0.12}_{-0.11}$&$0.05^{+0.11}_{-0.11}$\\
		    &$\bar{\zeta}$& $-0.25\pm0.12$ & $0.05\pm0.11$\\
		    	&$\chi^2_{\nu}$ &$2.7$ &$2.9$ \\
		    &$b$  &$-0.31^{+0.09}_{-0.10}$ & $0.10^{+0.05}_{-0.05}$ \\
			\cline{1-4} 
			\multirow{4}{*}{$\ln(\Tjump)$}&$\bar{\corl}$&$-0.17^{+0.13}_{-0.12}$\\
			&$\bar{\zeta}$& $-0.17\pm0.13$\\
				&$\chi^2_{\nu}$ &$2.9$\\
			&$b$ &$-0.24^{+0.02}_{-0.02}$\\
			\cline{1-3} 
		\end{tabular}}\label{tab:nTZ_Crl_delta_mock}\vspace{0.2cm}
		\begin{tablenotes}
			\item	
Rows: same as Table~\ref{tab:nTZ_Crl_delta}.
	\end{tablenotes}\vspace{0.2cm}
	\end{table}

The strongest correlation seen in Table \ref{tab:nTZ_Crl_delta}, especially when compared with the mock CF Table \ref{tab:nTZ_Crl_delta_mock},
is the rather strong, positive correlation between $\njump$ and $\Tjump$, with a significantly positive slope of $b=0.7\pm0.2$.
Such a correlation is the defining property of CFs: a large density drop, associated with a comparably large temperature jump.
However, as the tables show, the results are quite noisy; even this $\njump$--$\Tjump$ correlation presents at only a marginal significance level according to its $\bar{\zeta}$ uncertainty.
No significant correlations are found even when $\njump$ and $\Tjump$ are replaced by the mean contrast parameter, $q\equiv(\njump\Tjump)^{1/2}$ (see \S\ref{subsec:qTr500}).

Comparing the two tables suggests a possible positive correlation between $\njump$ and $\Mydelta$, anticipated if a strong shear is needed to sustain a large contrast, but the confidence is very low.
There is a similar hint of a positive correlation between $\njump$ and $\Zjump$, anticipated if the dense phase below the CF originated from the GC centre, but the significance is again very low.
Better data is needed in order to test such correlations.

\bibliography{\mybib{Pjump}}
	
\label{lastpage}	

\end{document}